\documentclass[10pt]{article}

\pdfoutput=1

\usepackage[latin1]{inputenc}
\usepackage{mystyle}
\usepackage[bookmarks,bookmarksdepth=2, colorlinks=true, linkcolor=blue,citecolor=red, urlcolor=blue]{hyperref}
\usepackage{comment}

\usepackage{mystylebis}
\usepackage{enumitem}
\usepackage{mathtools}

\usepackage{fullpage}

\title{The geometry of off-the-grid compressed sensing} 

\author{%
Clarice Poon\footnote{Department of Mathematical Sciences, University of Bath, \url{cmhsp20@bath.ac.uk}},
\quad%
Nicolas Keriven\footnote{CNRS and GIPSA-lab, Univ. Grenoble Alpes, Grenoble INP \url{nicolas.keriven@gipsa-lab.grenoble-inp.fr}}
 \quad%
Gabriel Peyr\'e\footnote{CNRS and DMA, Ecole Normale Sup\'erieure, 45 rue d'Ulm, F-75230 PARIS cedex 05, FRANCE, \url{gabriel.peyre@ens.fr} }%
}

\begin{document}

\maketitle

\begin{abstract}
	This paper presents a sharp geometric analysis of the recovery performance of sparse regularization. More specifically, we analyze the BLASSO method which estimates a sparse measure (sum of Dirac masses) from randomized sub-sampled measurements. This is a ``continuous'', often called off-the-grid, extension of the compressed sensing problem, where the $\ell^1$ norm is replaced by the total variation of  measures. This extension is appealing from a numerical perspective because it avoids to discretize the the space by some grid. But more importantly, it makes explicit the geometry of the problem since the positions of the Diracs can now freely move over the parameter space. 
On a methodological level, our contribution is to propose the Fisher geodesic distance on this parameter space as the canonical metric to analyze super-resolution performances in a way which is invariant to reparameterization of this space. While previous works express recovery conditions using a flat Euclidean distance, switching to the Fisher metric allows us to take into account measurement operators which are not translation invariant, which is crucial for applications such as Laplace inversion in imaging, Gaussian mixtures estimation and training of multilayer perceptrons with one hidden layer.      
On a theoretical level, our main contribution shows that if the Fisher distance between spikes is larger than a Rayleigh separation constant, then the BLASSO recovers in a stable way a stream of Diracs, provided that the number of measurements is proportional (up to log factors) to the number of Diracs. We measure the stability using an optimal transport distance constructed on top of the Fisher geodesic distance. Our result is (up to log factor) sharp and does not require any randomness assumption on the amplitudes of the underlying measure. Our proof technique relies on an infinite-dimensional extension of the so-called ``golfing scheme'' which operates over the space of measures and is of general interest. 
\end{abstract}




\section{Introduction}

Sparse regularization, and in particular convex approaches based on $\ell^1$ minimization, is one of the workhorses to  ill-posed linear  inverse models. It finds numerous applications ranging from signal processing~\cite{chen-atomic1998} to machine learning~\cite{tibshirani1996regression}. In many situations, it makes sense to consider a ``continuous'' counterpart to $\ell^1$ minimization, which avoids gridding the parameter space, thus enabling more efficient solvers and a sharper theoretical analysis. 
The most natural continuous extension encodes the positions and amplitudes of the sought after solution into a Radon measure, so that the $\ell^1$ norm is replaced by the total variation (total mass) of the measure. A measure is then naturally said to be ``sparse'' when it is a \emph{sum of Diracs} at the desired positions and amplitudes. The corresponding infinite dimensional optimization problem is called BLASSO in~\cite{deCastro-exact2012} following theoretical works on spectral extrapolation~\cite{beurling1938integrales}. This setting of optimization on measures has also been considered in the inverse problem community~\cite{bredies-inverse2013}. 
Successful examples of applications of such ``off-the-grid methods'' include single-molecule fluorescent imaging~\cite{boyd2017alternating}, spikes sorting in neurosciences~\cite{ekanadham2014unified}, mixture model estimation~\cite{gribonval2017compressive} and training shallow neural networks~\cite{bach2017breaking}.

\subsection{Sparse spikes recovery using the BLASSO}

\paragraph{Observation model.}

We consider the general problem of estimating a complex-valued unknown Radon measure $\mu_0 \in \Mm(\Xx)$ defined over some metric space $\Xx$ from a small number $m$ of randomized linear observations $y \in \CC^m$. In this paper, $\Xx$ will either be a connected bounded open subset of $\RR^d$ or the $d$-dimensional torus $\TT^d$, even though some of our results extend beyond this case. We define the  product between a complex-valued continuous function $f \in \Cder{}(\Xx)$ and complex-valued measure $\mu \in \Mm(\Xx)$ as $\dotpmeas{f}{\mu} \eqdef \int_\Xx \overline{f(x)} \mathrm{d} \mu(x)$. The (forward) \emph{measurement operator}
$
\Phi: \Mm(\Xx) \mapsto \CC^m$ that we consider in this paper is of the form
\begin{equation}\label{eq:measurements}
\Phi \mu \eqdef \tfrac{1}{\sqrt{m}} \pa{\dotpmeas{\phi_{\om_k}}{\mu}}_{k=1}^m 
\end{equation}
where  $(\om_1,\ldots,\om_m)$ are parameters identically and independently distributed according to a probability distribution $\Lambda(\om)$ over some space $\Om$, and $\phi_\om : \Xx \rightarrow \CC$ are smooth functions parameterized by $\omega$. We further assume that $\phi_\om$ is normalized, that is
$
 \EE_{\om\sim \Lambda}[\abs{\varphi_\om(x)}^2]   = 1$ for all $x\in \Xx$.
Our observations are of the form
\begin{equation}\label{eq:meas}
y = \Phi (\mu_0 + \tilde \mu_0) + w\, ,
\end{equation}
where $\mu_0=\sum_{i=1}^s a_i \de_{x_i}$ with $ (x_i,a_i) \in \Xx \times \CC$ is the $s$-sparse measure we are interested in, $\tilde \mu_0\in\Mm(\Xx)$ accounts for modelling error, and $w\in \CC^m$ is measurement noise. In the rest of the paper, we naturally assume that the support of $\tilde \mu_0$ does not include any of the $x_i$.

\paragraph{BLASSO.}

%
An increasingly popular  method to estimate such a sparse measure corresponds to solving a infinite-dimensional analogue of the Lasso regression problem
\eql{\tag{$\Pp_\la(y)$}
\label{eq:blasso}
	\umin{\mu \in \Mm(\Xx)} \frac{1}{2} \norm{\Phi \mu - y}_2^2 + \la |\mu|(\Xx).
}
Following~\cite{deCastro-exact2012}, we call this method the BLASSO (for Beurling-Lasso).
Here $|\mu|(\Xx)$ is the so-called total variation (or total mass) of the measure $\mu$, and is defined as
\eq{
	|\mu|(\Xx) \eqdef \sup \enscond{ \rep{\dotpmeas{f}{\mu}} }{f \in \Cder{}(\Xx), \normi{f} \leq 1}.
}
Note that on unbounded $\Xx$, one needs to impose that $f$ vanishes at infinity.
If $\Xx=\{x_i\}_i$ is a finite space, then this would correspond to the classical finite-dimensional Lasso problem~\citep{tibshirani1996regression}, because $|\mu|(\Xx)=\norm{a}_1 \eqdef \sum_i |a_i|$ where $a_i=\mu(\{x_i\})$. Similarly, when $\Xx$ is possibly infinite but $\mu=\sum_i a_i \de_{x_i}$, one also has that $|\mu|(\Xx)=\norm{a}_1$.

\subsection{Previous works}
\label{sec-pw}

From a theoretical perspective, understanding the performance of this approach corresponds to establishing a ``Rayleigh criterion'', which is the minimum allowable separation distance between two spikes $\min_{i\neq j}\norm{x_i - x_j}$ for the method to recover them from linear measurements. 
The first result in this direction is due to Cand\`es and Fernandez-Granda~\cite{candes-towards2013}, who prove that for Fourier measurements, this separation distance is (almost) equal to the inverse of the maximum sample frequency. This results has been extended to provide robustness to noise~\cite{candes-superresolution2013,fernandez-support2013,azais-spike2014,duval2015exact} and to cope with more general measurement operators \cite{bendory2015recovery}.  
All these previous theoretical works however strongly rely on the translation invariance of the linear operator (Fourier measurements or convolutions) and  the underlying domain (either Euclidean space or the periodic torus). Applying these results to spatially varying operator (such as for instance when imaging with non-stationary point spread functions) generally leads to overly pessimistic minimum separation condition. 
%

In parallel, it has been shown by Tang et al~\citep{tang2013compressed} that the recovery guarantees of Cand\`es and Fernandez-Granda~\cite{candes-towards2013} remain valid with high probability when only a small number of (Fourier) measurements are \emph{randomly selected}, of the order (up to log factors) of the sparsity of the underlying measure. However, this result is only valid under a \emph{random signs assumption} on the amplitudes of the sought-after Dirac masses, which is a well-known assumption in classical discrete compressed sensing~\citep{candes2006robust,donoho2006compressed} but appears somewhat unrealistic. While more detailed conditions can be derived when the amplitudes $a_i$ are all assumed real and positive \cite{denoyelle2017support}, in the general case $a_i \in \CC$ it was not known until this paper whether the random signs assumption could be removed. 

%

Although it is not the topic of this paper, let us note that lifting the minimum separation condition requires to impose positivity of the weights~\cite{deCastro-exact2012,schiebinger2015superresolution} and the price to pay is an explosion of the instabilities as spikes cluster together~\cite{denoyelle2017support}.

\paragraph{Numerical solvers and alternative approaches.} 

The focus of this paper is on the theoretical analysis of the performance BLASSO method, not on the development and analysis of efficient numerical solvers. Although the BLASSO problem is infinite dimensional, there are efficient numerical solvers that use the fact that the sought-after sparse solution is parameterized by a small number of parameters (positions and amplitudes of the spikes). This open the door to algorithms which do not scale with some grid size, and hence can scale beyond 1-D and 2-D problems. Let us mention in particular: (i) discretization on a grid~\cite{tang2013sparse,2017-Duval-IP-lasso}, (ii) semi-definite programming (SDP) relaxation using Lasserre Hierarchy \cite{candes-towards2013,de2016exact}, (iii) Frank-Wolfe and its variants~\cite{bredies-inverse2013,boyd2017alternating}, (iv) non-convex particle flows~\cite{chizat2018global}.
%
%
We also emphasize that the BLASSO is by no means the only  method for  estimating sparse measures in an off-the-grid setup. Let us, among other approaches, cite Prony-type spectral methods such as MUSIC and ESPRIT~\citep{schmidt-multiple1986,roy1989esprit,liao-music2014} and non-convex approaches for instance based on $\ell^0$ or greedy minimization (see for instance~\cite{gribonval2017compressive,soubies2015continuous} for recent contributions in this direction). In practice, these methods often surpasses BLASSO in term of performance, in particular when the noise is small and the spikes tends to cluster so that the minimum separation distance condition does not hold. A rule of thumb is that $\ell^1$-regularization is however good baseline, which benefit from both efficient and stable numerical solvers and an in-depth theoretical analysis which leverage the convexity of the problem.

\subsection{Contributions}

The goal of this paper is twofold: remove the random signs assumption of Tang et al~\cite{tang2013compressed} while still keeping a sharp number of random measurements, and extend the framework to encompass non-translation invariant operators in a natural manner with improved separation conditions. The former is achieved by extending the so-called golfing scheme~\cite{gross2011recovering, candes2011probabilistic} to the infinite-dimensional setting, while the latter is done through a particular Riemannian geometric framework, defined by the metric tensor associated to the covariance kernel of the measurement operator. We will show that, by imposing a minimal separation between Diracs with respect to the geodesic distance associated to this tensor, previous strategies can be naturally extended. 

Informally, our main result reads as follows. Define the \emph{limit covariance kernel} $\fullCov(x,x') \eqdef \EE_\om \overline{\phi_\om(x)}\phi_\om(x')$, which measures how much two Diracs at $x$ and $x'$ \emph{interact} with each other in the large samples  limit as $m\to \infty$, and assume that $K$ is real-valued (primary examples include the Gaussian kernel, or the so-called Jackson kernel used in \cite{candes-towards2013}). Define the metric tensor $\met_x \eqdef \nabla_1\nabla_2 \fullCov(x,x) \in \RR^{d\times d}$, where $\nabla_i$ indicates the gradient with respect to the $i$th variable, and assume that for all $x\in \Xx$ it is a positive definite matrix. Finally, define the associated geodesic distance $\dsep_\met(x,x') = \inf_\gamma \int_0^1 \sqrt{\gamma'(t)^\top \met_{\gamma(t)}\gamma'(t)}\mathrm{d}t$, where the infimum is taken over all continuous path $\gamma:[0,1] \to \Xx$ such that $\gamma(0)=x$ and $\gamma(1)=x'$ (more details about this geodesic distance are given in Section~\ref{sec-fisher-ot}). Denote by $\Bb_{\dsep_\met}(x;r)$ the ball of radius $r$ centered on $x$, for the metric $\dsep_\met$. The main result of the paper, here stated in an informal way, is the following.

\begin{thm}[Main result, informal]\label{thm:main-short}
Let $R_\Xx\eqdef \sup_{x,x'} \dsep_\met(x,x')$. Under some assumptions on the kernel $\fullCov$ (see Assumption 1 in Sec. \ref{sec:limit-cert}) and features $\phi_\om$ (see Assumption 2 in Sec. \ref{sec:m-finite}), there are constants $r, \Delta>0$, that only depends on $\fullCov$, and $C_1, C_2>0$ which depend on $K$ and the regularity of $\phi_{\omega_k}$ (up to 2nd order), such that the following holds.
Suppose that $y$ is of the form \eqref{eq:meas} with $\min_{i\neq j}\dsep_\met(x_i,x_j) \geq \Delta$ and
\begin{equation}\label{eq:number-m-main-short}
m\geq C_1 \cdot s \cdot \log(s) \log((C_2 R_\Xx)^d/\rho).
\end{equation}
Then with probability $1-\rho$, when $\norm{w}\leq \delta$ and $\lambda \sim \frac{\delta}{\sqrt{s}}$, any solution $\hat \mu$ to \eqref{eq:blasso} satisfies
\begin{equation}\label{eq:recovery-bound-main-short}
\Tt_{\dsep_\met}^2\pa{\sum_{j=1}^s \hat A_j \delta_{x_j}, \; \abs{\hat \mu}} \lesssim \sqrt{s}\delta + \abs{\tilde \mu_0}(\Xx)
\qandq 
\max_{j=1}^s\abs{\hat a_j - a_j}\lesssim \sqrt{s}\delta + \abs{\tilde \mu_0}(\Xx),
\end{equation}
where $\hat A_j \eqdef \abs{\hat \mu}(\Bb_{\dsep_\met}(x_j;r))$, $\hat a_j \eqdef \hat \mu(\Bb_{\dsep_\met}(x_j;r))$, and $\Tt_{\dsep_\met}$ is the partial optimal transport distance associated to $\dsep_\met$ (see Def. \ref{def:wasserstein}).
\end{thm}

Let us comment a bit on this result. On a technical level, the most salient feature of Theorem \ref{thm:main-short} is that, up to log factors, the bound \eqref{eq:number-m-main-short} is \emph{linear} in the sparsity of the underlying measure. This improves over the best known result of Tang et al \cite{tang2013compressed}, since in our case we do not require the random signs assumption.

On a methodological level, the assumptions on the kernel $\fullCov(x,x')$ mainly state that it must \emph{decrease} sufficiently when $x$ and $x'$ are far apart, or, in other words, that the \emph{coherence} between $\Phi\delta_x$ and $\Phi\delta_{x'}$ must be low. The main novelty of our approach is that we measure this separation in term of the geodesic metric $\dsep_\met$, which allows to account for non-translation invariant kernels in an intrinsic and natural manner. The assumptions on the features $\phi_\om$, which are more technical in nature, relates to their regularity and the boundedness of their various derivatives.

Concerning the recovery bound \eqref{eq:recovery-bound-main-short}, the first part states that the measure $\hat \mu$ concentrates around the true positions of the Diracs, while the second part guarantees that the complex amplitudes of $\hat \mu$ around the Diracs are close to their true values. The discrepancy in the first part is measured in terms of a \emph{partial optimal transport} distance associated to $\dsep_\met$ (Def.~\ref{def:wasserstein} in Sec.~\ref{sec:duality}).

Finally, the constants $C_1,C_2$ that appear in \eqref{eq:number-m-main-short} can depend (generally polynomially) on the dimension $d$ but not on the sparsity $s$. As we will see in Section \ref{sec:m-finite} and the detailed version of Theorem \ref{thm:main-short} (Theorem \ref{thm:main}), the bound \eqref{eq:number-m-main-short} is actually valid when we suppose the features $\phi_\om$ and their derivatives to be \emph{uniformly bounded} for all $x$ and $\om$. When this is not the case, we will be able to relax this assumption, similar to the notion of stochastic incoherence \cite{candes2011probabilistic} in compressed sensing. As a result, $m$ can actually appear in $C_1,C_2$, generally in a logarithmic form (see examples in Section \ref{sec:examples}), which only adds logarithmic terms in $s$ and $d$ in the final number of measurements.


\paragraph{Outline of the paper.} 

The paper is organized as follows. In Section \ref{sec:examples} we give example applications of Theorem \ref{thm:main-short}, including non-translation invariant frameworks such as Laplace measurements used in microscopy \cite{denoyelle2018microscopy}. In Section \ref{sec:duality}, we introduce our Riemannian geometry framework and prove intermediate recovery results based on the existence of a so-called \emph{non-degenerate dual certificate}, which is known in the literature to be the key object in the analysis of the BLASSO model. In Section \ref{sec:limit-cert}, we study in more detail the relationship between the minimal separation condition and the covariance kernel. We prove that, under some conditions on $\fullCov$, in the limit $m\to \infty$, one can indeed prove the existence of a non-degenerate dual certificate when minimal separation is imposed with respect to $\dsep_\met$. Finally, in Section \ref{sec:m-finite}, we state our main result with finite number of measurements $m$ (Theorem \ref{thm:main}, which is a detailed version of Theorem \ref{thm:main-short}). Section \ref{sec:main-proof} is dedicated to its proof using an infinite-dimensional extension of the celebrated golfing scheme \cite{candes2011probabilistic}, with technical computations in the appendix.


\paragraph{Relationship to our previous work~\cite{2019-Poon-aistats}}

This article is a substantially extended version of the conference publication~\cite{2019-Poon-aistats}.
The results of Section \ref{sec:limit-cert} are in most part already published (under slightly more restrictive assumptions) in this conference paper. 
The remainder of the paper is however entirely novel. 
We remove the random signs assumption of~\cite{2019-Poon-aistats} thanks to a new proof technique with the golfing scheme.
Furthermore, the results in~\cite{2019-Poon-aistats} are restricted to the small noise setting and focus on exact support stability, while we study here arbitrary noise levels and establish more general stability bounds in terms of optimal transport distances. 


\paragraph{Notations.}

Given $n\in \NN$, we denote by $[n]\eqdef \{1,2,\ldots, n\}$ the first $n$ integers. We write $1_n$ to denote the vector of length $n$ whose entries are all $1$'s, and $0_n$ to denote the vector of length $n$ whose entries are all $0$'s.
Given two matrices $A$ and $B$, we write $A\prec B$ to mean that $B-A$ is positive definite and $A\preceq B$ to mean that $B-A$ is semi-positive definite. Given two positive numbers $a,b$, we write $a\lesssim b$ to mean that there exists some universal constant $C>0$ so that $a \leq C b$. Given $(\Xx, \dsep)$ a metric space, $x\in \Xx$ and $r>0$, we define $\Bb_\dsep(x;r)\eqdef \enscond{z\in \Xx}{\dsep(x,z)<r}$ the ball centered on $x$ of radius $r$, or just $\Bb_{\norm{\cdot}}(r) =\enscond{z\in \Xx}{\norm{z}<r}$ the ball centered on $0$ for a norm $\norm{\cdot}$.

We write $\norm{\cdot}_p$ to denote the $\ell_p$ norm, and $\norm{\cdot}$ without any subscript denotes the spectral norm for matrices or $\ell_2$ norm for vectors. For any norm $\norm{\cdot}_X$ on vectors, the corresponding matrix norm is $\norm{A}_{X \to Y} = \sup_{\norm{x}_X = 1} \norm{Ax}_Y$ and $\norm{A}_X = \norm{A}_{X \to X}$ for short.
Given a vector $x \in \CC^{sd}$ decomposed in blocks $x = [x_1^\top,\ldots,x_s^\top]^\top$ with $x_i \in \CC^d$, where $s$ and $d$ will always be defined without ambiguity, we define the block norm $\normblock{x}\eqdef \max_{1\leq i\leq s} \norm{x_i}$.
Given a vector $x\in \CC^{s(d+1)}$ decomposed as $x = [x_0^\top, X_1^\top, \ldots, X_s^\top]^\top$ where $x_0 \in \CC^s$ and $X_j\in \CC^d$, we define
$
\nB{x} \eqdef \max\pa{\norm{x_0}_\infty, \max_{j=1}^s\norm{X_j}_2}
$.

For a complex number $a$, its sign is denoted by $\sign(a) = \frac{a}{\abs{a}}$.
Given a complex-valued measure $\mu \in \Mm(\Xx)$ and complex-valued continuous function $f \in \Cder{}(\Xx)$, we recall that $\dotpmeas{f}{\mu} \eqdef \int_\Xx \overline{f(x)} \mathrm{d} \mu(x)$. For two complex vectors $v$ and $w$, $\dotp{v}{w}_2 \eqdef v^* w$, where $v^* = \overline{v}^\top$ denotes conjugate transpose.


\section{Examples}\label{sec:examples}

In this section, we illustrate Theorem \ref{thm:main-short} for some special cases of practical interest in imaging and machine learning. The following statements are obtained by bounding the constants in Theorem \ref{thm:main} in Section \ref{sec:m-finite} (the detailed version of Theorem \ref{thm:main-short}). These computations, which can be somewhat verbose, are delayed to Appendices \ref{app-discretefourier}, \ref{app-gaussian} and \ref{app-laplace}.

\paragraph{Off-the-grid Compressed Sensing.} Off-the-grid Compressed sensing, initially introduced in the special case of 1-D Fourier measurements on $\Xx=\TT=\RR/\ZZ$ by~\citep{tang2013compressed}, corresponds to Fourier measurements of the form~\eqref{eq:measurements}. This is a ``continuous'' analogous of the celebrated compressed sensing line of works~\citep{candes2006robust,donoho2006compressed}. We give a multi-dimensional version below.

Let $f_c \in \NN$ with $f_c\geq 128$ (for simplicity) and $\Xx = \TT^d$ the $d$-dimensional torus.
Let  $\phi_\om(x) \eqdef  e^{\mathrm{i} 2\pi \omega^\top x}$, $\Omega \eqdef \enscond{\om\in \ZZ^d}{\norm{\om}_\infty\leq f_c }$, and  $\Lambda(\om) = \prod_{j=1}^d g(\omega_j)$ where $g(j) = \frac{1}{f_c } \sum_{k=\max(j-f_c ,-f_c )}^{\min(j+f_c ,f_c )}(1-\abs{k/f_c })(1-\abs{(j-k)/f_c })$. The Fisher metric is, up to a constant $C$, the Euclidean metric $\dsep_\met(x,x') = C f_c \norm{x-x'}$. Provided that $\min_{i\neq j}\norm{x_i-x_j} \gtrsim \frac{d^{\frac12} s^{\frac{1}{4}}}{f_c}$,
stable recovery is guaranteed with
$$
m\gtrsim d^2 s  \pa{ \log(s) \log\pa{\frac{s}{\rho}} + \log\pa{\frac{(s f_c d)^d}{\rho}}}.
$$
Note that, compared to the uni-dimensional case, the minimal separation $\Delta$ depends on $s$: this could actually be replaced by a bound exponential in the dimension $d$, which we prefer not to do here. Indeed, during the proof, one must bound a quantity of the form $\sum_{i=2}^s \norm{x_1-x_i}^{-4}$, for $\Delta$-separated Diracs. Since in one dimension only $2$ Diracs can be situated at distance $k\Delta$ from $x_1$ for each integer $k>0$, this can be easily bounded by a global bound $\Delta^{-4}\sum_{k=1}^\infty k^{-4}$ that does not depend on $s$. In the multidimensional case however, an \emph{exponential} number of Diracs can be packed around $x_1$, and applying the same strategy would lead to a bound on $\Delta$ which is exponential in the dimension.

\paragraph{Continuous sampling Fourier transform}

A variant of the previous framework is \emph{continuous} Fourier sampling, for instance with frequencies distributed according to a Gaussian distribution. Let $\Xx \subset \RR^d$ be a bounded open subset of $\RR^d$. 
The space of frequencies is $\Omega = \RR^d$, $\phi_\om(x) \eqdef  e^{\mathrm{i} \omega^\top x}$, and  $\Lambda(\om) = \Nn(0,\Sigma^{-1})$ for some known symmetric positive definite matrix $\Sigma$. Note that, for simplicity, the frequencies are drawn according to a Gaussian with \emph{precision matrix} $\Sigma$ (the inverse of the covariance matrix), such that the kernel $\fullCov$ is the classical \emph{Gaussian kernel} $\fullCov(x,x') = e^{-\frac12 \norm{\Sigma^{-\frac12}(x-x')}^2}$. The Fisher metric is $\dsep_\met(x,x') = \norm{\Sigma^{-\frac12}(x-x')}$. In this case, provided that $\min_{i\neq j}\dsep_\met(x_i,x_j) \gtrsim \sqrt{\log(s)}$,
stable recovery is guaranteed with
$$
m\gtrsim s  \pa{ L \log(s) \log\pa{\frac{s}{\rho}} + L^2 \log\pa{\frac{(s L \Rr_\Xx)^d}{\rho}}}.
$$
where $L = d + \log^2\pa{\frac{dm}{\rho}}$. Note that $\log(m)$ appears in $L$ in the r.h.s. of the expression above, which only incurs additional logarithmic terms in the bound on $m$, as mentioned in the introduction.

\paragraph{Learning of Gaussian mixtures with fixed covariances}

An original framework for continuous sparsity is \emph{sketched learning} of mixture models \cite{gribonval2017compressive}, and in particular Gaussian mixture models (GMM), for which we can exploit the computations of the previous case of  Fourier measurements sampled in accordance to a Gaussian distribution. Assume that we have data samples $z_1,\ldots,z_n \in \RR^d$ drawn $i.i.d.$ from a mixture of Gaussians $\xi \eqdef \sum_{i=1}^s a_i \Nn(x_i, \Sigma)$ with known covariance $\Sigma$. The means $x_1,\ldots,x_s \in \Xx \subset \RR^d$ and weights $a_1,\ldots,a_s>0$ are the objects which  we want to estimate. Sample frequencies $\om_1,\ldots,\om_m \in \RR^d$ $i.i.d.$ from a Gaussian $\Lambda = \Nn(0, \Sigma^{-1}/d)$, and construct the following \emph{linear sketch} \cite{cormode2011book} of the data:
\begin{equation}\label{eq:sketch}
y = \frac{C}{n} \sum_{i=1}^n (e^{-\mathrm{i} \om_k^\top z_i})_{k=1}^m
\end{equation}
where the constant $C = (1+\frac{2}{d})^{\frac{d}{4}} \leq e^\frac12$ is here for normalization purpose. Linear sketches are mainly used for computational gain: they are easy to compute in a streaming of distributed context, and are much smaller to store in memory than the whole database \cite{cormode2011book, gribonval2017compressive}. It is easy to see that the sketch can be reformulated as \eqref{eq:measurements}, by writing
\begin{equation}
y \approx \EE_z (Ce^{-\mathrm{i} \om_k^\top z})_{k=1}^m= \Phi \mu_0
\end{equation}
where $\mu_0 = \sum_i a_i \delta_{x_i}$, and $\Phi$ is defined using the feature functions $$\phi_\om(x) = \EE_{z\sim \Nn(x,\Sigma)} C e^{\mathrm{i} \om^\top z} = Ce^{\mathrm{i} \om^\top x}e^{-\frac12 \om^\top \Sigma \om}.$$ The ``noise'' $w\eqdef y - \EE_z (Ce^{-\mathrm{i} \om_k^\top z})_{k=1}^m$ is simply the difference between empirical and true expectations, and using simple concentration inequalities that we skip here for simplicity, it is possible to show that with high probability, $\norm{w} \leq \Oo\pa{n^{-\frac12}}$. Applying the previous computations we obtain the following result: provided that $\min_{i\neq j} \norm{\Sigma^{-\frac12}(x_i-x_j)}_2 \gtrsim \sqrt{d \log(s)}$,
stable recovery of $\mu_0$ is guaranteed when
$$
m\gtrsim s  \pa{ d \log(s) \log\pa{\frac{s}{\rho}} + d^2 \log\pa{\frac{(s d \Rr_\Xx)^d}{\rho}}}
$$
and the concentration in the recovery bound \eqref{eq:recovery-bound-main-short} is given by $\delta =\norm{w}=\Oo\pa{n^{-\frac12}}$.

\paragraph{Gaussian mixtures with varying covariances}

The case of simultaneously recovering both  the means and covariance matrices is an interesting venue for future research. We simply describe here the associated metric and distance in the univariate case.  The geodesic distance between univariate  Gaussian distributions is well known \cite{costa2015fisher}: Given $x=(m,u)$ and $ x'=(n,v)$ with $m,n\in\RR$ and $u,v\in\RR_+$,
let $\phi(x) \eqdef \frac{1}{\sqrt[4]{\pi} \sqrt{u}} e^{-(m-\cdot)/(2u^2)}$, then the covariance kernel is $$K_0(x,x') = \dotp{\phi(x)}{\phi(x')}_{L^2} = \frac{\sqrt{2uv}}{\sqrt{u^2+v^2}} e^{-\frac{(m-n)^2}{2(u^2+v^2)}}.$$ The associated metric at $x=(m,u)$ is $\met_x=\frac{1}{2u^2} \Id_2$ and the Fisher-Rao distance is the Poincar\'e half plane distance
\begin{equation}\label{eq:poincare_half}
\dsep_0(x,x') = 2\mathrm{arsinh}\pa{\frac{\norm{x-x'}}{2\sqrt{uv}}   }, \qwhereq \mathrm{arsinh}(x) \eqdef \ln\pa{x+\sqrt{x^2-1}}.
\end{equation}
Consider now the case of Gaussian mixture $\xi \eqdef \sum_{i=1}^s a_i \Nn(x_i, v_i^2)$, where the unknowns are $a_i>0$, $x_i\in\RR$ and $v_i>0$, and we are given data $\ens{z_i}_{i=1}^n$ drawn iid from $\xi$ and we construct the linear sketch \eqref{eq:sketch} as before, where $\omega_k \in\RR$ are iid from $\Nn(0,\sigma^2)$.  This corresponds to the normalised random features
$$
\phi_\om(m,u) = \pa{2 u^2 \sigma^2 +1 }^{\frac{1}{4}} e^{-\mathrm{i}m \om } e^{-\frac12 u^2 \om^2},
$$
and
\begin{equation}\label{eq:kernel_gsketch1d}
K((m,u), (n,v)) = \frac{\sqrt{2u_\sigma v_\sigma}}{\sqrt{u_\sigma^2 + {v_\sigma}^2}} e^{-\frac{(m-n)^2}{2(u_\sigma^2 + {v_\sigma}^2)}}
\end{equation}
where $u_\sigma^2 = \frac{1}{2\sigma^2}+u^2$ and $v_\sigma^2 = \frac{1}{2\sigma^2}+v^2$. 
The metric at $x=(m,u)$ is $\met_x=\frac{1}{2u_\sigma^2} \Id_2$. Note that since  \eqref{eq:kernel_gsketch1d} also corresponds the the kernel between Gaussian distributions with mean and standard deviation as $x_\sigma \eqdef (m,u_\sigma)$ and $x_\sigma'\eqdef (n,v_\sigma)$,  the associated geodesic distance is therefore  $\dsep_0(x_\sigma,x_\sigma')$ where $\dsep_0$ is the Poincar\'e half plane distance described in \eqref{eq:poincare_half} (as mentioned in \eqref{eq:equiv_dist_randfeatures},  geodesic distances on random features and parameter space are equivalent).

\paragraph{Sampling the Laplace transform.} 

In some fluorescence microscopy applications (see~\cite{denoyelle2018microscopy} and the references therein), depth measurements are obtained from the Laplace transform of the signal. Contrary to Fourier measurements, this gives rise to a non-translation invariant kernel $\fullCov$, and was therefore not covered by existing theory. Using the proposed Riemannian geometry framework, we can cover this setting.

Let $\Xx= (0,1)^d \subset \RR^d_+$. Let $\Omega = \RR_+^d$. Define for $x\in \Xx$ and $\omega\in \Omega$,
$$
\phi_\om(x) \eqdef \exp\pa{-x^\top \omega} \prod_{i=1}^d \sqrt{\frac{x_i+\al_i}{\al_i}} \qandq \Lambda(\omega) = \exp(-2\al^\top\om) \prod_{i=1}^d(2\alpha_i).
$$
where $\alpha_i  \sim d$ are positive and distinct real numbers. The Fisher metric is $$
\dsep_\met(x,x') = \sqrt{\sum_{i=1}^d \abs{\log\pa{\frac{x_i+\al_i}{x_i'+\al_i}}}^2},
$$
and provided that $\min_{i\neq j}\dsep_\met(x_i,x_j) \gtrsim d + \log(d^{3/2}s)$, stable recovery is guaranteed with
$$
m \gtrsim  s  \pa{C \log(s) \log\pa{\frac{s}{\rho}} +C^2 \log\pa{\frac{C^d}{\rho}}}
$$
where 
$C \eqdef  d^2 \pa{d + \log^2(m) + \log^2\pa{\frac{d}{\rho}}}$. Similar to the Gaussian example, $\log(m)$ appears in $C$.


\paragraph{Training a two-layers neural network}

Following~\cite{bach2017breaking}, in the large number of neurons limit, regression using a neural network with a single hidden layer can be formulated using our framework. 
Given a set of $m$ training samples $(\om_k,y_k)_{k=1}^m$, one aims to predicts the values $y_k \in \RR$ from the features $\om_k \in \Om$  using a continuous dictionary of functions $\om \mapsto \phi_\om(x)$ (here $x \in \Xx$ parameterizes the dictionary), as  $y_k \approx \int_\Xx \phi_{\om_k}(x) \mathrm{d}\mu_0(x) = \sum_{i=1}^s a_i \phi_{\om_k}(x_i)$. In the context of neural networks, one uses ridge functions of the form $\phi_\om(x) = \xi(\dotp{x}{\om})$, For instance, one can consider $\xi(u) \eqdef \xi_n(u) \eqdef \max(u,0)^n$, where $n=1$ is the ReLu non-linearity.  Detailed treatment of this example will be left for future work, however, we simply mention that the associated kernel and metric was studied in~\cite{cho2010large,cho2011analysis}. We recall their results here. Suppose that $\om_k \sim \Nn(0,\Id_d)$, then the associated kernel is
\begin{align*}
k_n(x,y) &= 2\int \frac{e^{-\norm{\om}^2/2}}{(2\pi)^{d/2}} \xi_n(\dotp{\om}{ x})  \xi_n(\dotp{\om}{ y}) \mathrm{d}\om
= \frac{1}{\pi}\norm{x}^n\norm{y}^n J_n(\theta), 
\end{align*}
$$
\qwhereq \theta = \cos^{-1}\pa{\frac{x^\top y}{\norm{x}\norm{y}}}
\qandq J_n(\theta) = (-1)^n (\sin\theta)^{2n+1} \pa{\frac{1}{\sin\theta} \frac{\partial}{\partial \theta}}^n \pa{\frac{\pi - \theta}{\sin\theta}}.
$$
Since $J_n(0) = \pi (2n-1)!!$, the normalised random features  the normalised kernel are therefore
$$
\phi_\om(x) = \frac{\pi}{\sqrt{J_n(0)}} \xi_n\pa{ \frac{ \dotp{\om}{x}}{\norm{x}}}\qandq	 K_n(x,y) = \frac{J_n(\theta)}{J_n(0)}, 
$$
and the associated metric is
$$
\met_x = \nabla_1\nabla_2 K_n(x,x) = \frac{n^2}{\norm{x}^2 (2n-1)} \pa{\Id_d +(4n-3) \frac{xx^\top}{\norm{x}^2}}.
$$
Note that since $\phi_\om(x) = \phi_\om(x/\norm{x})$ for all $x$, the geodesic path between any $x$ and $y$ must lie on the unit sphere, moreover, given any $\gamma:[0,1]\to \Xx$  with $\norm{\gamma(t)} =1$ for all $t\in[0,1]$, we have
$$
  \int_0^1\sqrt{ \dotp{\met_{\gamma(t)} \gamma'(t)}{\gamma'(t)}}  \mathrm{d}t = \int_0^1  \sqrt{ \frac{n^2}{ (2n-1)} \pa{\norm{\gamma'(t)}^2 +(4n-3) \abs{\dotp{\gamma'(t)}{\gamma(t)}}^2} } \mathrm{d}t =\sqrt{ \frac{n^2}{2n-1}} \int_0^1 \norm{\gamma'(t)}\mathrm{d}t,
$$
therefore, $\dsep_\met(x,y)  =\sqrt{ \frac{n^2}{2n-1}} d_{\mathbb{S}}\pa{\frac{x}{\norm{x}}, \frac{y}{\norm{y}}}$, where $d_{\mathbb{S}}$ is the geodesic distance on the sphere.



\section{Stability and the Fisher information metric}\label{sec:duality}

In this section, we introduce the proposed Riemmanian geometry framework, and give intermediate recovery guarantees which constitute the first building block of our main result. Namely, we introduce so-called \emph{dual certificates}, which are known to be key objects in the study of the BLASSO, and show how they lead to sparse recovery guarantees in our Riemannian framework.


\subsection{Fisher and Optimal Transport Distances}
\label{sec-fisher-ot}

Let us first introduce the proposed Riemannian geometry framework, and define objects related to it.

\subsubsection{The covariance kernel and the Fubini-Study metric}

A natural property to analyse in our problem is the way two Diracs \emph{interact} with each other, which is linked to the well-known notion of \emph{coherence} (or, rather, incoherence) between measurements in compressive sensing \cite{foucart2013mathematical}. This is done through what we refer to as the \emph{covariance kernel} $\Cov:\Xx\times \Xx \to \CC$, defined as
\begin{equation}\label{eq:cov}
\Cov(x,x') \eqdef \dotp{\Phi \delta_x}{\Phi \delta_{x'}}_2 = \frac{1}{m}\sum_{j=1}^m \overline{\phi_{\omega_k}(x)} \phi_{\omega_k}(x'), \qquad \forall x,x'\in \Xx\, .
\end{equation}
In the limit case $m \to \infty$, the law of large number states that $\Cov$ converges almost surely to the \emph{limit} covariance kernel:
\begin{equation}\label{eq:fullCov}
\fullCov(x,x') \eqdef \EE_\om \overline{\phi_{\om}(x)} \phi_{\om}(x')
\end{equation}
where we recall that $\om \sim \Lambda$. This object naturally governs the geometry of the space, and we use it to define our Riemmanian metric, which as we will see is linked to a notion of Fisher information metric. In the rest of the paper, we assume throughout that $\fullCov$ is real-valued, even though $\Cov$ may be complex-valued.

Given the normalisation $\EE_\omega \abs{\phi_\om(x)}^2 = 1$ for all $x\in\Xx$, $\phi_\omega(x)$ can be interpreted as a complex-valued \emph{probability amplitude} with respect to $\om$ (parameterized by $x$), a classical notion in quantum mechanics (see \cite{Griffiths2004book}). When $x$ varies, a natural metric between probability amplitudes is the so-called Fubini-Study metric, which is the complex equivalent of the well-known Fisher information metric. Writing $\phi_\omega(x) = \sqrt{p(\omega,x)} e^{\mathrm{i}\alpha(\omega,x)}$ where $p(\omega,x) \eqdef \abs{\phi_\omega(x)}^2 $ and $\alpha(\omega,x) \eqdef \mathrm{arg}(\phi_\omega(x))$, the Fubini-Study metric is defined by the following metric tensor in $\CC^{d\times d}$ \cite{Facchi2010}:
\begin{equation}\label{eq:fubini}
\begin{split}
\met_x \eqdef& \frac{1}{4} \EE_p[\nabla_x \log(p) \nabla_x \log(p)^\top]+ \EE_p[\nabla_x\al \nabla_x\al^\top] - \EE_p[\nabla_x \alpha] \EE_p[\nabla_x \alpha]^\top \\
&- \frac{\ii}{2} \EE_p[\nabla_x \log(p) \nabla_x \alpha- \nabla_x \alpha \nabla_x \log(p)^\top]. 
\end{split}
\end{equation}
where we use the notation $\EE_p[f] = \int f(\om) p(\om, x) \mathrm{d}\Lambda(\om)$. If $\phi_\omega$ is real-valued, then $\alpha = 0$ and this is indeed the Fisher metric up to a factor of $\frac{1}{4}$. The following simple Lemma shows the link between this metric and the derivatives of the covariance kernel $\fullCov$.
\begin{lem}\label{lem:fubini-study}
For any kernel $\fullCov(x,x') \eqdef \EE_\om \overline{\phi_\om(x)}\phi_\om(x')$, the Fubini-Study metric defined in \eqref{eq:fubini} satisfies
\begin{equation}
\met_x = \nabla_1 \nabla_2 \fullCov(x,x) - \EE_p[\nabla_x \alpha] \EE_p[\nabla_x \alpha]^\top
\end{equation}
If furthermore $\fullCov(x,x')$ is assumed real-valued, then $\EE_p[\nabla_x \alpha] = 0$, and $\met_x = \nabla_1 \nabla_2 \fullCov(x,x)$.
\end{lem}

\begin{proof}
Using $p = \abs{\phi_\om}^2$ and $\nabla \phi_\om = \pa{\frac{\nabla p}{2p} + i \nabla \alpha}\phi_\om$, a direct computation shows that
\begin{align}\label{eq:nabla_logp_alpha}
\nabla_x \log(p) = \frac{2}{p}\rep{\overline{\phi_\om}\nabla \phi_\om}\qandq \nabla_x \alpha = \frac{1}{p} \imp{\overline{\phi_\om}\nabla \phi_\om}
\end{align}
Therefore, 
\begin{align*}
\frac{1}{4}\EE_p[&\nabla_x \log(p) \nabla_x \log(p)^\top]+ \EE_p[\nabla_x\al \nabla_x\al^\top] \\
&= \int \frac{1}{p^2} \pa{\rep{\overline{\phi_\om}\nabla \phi_\om}\rep{\overline{\phi_\om}\nabla \phi_\om}^\top + \imp{\overline{\phi_\om}\nabla \phi_\om}\imp{\overline{\phi_\om}\nabla \phi_\om}^\top} p \mathrm{d}\Lambda \\
&=\int \frac{1}{p} \rep{\abs{\phi_\om}^2 \overline{\nabla \phi_\om} \nabla \phi_\om^\top} \mathrm{d}\Lambda =\int \rep{\overline{\nabla \phi_\om} \nabla \phi_\om^\top} \mathrm{d}\Lambda = \rep{\nabla_1\nabla_2\fullCov(x,x)}
\end{align*}
Similarly,
\begin{align*}
- \frac{\ii}{2} \EE_p[&\nabla_x \log(p) \nabla_x \alpha- \nabla_x \alpha \nabla_x \log(p)^\top] \\
&= -\ii \int \frac{1}{p^2} \pa{\rep{\overline{\phi_\om}\nabla \phi_\om}\imp{\overline{\phi_\om}\nabla \phi_\om}^\top + \imp{\overline{\phi_\om}\nabla \phi_\om}\rep{\overline{\phi_\om}\nabla \phi_\om}^\top} p \mathrm{d}\Lambda \\
&=-\ii \int \frac{1}{p} \imp{\abs{\phi_\om}^2 \overline{\nabla \phi_\om} \nabla \phi_\om^\top} \mathrm{d}\Lambda =\ii \int \imp{\overline{\nabla \phi_\om} \nabla \phi_\om^\top} \mathrm{d}\Lambda = \ii \cdot \imp{\nabla_1\nabla_2\fullCov(x,x)}
\end{align*}
which proves the first claim. The second claim is immediate by noticing from \eqref{eq:nabla_logp_alpha} that $\nabla_p \alpha = \Im\pa{\nabla_2 \fullCov(x,x)}$, which cancels when $\fullCov(x,x')$ is real (in particular in a neighborhood around $x=x'$).
%
\end{proof}


Since in this paper the limit covariance kernel \eqref{eq:fullCov} is assumed real-valued, the previous Lemma justifies the definition $\met_x = \nabla_1\nabla_2 \fullCov(x,x)$ that we adopt in the rest of the paper. For two vectors $u,v \in \CC^d$, we define the corresponding inner product
\begin{equation}\label{eq:dotpx}
\dotp{u}{v}_x \eqdef {u}^* \met_x v \qandq \norm{u}_x \eqdef \sqrt{\dotp{u}{u}_x}
\end{equation}

As described in the introduction, this induces a geodesic distance on $\Xx$:
\begin{equation}\label{eq:}
\dsep_\met(x,x') \eqdef \inf\enscond{\int_0^1 \norm{\gamma'(t)}_{\gamma(t)} \mathrm{d} t}{\gamma: [0,1] \to \Xx \text{ smooth},~\gamma(0)=x,~\gamma(1)=x'}
\end{equation}
and in the case where $\phi_\omega(x)$ is real-valued, this coincides with the ``Fisher-Rao'' geodesic distance~\citep{rao1945information} which is used extensively in information geometry for estimation and learning problems on parametric families of distributions~\citep{amari2007methods}. 

\begin{rem}[As a distance on the feature space]
The  geodesic distance induced by $\met$ is the natural distance between the random features $\phi_{\cdot}(x)$. Indeed, as discussed in \cite{burges1999geometry}, the manifold $(\Xx, \met)$ as an embedded submanifold of the sphere in Hilbert space $L_2(\mathrm{d}\Lambda)$ with embedding $x\mapsto \phi_{\cdot}(x)$, and given any $x,x'\in\Xx$, we have
\begin{equation}\label{eq:equiv_dist_randfeatures}
\inf_{\gamma\in\Gamma_{x,x'}} \int_0^1 \norm{\gamma'(t)}_{L_2(\mathrm{d}\Lambda)} \mathrm{d}t = \dsep_\met(x,x'),
\end{equation}
 where $\Gamma_{x,x'}$ consists of all piecewise smooth paths $\gamma: [0,1]\to \enscond{\phi(x)}{x\in\Xx}$ with $\gamma(0) = \phi(x)$ and $\gamma(1) = \phi(x')$.
\end{rem}

\begin{rem}[Fisher metric and invariances] The Fisher-Rao metric $\dsep_\met$ is ``canonical'' in the sense that it is the only (up to scalar multiples) geodesic distance which satisfies the natural invariances of the BLASSO problem. Indeed, the solutions to~\eqref{eq:blasso}, in the large sample limit $m \rightarrow +\infty$, are (i) invariant by the multiplication of $\phi(x) \eqdef (\phi_\om(x))_{\om \in \Om}$ by an arbitrary orthogonal transform $U$ (orthogonality on $L_2(\mathrm{d}\Lambda)$), i.e. invariance to $\phi(x) \mapsto U \phi(x)$, (ii) covariance under any change of variable $\phi \mapsto \phi \circ h$ where $h$ is a diffeomorphism between two $d$-dimensional parameter spaces. 
The covariance (ii) means that if $\mu = \sum_i a_i \de_{x_i}$ is a solution associated to $\phi$,
then the push-forward measure $(h^{-1})_\sharp \mu \eqdef \sum_i a_i \de_{h^{-1}(x_i)}$
is a solution associated to $\phi \circ h$. 
Note that the invariance (i) is different from the usual invariance under ``Markov morphisms'' considered in information theory~\cite{cencov1982statistical,campbell1986extended}.
When considering $\dsep_{\met} = \dsep_{\met_\phi}$ as a Riemannian distance depending solely on $\phi$, the invariance under any diffeomorphism $h$ reads  
\eql{\label{eq-invariance-fisher}
	\dsep_{\met_\phi}(x,x') = \dsep_{\met_{\phi \circ h}}(h^{-1}(x),h^{-1}(x')).
}
Assuming for simplicity that $\phi$ is injective, this invariance~\eqref{eq-invariance-fisher} is equivalent to the fact that the formula  
\eq{
	\foralls (q,q') \in \Mm^2, \quad 
	d_\Mm(q,q') \eqdef \dsep_{\met_\phi}(\phi^{-1}(q),\phi^{-1}(q'))
}
defines a proper (i.e. parameterization-independent) Riemannian distance $d_\Mm$ on the embedded manifold $\Mm \eqdef ( \phi(x) )_x \subset L_2(\mathrm{d}\Lambda)$.
%
Among all possible such Riemannian metrics on $\Mm$, the only ones being invariant by orthogonal transforms $\phi \mapsto U \phi$ are scalar multiples of the hermitian positive tensor $\partial \phi(x)^* \partial \phi(x) \in \CC^{d \times d}$, which is equal to $\met_\phi$ (here $\partial \phi(x)^*$ refers to the adjoint in $L_2(\mathrm{d}\Lambda)$ for the inner product defined by the measure $\La(\om)$). 
\end{rem}

\begin{rem}[Tangent spaces] Formally, in Riemannian geometry, one would use the notion of \emph{tangent space} $\Tt_x$, and for instance the inner product $\dotp{\cdot}{\cdot}_x$ would only be defined between vectors belonging to $\Tt_x$. However, in our case, since the considered ambient ``manifold'' is just $\RR^d$, in the sense that $\Xx$ is not a low-dimensional sub-manifold of $\RR^d$ but an open set of $\RR^d$, each tangent space can be identified with $\RR^d$, and we extend the definitions to complex vectors for our needs.
\end{rem}


\subsubsection{Optimal Transport metric}

In order to state quantitative performance bounds, one needs to consider a geometric distance between measures. The canonical way to ``lift'' a ground distance $\dsep_\met(x,x')$ between parameter to a distance between measure is to use optimal transport distances~\cite{santambrogio2015optimal}.

\begin{defn}[Wasserstein distance]\label{def:wasserstein}
Given $\mu,\nu\in \Mm_+(\Xx)$ with $\abs{\mu}(\Xx) = \abs{\nu}(\Xx)$, the Wasserstein distance between $\mu$ and $\nu$, relative to the metric $\dsep$ on $\Xx$ is defined by
\begin{equation*}
W_\dsep^2(\mu,\nu) \eqdef \inf_{\gamma\in \Pi(\mu,\nu)} \int_{\Xx^2} \dsep(x,x')\mathrm{d}\gamma(x,x'),
\end{equation*}
where $\Gamma(\mu,\nu) \subset \Mm_+(\Xx^2)$ is the set of all transport plans with marginals $\mu$ and $\nu$. Given $\mu,\nu \in \Mm_+(\Xx)$ (not necessarily of equal total mass), the optimal partial distance between $\mu$ and $\nu$ is defined as
\begin{equation*}
\Tt_\dsep^2(\mu,\nu) \eqdef \inf_{\tilde \mu,\tilde \nu} \ens{ W_\dsep^2(\tilde \mu,\tilde \nu) + \abs{\mu - \tilde \mu}(\Xx) + \abs{\tilde \nu -\nu}(\Xx)  }.
\end{equation*}
\end{defn}
Note that the distance $W_\dsep(\mu,\nu)$ is actually an hybridation (an inf-convolution) between the classical Wasserstein distance between probability distributions and the total variation norm. It is often called ``partial optimal transport'' in the literature (see for instance~\cite{caffarelli2010free}), and belongs to the larger class of unbalanced optimal transport distances~\cite{liero2018optimal,chizat2018unbalanced}.

\subsection{Nondegenerate  certificates, uniqueness and stability for sparse measures}
\label{sec:dual-nondegen}

We now introduce the notion of a dual certificate and prove recovery guarantees under certain non-degeneracy conditions, which is the first step toward our main result.

\subsubsection{Dual certificates}

The minimisation problem \eqref{eq:blasso} is a convex optimisation problem  and a natural way of studying their solutions are via their corresponding Fenchel-dual problems. It is well known that, in the limit as $\la\to 0$, its solutions cluster in a weak-* sense around minimisers of
\begin{equation}\label{eq:blasso0}
\min_{\mu\in \Mm(\Xx)} \abs{\mu}(\Xx) \text{ subject to } \Phi \mu = y \tag{$\Pp_0(y)$}\, ,
\end{equation}
and that properties of the dual solutions to \eqref{eq:blasso0} with $y = \Phi \mu_0$ can be used to derive stability estimates for \eqref{eq:blasso} under noisy measurements. In this section, we recall some of these results (see \cite{bredies-inverse2013, duval2015exact} for further details).
The (pre)dual of \eqref{eq:blasso} is
\begin{equation}\label{eq:dual}
\sup \enscond{\dotp{p}{y}_2 - \frac{\la}{2} \norm{p}_2^2}{{p\in \CC^m, \norm{\Phi^* p}_\infty \leq 1}} \tag{$\Dd_\la(y)$}
\end{equation}
where we remark that  the adjoint operator $\Phi^*:\CC^m \to \Cder{}(\Xx)$ is defined by $(\Phi^* p)(x) = \frac{1}{\sqrt{m}}\sum_{i=1}^m p_i \phi_{\om_i}(x)$.
Note that for $\la>0$, this is the projection of $y/\la$ onto the closed convex set $\enscond{p}{\norm{\Phi^* p}_\infty \leq 1}$ and the solution $p_\la$ is hence unique. The dual solution $p_\la$ is related to any primal solution $\mu_\la$ of \eqref{eq:blasso} by the condition 
\begin{equation}\label{eq:primal_dual}
\Phi^* p_\la \in \partial\abs{\mu_\la}(\Xx) \qandq p_\la =\frac{1}{\la}\pa{ y - \Phi \mu_\la}.
\end{equation}
Conversely, any pair $p_\la$ and $\mu_\la$ which satisfy this equation \eqref{eq:primal_dual} are necessarily dual and primal solutions of \eqref{eq:dual} and \eqref{eq:blasso} respectively.
In the case where $\la = 0$, a dual solution need not be unique, although existence is guaranteed (since in our setting, the dual variable belongs to a finite dimensional space).  In this case, $p_0$ and $\mu_0$ solve \eqref{eq:dual} with $\la =0$ and \eqref{eq:blasso0}, respectively, if and only if
\begin{equation}\label{eq:primal_dual0}
\Phi \mu_0 = y \qandq \Phi^* p_0 \in \partial\abs{\mu_0}(\Xx).
\end{equation}
Following the literature, we call any element $\eta \in \Im(\Phi^*) \cap \partial\abs{\mu_0}(\Xx)$ a \emph{dual certificate} for $\mu_0$. For $\mu_0 = \sum_{j=1}^s a_j \delta_{x_j}$, the condition $\eta \in \partial\abs{\mu_0}(\Xx)$ imposes that $\eta(x_j) = \sign(a_j)$ and $\norm{\eta}_\infty \leq 1$. Furthermore, it is known that in the noiseless case, $\mu_0$ is the unique solution to \eqref{eq:blasso0} if: the operator $\Phi_x: \CC^s \to \CC^m$ defined by $\Phi_x b = \sum_{j=1}^s b_j \Phi \delta_{x_j}$ is injective, and there exists $\eta \in \Im(\Phi^*) \cap \partial\abs{\mu_0}(\Xx)$ such that $\abs{\eta(x)}<1$ for all $x\not \in \{x_j\}$. In order to quantify the latter constraint and provide quantitative stability bounds, we impose even stronger conditions on $\eta$ and make the following definition.
\begin{defn}[Non-degenerate dual certificate]\label{def:nondegen}
Given $(a_i,x_i)_{i=1}^s$, we say that  $\eta\in \Im(\Phi^*)$ is an $(\epsilon_0,\epsilon_2,r)$-nondegenerate dual certificate  if: 
\begin{enumerate}[label=(\roman*)]
\item $\eta(x_i) = \sign(a_i)$ for all $i=1,\ldots,s$,
\item $\abs{\eta(x)} \leq 1- \epsilon_0$ for all $x\in \xf$,
\item $\abs{\eta(x)} \leq 1- \epsilon_2 \dsep_\met(x,x_i)^2$ for all $x\in \xn_i$,
\end{enumerate}
where $\xn_i \eqdef \Bb_{\dsep_\met}(x_i;r)$ and $\xf\eqdef \Xx\setminus \bigcup_i \xn_i$.
\end{defn}
In other words, there are neighborhoods of the $x_j$ such that, outside of these neighborhoods, $\eta$ is strictly bounded away from $1$, and inside, $\abs{\eta}$ has quadratic decay. In the next section we prove stable recovery results from the existence of non-degenerate dual certificates.

\subsubsection{Stable recovery bounds}

The following two propositions describe stability guarantees under the nondegeneracy condition. Proposition \ref{prop:robustness} quantifies how the recovered measure is approximated by a sparse measure supported on $\{x_j\}_j$, and Proposition \ref{prop:stab_near} describes the error in measure around small neighbourhoods of the points $\{x_j\}_j$.

\begin{prop}[Stability away from the sparse support]\label{prop:robustness}
 Suppose that there exists $\epsilon_0,\epsilon_2>0$, $\eta \eqdef \Phi^* p$ for some $p\in \CC^m$  such that $\eta$ is $(\epsilon_0,\epsilon_2,r)$-nondegenerate.
Assuming the measurement model \eqref{eq:measurements}, any minimiser $\hat \mu$ of \eqref{eq:blasso},  with $\norm{w}\leq \delta$ and $\lambda\sim \delta/\norm{p}$ is approximately sparse:
 by defining 
 $\hat A_j = \abs{\hat \mu}(\xn_j)$, we have
\begin{equation}\label{eq:stability-away}
\Tt_{\dsep_\met}^2\pa{\abs{\hat\mu}, \sum_{j=1}^s \hat A_j \delta_{x_j}} \lesssim \frac{1}{\min\pa{\bep,\bla}} \pa{\abs{\tilde \mu_0}(\Xx)+ \delta\norm{p}}.
\end{equation}
\end{prop}

\begin{proof}

To prove this proposition, we first establish  the following bound
\begin{equation}\label{eq:main-noise-stability}
\epsilon_0 \abs{\hat \mu}(\xf) + \epsilon_2 \sum_{i=1}^s \int_{\xn_i} {\dsep_\met}(x,x_i)^2 \mathrm{d}\abs{\hat \mu}(x) \lesssim \delta \norm{p} + \abs{\tilde \mu_0}(\Xx).
\end{equation}
As we will see, the optimal partial transport bound above is then a  consequence of this bound.

For $i=1,\ldots,s$, let $\xn_i\subset\Xx$ and $\xf = \Xx \setminus \bigcup_{j=1}^s \xn_j$ be as in Definition \ref{def:nondegen}. Recall the measurement model $y = \Phi(\mu_0 + \tilde \mu_0) +w$, and define $\bar\mu_0 = \mu_0 + \tilde \mu_0$ for simplicity.
We first adapt the proof of \cite[Thm. 2]{burger2004convergence} to derive an upper bound for  $\abs{\hat\mu} - \abs{\bar\mu_0} - \rep{\dotpmeas{\eta}{\hat \mu-\bar\mu_0}}$. By minimality of $\hat \mu$ and since $\norm{w}\leq \delta$,
$$
\la \abs{\hat \mu}(\Xx) + \frac{1}{2}\norm{\Phi \hat \mu - y}^2 \leq \la \abs{\bar \mu_0}(\Xx) + \frac{1}{2}\norm{\Phi \bar \mu_0 - y}^2 \leq \la \abs{\bar \mu_0}(\Xx)  + \frac{\delta^2}{2}
$$
Using $\eta = \Phi^* p$, and by adding and subtracting $\rep{\dotpmeas{\eta}{\hat \mu - \bar\mu_0}} = \rep{\dotp{p}{\Phi \hat \mu - y}_2} + \rep{\dotp{p}{w}_2}$, we obtain
\begin{equation}\label{eq:bregman}
\begin{split}
&\la \pa{\abs{\hat \mu}(\Xx) - \abs{\bar \mu_0}(\Xx)  - \rep{\dotpmeas{\eta}{\hat \mu - \bar\mu_0}}} +  \rep{\dotp{\la p}{\Phi(\hat \mu - \bar \mu_0)}_2}  + \frac{1}{2}\norm{\Phi \hat \mu - y}^2 \leq \frac{\delta^2}{2}
\\
&\implies \la \pa{\abs{\hat \mu}(\Xx) - \abs{\bar \mu_0}(\Xx)  - \rep{\dotpmeas{\eta}{\hat \mu - \bar\mu_0}}}   + \frac{1}{2}\norm{\Phi \hat \mu - y + \la p}^2 \leq \frac{\delta^2}{2}+  \frac{\la^2 \norm{p}^2}{2} - \rep{\dotp{\la p}{w}_2}
\\
&\implies  \abs{\hat \mu}(\Xx) - \abs{ \bar \mu_0}(\Xx)  - \rep{\dotpmeas{\eta}{\hat \mu - \bar \mu_0}}  \leq \frac{1}{2\la}\pa{\delta+  \la \norm{p}}^2 \lesssim \delta\norm{p}
\end{split}
\end{equation}
using $\la \sim \delta / \norm{p}$.
We now derive a lower bound for $\abs{\hat\mu} - \abs{\mu_0} - \rep{\dotpmeas{\fulleta}{\hat \mu-\bar \mu_0}}$. Since $\eta$ is a dual certificate, we have $\dotpmeas{\eta}{\bar \mu_0} = \abs{\mu_0}(\Xx)$ and $\abs{\dotpmeas{\eta}{\mu}} \leq \abs{\mu}(\Xx)$. By further exploiting the nondegeneracy assumptions (ii) and (iii) on $\eta$, we have
\begin{align*}
\abs{\hat\mu}(\Xx) -& \abs{\bar\mu_0}(\Xx) - \rep{\dotp{\fulleta}{\hat \mu-\bar \mu_0}} \geq \abs{\hat\mu}(\Xx) - \rep{\dotp{\eta}{\hat\mu}} - 2\abs{\tilde \mu_0}(\Xx)\\
&\geq \abs{\hat\mu}(\Xx) - \sum_{i} \int_{\xn_i} \abs{\eta}\mathrm{d}\abs{\hat\mu} - \int_{\xf} \abs{\eta}\mathrm{d}\abs{\hat\mu} - 2\abs{\tilde \mu_0}(\Xx)\\
&\geq \abs{\hat\mu}(\Xx) - \sum_i \int_{\xn_i} \pa{1-\epsilon_2 {\dsep_\met}(x,x_i)^2}\mathrm{d}\abs{\hat\mu}(x) - (1-\epsilon_0) \abs{\hat\mu}\pa{\xf} - 2\abs{\tilde \mu_0}(\Xx)\\
&= \epsilon_0 \abs{\hat\mu}\pa{\xf}   + \epsilon_2 \sum_i \int_{\xn_i}  {\dsep_\met}(x,x_i)^2\mathrm{d}\abs{\hat\mu}(x) - 2\abs{\tilde \mu_0}(\Xx)
\end{align*}
which proves \eqref{eq:main-noise-stability}. Note also that by combining this with \eqref{eq:bregman}, we  obtain the following bound that we will use later:
\begin{align}
\norm{\Phi \hat \mu - y + \la p}^2 \leq (\delta + \la \norm{p})^2 + 4 \la \abs{\tilde \mu_0}(\Xx) \implies \norm{\Phi \hat \mu - y } \leq \delta + 2\la \norm{p} + 2\sqrt{\la \abs{\tilde \mu_0}(\Xx)}  \label{eq:discreploss}
\end{align}

It remains to show that the bound \eqref{eq:main-noise-stability} yields an upper bound on the partial optimal transport distance between the recovered measure $\abs{\hat \mu}$ and $\rho\eqdef \sum_i \abs{\hat \mu}(\xn_i) \delta_{x_i}$, its ``projection" onto the positions $\{x_j\}_j$. To see this, first note that
the Kantorovich dual formulation \cite{santambrogio2015optimal} of the Wasserstein distance in Def. \ref{def:wasserstein} is
$$
\sup\enscond{ \int_\Xx \varphi \mathrm{d}\mu + \int_\Xx \psi\mathrm{d}\nu}{\varphi, \psi\in C_b(\Xx), \;\forall x,y\in \Xx, \; \varphi(x)+\psi(y) \leq {\dsep_\met}(x,y)^2}
$$
 Given any $\varphi, \psi\in C_b(\Xx)$ satisfying $\varphi(x)+\psi(y) \leq {\dsep_\met}(x,y)^2$ for all $x,y\in \Xx$, we have
\begin{align*}
W^2_\met(\rho, \abs{\hat\mu}_{\Sn}) &\leq \int \varphi \mathrm{d}\abs{\hat \mu}_{\Sn} + \int\psi \mathrm{d}\rho \\
&= \sum_j \pa{\int_{\Sn_j} ( \varphi(x) + \psi(x_j)) \mathrm{d}\abs{\hat \mu}(x) -   \psi(x_j) \int_{\Sn_j} \mathrm{d}\abs{\hat \mu}(x)  + \psi(x_j) \abs{\hat \mu}(\xn_j) }\\
&= \sum_j \int_{\Sn_j} ( \varphi(x) + \psi(x_j)) \mathrm{d}\abs{\hat \mu}(x) \leq \sum_j \int_{\Sn_j} {\dsep_\met}(x,x_j)^2 \mathrm{d}\abs{\hat \mu}(x)
\end{align*}
So, $$
\bla W^2_\met(\rho, \abs{\hat\mu}_{\Sn}) \lesssim \abs{\tilde \mu_0}(\Xx)+ \delta\norm{p}$$
So, since $\bep \abs{\hat \mu}_{\xf}(\Xx) \lesssim \abs{\tilde \mu_0}(\Xx)+ \delta\norm{p}$, we have
$$
\Tt^2_{\met}(\abs{\hat\mu}, \rho) \lesssim \frac{1}{\min\pa{\bep,\bla}} \pa{\abs{\tilde \mu_0}(\Xx)+ \delta\norm{p}}.
$$
\end{proof}

We now give stability bounds around the sparse support, under some additional assumptions.
\begin{prop}[Stability around the sparse support] \label{prop:stab_near}
Under the assumptions of Proposition \ref{prop:robustness}, let $\hat\mu$ be a solution of \eqref{eq:blasso}, and let  $\hat a = (\hat \mu(\xn_j))_{j=1}^s$. 
Suppose in addition that for $j=1,\ldots, s$, there exists $\eta_j = \Phi^* p_j$ which satisfies
\begin{enumerate}[label=(\roman*)]
\item $\eta_j(x_j) = 1$ and $\eta_j(x_\ell) = 0$ for all $\ell\neq j$
\item $\abs{1-\eta_j(x)} \leq \epsilon_2 {\dsep_\met}(x,x_j)^2$ for all  $x\in \xn_j$,
\item $\abs{\eta_j(x)} \leq \epsilon_2 {\dsep_\met}(x,x_\ell)^2$ for all $x\in \xn_\ell$ and $\ell\neq j$,
\item $\abs{\eta_j(x)} \leq 1-\epsilon_0$ for all $x\in \xf$. 
\end{enumerate} 
Then 
\begin{equation}\label{eq:stability-close}
\forall j=1,\ldots,s,\quad \abs{\hat a_j - a_j} \lesssim
\norm{p_j}\pa{\delta + \lambda \norm{p_j}} + \bep^{-1}\pa{\delta\norm{p} + \abs{\tilde \mu_0}(\Xx)}
\end{equation}
where $p$ is as in Proposition \ref{prop:robustness}.
\end{prop}

\begin{proof}
First observe that  writing $\nu= \hat \mu-\mu_0$, we have
\begin{align*}
\abs{\hat a_j - a_j} = &\abs{\int_{\xn_j} \mathrm{d}\nu(x)} =\abs{ \int_{\Xx}\fulleta_j(x)\mathrm{d}\nu(x) + \int_{\xn_j} (1-\fulleta_j(x))\mathrm{d}\nu(x)  - \sum_{\ell \neq j}\int_{\xn_\ell}\fulleta_j(x)\mathrm{d}\nu(x) -  \int_{\xf} \fulleta_j(x) \mathrm{d}\nu(x)}\\
&\leq \abs{ \int_{\Xx}\fulleta_j(x)\mathrm{d}\nu(x)} + \bla \sum_{j=1}^s \abs{ \int_{\xn_j} {\dsep_\met}(x,x_j)^2 \mathrm{d}\nu(x)} + (1-\bep) \abs{\nu}(\xf).
\end{align*}
Using \eqref{eq:main-noise-stability}, we have
$
\abs{\nu}(\xf) = \abs{\hat \mu}(\xf) \lesssim \bep^{-1}\pa{\delta\norm{p} + \abs{\tilde\mu_0}(\Xx)}
$ and
\begin{align*}
\bla \sum_{j=1}^s \abs{ \int_{\xn_j} {\dsep_\met}(x,x_j)^2 \mathrm{d}\nu(x)} &= \bla \sum_{j=1}^s \abs{ \int_{\xn_j} {\dsep_\met}(x,x_j)^2 \mathrm{d}\hat \mu(x)} \leq \delta\norm{p} + \abs{\tilde\mu_0}(\Xx)
\end{align*}
Finally, by \eqref{eq:discreploss},
\begin{align*}
\abs{ \int_{\Xx}\fulleta_j(x)\mathrm{d}\nu(x)}& \leq \abs{ \dotpmeas{\eta_j}{\hat\mu - \bar\mu_0}} + \abs{\tilde\mu_0}(\Xx) \leq \norm{p_j} \norm{\Phi (\hat \mu - \bar\mu_0)} + \abs{\tilde\mu_0}(\Xx)\\
&\leq \norm{p_j}(\delta + \norm{\Phi \hat \mu - y})+ \abs{\tilde\mu_0}(\Xx) \leq \norm{p_j}\pa{2\delta + 2\la \norm{p} + 2\sqrt{ \la \abs{\tilde \mu_0}(\Xx)}}+ \abs{\tilde\mu_0}(\Xx)\\
&\leq  {2\delta\norm{p_j} + 2\la \norm{p}\norm{p_j} +  \la\norm{p_j}^2 + 2\abs{\tilde \mu_0}(\Xx)}
\end{align*}
using $\sqrt{ab}\leq (a+b)/2$. Therefore, we obtain
\begin{equation*}
 \abs{\hat a_j - a_j} \lesssim
\norm{p_j}\pa{\delta + \lambda \norm{p_j}} + \bep^{-1}\pa{\delta\norm{p} + \abs{\tilde \mu_0}(\Xx)}
\end{equation*}

\end{proof}

\paragraph{Additional certificates.} Proposition \ref{prop:stab_near} assumes the construction of additional functions $\eta_j \in \Im(\Phi^*)$, which are essentially similar to non-degenerate certificates but with all ``signs'' to interpolate put to $0$ except for one. As we will see, they are even simpler to construct than $\eta$: indeed, the reason one has to resort to the random signs assumption (as in \cite{tang2013compressed}) or to the golfing scheme (as in this paper) is that the Euclidean norm of the vector of signs $(\sign(a_i))_{i=1}^s$ appears in the proof, which results in a spurious term $\sqrt{s}$. When constructing the $\eta_j$, this problem does not occur, since only one sign is non-zero. 

%
%

\paragraph{Relation to previous works.}
Note that \eqref{eq:main-noise-stability} and \eqref{eq:stability-close}, without the inexact sparsity term $ \abs{\tilde \mu_0}(\Xx)$, were previously presented in \cite{fernandez-support2013} in the context of sampling Fourier coefficients and in a  more general setting in \cite{azais-spike2014}. However, the statement in \cite{azais-spike2014} is given in terms of orthonormal systems, and the so-called Bernstein Isolation Property which imposes that $\abs{P'(x)} \leq C m^2 \norm{P}_\infty$ for all $P\in \Im(\Phi^*)$. These conditions can be difficult to check in our setting of random sampling and were imposed only to ensure the existence of nondegenerate dual certificates, and to have explicit control on the constant $C$. For completeness, we still present the proof of \eqref{eq:main-noise-stability} under nondegeneracy assumptions, and we later establish that these nondegeneracy assumptions hold, under appropriate separation conditions imposed via $\dsep_\met$.

In \cite{candes-superresolution2013}, one could also obtain bounds $\sum_{j=1}^s \abs{\hat a_j - a_j} \lesssim \delta$ in the case of Fourier sampling, however, to prove such a statement, one is required to construct a trigonometric function (a dual certificate) which interpolates arbitrary sign patterns. In the case of subsampling, such an approach cannot lead to sharp dependency on $s$, since in the real setting, one is then required to show the existence of $2^s$ random polynomials corresponding to all possible sign patterns. We therefore settle for the bound \eqref{eq:stability-close} in this paper. We remark that being able to construct dual functions which interpolate arbitrary signs patterns lead to Wasserstein-1 error bounds, as opposed to Wasserstein-2 error bounds presented here.

Finally, we mention the more recent work of \cite{eftekhari2018sparse} which presents stability bounds for the sparse spikes problem where one restricts to positive measures and where the sampling functions form a T-systems. Under a positivity constraint (rather than total variation penalisation), they derive stability bounds in terms of optimal partial transport distances. 
We stress that since we consider more general measurement operators than T-systems in this work,  we consider transport distances under the Fisher metric as opposed to the Euclidean metric. Moreover, another difference is that our error bounds use  the Wasserstein-2 distance, whereas they use the Wasserstein-1 distance -- the reason is that since they do not consider random subsampling, their proofs in fact follow the work of \cite{candes-superresolution2013} to construct dual certificates which interpolate arbitrary sign patterns.


\newcommand{\normH}[1]{\norm{#1}_{\met}}

\section{Nondegenerate limit certificates}\label{sec:limit-cert}

In this section, we provide the second building block of our main theorem: a generic way to ensure the existence and construct non-degenerates dual certificates, when $m\to \infty$ and the sought-after Diracs satisfy a minimal separation condition with respect to the metric $\dsep_\met$.

\subsection{Notions of differential geometry}\label{sec:diff_geo}

We start with additional definitions in differential Riemannian geometry. All these notions can be found in the textbook \cite{absil2014book}, to which we refer the reader for further details. In many instances, we extend classical definitions to the complex case in a natural way.

\paragraph{Riemannian gradient and Hessian.} Let $f: \RR^d \to \CC$ be a smooth function. The Riemannian gradient $\grad f(x) \in \CC^d$ and Riemannian Hessian $\Hess f(x) : \CC^d \to \CC^d$, which is a linear mapping, can be defined as:
\begin{align*}
\grad f(x) &= \met_x^{-1} \nabla f(x) \\
\dotp{\Hess f(x)[e_i]}{e_j}_x &= \partial_i \partial_j f(x) - \Gamma_{ij}(x)^\top \nabla f(x)
\end{align*}
where $\nabla, \partial_i$ are the classical Euclidean gradient and partial derivatives, and the $\{e_i\}$ are the canonical basis of $\RR^d$. The $\Gamma_{ij}(x) = [\Gamma_{ij}^k(x)]_k \in \RR^d$ are the Christoffel symbols, here equal to:
\[
\Gamma_{ij}^k(x) = \frac12 \sum_\ell g^{k\ell}(x)\pa{\partial_i g_{\ell j}(x) + \partial_j g_{\ell i}(x) - \partial_\ell g_{i j}(x)}\, ,
\]
where $g_{ij}(x) = [\met_x]_{ij}$ and $g^{ij}(x) = [\met_x^{-1}]_{ij}$. Finally we denote by $\Hmtx f(x)\in \CC^{d\times d}$ the matrix that contains these terms: $\Hmtx f(x) \eqdef \Big(\dotp{\Hess f(x)[e_i]}{e_j}_x\Big)_{ij}$.

For $r=0,1,2$, the ``covariant derivative" $\diff{r}{f}(x): (\CC^d)^r \to \CC$ are mappings (or scalar in the case $r=0$) defined as:
\begin{align*}
\diff{0}{f}(x) &\eqdef f(x) \\
\diff{1}{f}(x)[v] &\eqdef \dotp{v}{\grad f(x)}_x = v^* \nabla f(x) \\
\diff{2}{f}(x)[v,v'] &\eqdef \dotp{\Hess f(x)[v]}{v'}_x = v^* \Hmtx f(x) v'
\end{align*}
We define associated operator norms
\begin{align*}
\norm{\diff{1}{f}(x)}_x &\eqdef \sup_{\norm{v}_x = 1} \diff{1}{f}(x)[v] = \norm{\met_x^{-\frac12} \nabla f(x)}_2 \\
\norm{\diff{2}{f}(x)}_x &\eqdef \sup_{\norm{v}_x = 1, \norm{v'}_x = 1} \diff{2}{f}(x)[v,v'] = \norm{\met_x^{-\frac12} \Hmtx f(x) \met_x^{-\frac12}}_{2}
\end{align*}
where we recall that $\norm{\cdot}_x$ is defined by \eqref{eq:dotpx}.

\paragraph{Covariant derivatives of the kernel.} Recall the definition of the limit covariance kernel \eqref{eq:fullCov}. Given $0\leq i,j \leq 2$, let $\fullCov^{(ij)}(x,x')$ be a ``bi''-multilinear map, defined  for $Q\in (\CC^d)^i$ and $V\in (\CC^d)^j$ as
\begin{equation}\label{eq:defKij}
[Q]\fullCov^{(ij)}(x,x')[V] \eqdef \EE[ \overline{\diff{i}{\phi_\om}(x)[Q]} {\diff{j}{\phi_\om}(x')[V]}].
\end{equation} 
In the case $i,j \leq 1$, note that these admits simplified expressions: $\fullCov^{(00)}(x,x') = \fullCov(x,x')$, $[v]\fullCov^{(10)}(x,x') = v^\top \nabla_1 \fullCov(x,x')$ and $[v]\fullCov^{(11)}(x,x')[v'] = v^\top \nabla_1 \nabla_2 \fullCov(x,x') \overline{v'}$.
Define the operator norm of $\fullCov^{(ij)}(x,x')$ as
\begin{equation}\label{eq:defKijnorm}
\norm{K^{(ij)}(x,x')}_{x,x'} \eqdef \sup_{Q,V}\abs{[Q]{K^{(ij)}(x,x')}[V]}
\end{equation}
where the supremum is over all $V = [v_1,\ldots, v_i]$ with $\norm{v_\ell}_{x}\leq 1$ for all $\ell \in [i]$, and all $Q = [q_1,\ldots, q_j]$ with $\norm{q_\ell}_{x'}\leq 1$ for all $\ell \in [j]$. We will sometimes overload the notations and write $\norm{\cdot}_{x}$ when the dependence is only on  $x$, i.e. for $K^{(ij)}$ where $j = 0$. Note that, in particular,
\begin{equation}\label{eq:matnorm_riemann2euclid}
\begin{split}
\norm{K^{(10)}(x,x')}_x = \norm{\met_x^{-\frac12}\nabla_1 K(x,x')}_2, \quad \norm{K^{(11)}(x,x')}_{x,x'} = \norm{\met_x^{-\frac12}\nabla_1 \nabla_2 K(x,x')\met_{x'}^{-\frac12}}_2\\
\qquad \qandq \norm{K^{(20)}(x,x')}_{x} = \norm{\met_x^{-\frac12}\Hmtx[K(\cdot,x')](x)\met_{x}^{-\frac12}}_2
\end{split}
\end{equation}
All these definitions are naturally extended to the covariance kernel $\Cov$ by replacing the expectation $\EE$ in \eqref{eq:defKij} by an empirical expectation over $\om_1,\ldots, \om_m$.

%

\subsection{Non-degenerate dual certificate with $m \to \infty$}\label{subsec:m_infty}

Recall the definition of the covariance kernel \eqref{eq:cov}. Following \cite{candes-towards2013}, a natural approach towards constructing a dual certificate is by interpolating the sign vector $\sign(a_j)$ using the functions $\hat K(x_j,\cdot)$ and $ \hat K^{(10)}(x_j,\cdot)$, since we have
$$
\enscond{\eta \eqdef \sum_{j=1}^s \alpha_{1,j}  \Cov(x_j,\cdot) + \sum_{j=1}^s [\alpha_{2,j}]  \Cov^{(10)} (x_j,\cdot)}{\alpha_{1,j} \in \CC,~\alpha_{2,j} \in \CC^d} \subset \Im(\Phi^*)\, .
$$
Using the gradients of the kernel allows to additionally impose that $\nabla \eta(x_i) = 0$, which is a necessary (but not sufficient) condition for the dual certificate to reach its maximum amplitude in $x_i$. Usual proofs then show that, under minimal separation, applying this strategy indeed yields a non-degenerate dual certificate. 

We first consider the case where one has access to arbitrarily many measurements ($m \to \infty$), and to this end, we  consider the limit covariance kernel $\fullCov$ defined in \eqref{eq:fullCov}. Let us introduce some handy notations that will be particularly useful in later proofs (Section \ref{sec:main-proof}). Our aim is to find coefficients $(\alpha_{1,j})_{j=1}^s \in \CC^s$ and $(\alpha_{2,j})_{j=1}^s \in (\CC^{d})^s$ such that
\begin{equation}\label{eq:certificate}
\eta \eqdef \sum_{j=1}^s \alpha_{1,j}  K(x_j,\cdot) + \sum_{j=1}^s [\alpha_{2,j}]  K^{(10)} (x_j,\cdot)
\end{equation} 
satisfies $
\eta(x_j) = \sign(a_j)$ and $ \nabla \eta(x_j) = 0$ for all $j=1,\ldots,s$. 
Note that these $s(d+1)$ constraints can be written as the linear system 
\begin{equation}\label{eq:linsys1}
\etaMat\binom{\alpha_1}{\alpha_2} = \binom{(\sign(a_i))_{i=1}^s}{0_{sd}} \eqdef \SignVecPad_s\, ,
\end{equation}
where $\etaMat \in \RR^{s(d+1) \times s(d+1)}$ is a real symmetric matrix defined as
\begin{equation}\label{eq:fullmtx_func}
\etaMat \eqdef \EE_\om [{\RFVec(\om)\RFVec(\om)^*}] \in \CC^{s(d+1) \times s(d+1)},
\end{equation}
with the vector $\RFVec(\om)\in \CC^{s(d+1)}$ defined as
\begin{equation}\label{eq:gamma_vec}
\RFVec(\om) \eqdef \pa{ \pa{\phi_\om(x_i)}_{i=1}^s,\pa{\nabla \phi_\om(x_i)^\top}_{i=1}^s}^\top.
\end{equation}
Assuming that $\etaMat$ is invertible, we can therefore rewrite \eqref{eq:certificate} as $\eta(x) = (\etaMat^{-1} \SignVecPad_s)^\top \etaFunc(x)$,
where
\begin{equation}\label{eq:etaFunc}
\etaFunc(x) \eqdef \EE_\om [\overline{\RFVec(\om)} \phi_\om(x)] = \pa{ \pa{\fullCov(x_i,x)}_{i=1}^s,\pa{\nabla_1 \fullCov(x_i,x)^\top}_{i=1}^s}^\top \in \RR^{s(d+1)}\, .
\end{equation}
We also define the block diagonal  normalisation matrix  $D_\met\in \RR^{s(d+1)\times s(d+1)}$ as
\begin{equation}\label{eq:matrix-D-block}
D_\met \eqdef \begin{pmatrix}
\Id_{s} \\
& \met_{x_1}^{-\frac12} \\
&&\ddots\\
&&& \met_{x_s}^{-\frac12}
\end{pmatrix}
\end{equation}
so that $\tilde \etaMat = D_\met \etaMat D_\met$ has constant value 1 along its diagonal.

We will prove in Theorem \ref{thm:admiss-kernel} below that $\eta$ of the form \eqref{eq:certificate} is indeed nondegenerate, provided that there is  sufficient curvature on $K(x,\cdot)$ in a small neighbourhood around $x$  and  $\min_{k\neq j}\dsep_\met(x_j,x_k)\geq \Delta$ where $\Delta$ is the distance at which the kernel and its partial derivatives are sufficiently small (to allow for interpolation with $K(\cdot,x_j)$ with minimal inference between the point sources). To do so we need the following definition.

\begin{defn}\label{def:constker}
Given $r>0$, the local curvature constants $\constker_0(r)$ and $\constker_2(r)$ of $\fullCov$ are defined as
\begin{align*}
\constker_0(r)&\eqdef \sup\enscond{\epsilon}{\fullCov(x,x')\leq 1-\epsilon, \; \forall x,x'\in\Xx \text{ s.t. } \dsep_\met(x,x')\geq r}\\
\constker_2(r)&\eqdef \sup\enscond{\epsilon}{-K^{(02)}(x',x)[v,v] \geq \epsilon \norm{v}_{x}^2, \;\forall x,x'\in\Xx \text{ s.t. }\dsep_\met(x,x') <r, \forall v \in \RR^d }
\end{align*}
Given $h>0$ and $s\in\NN$, the kernel width of $\fullCov$ is defined as
$$
\Delta(h,s) \eqdef \inf \enscond{\Delta}{ \sum_{k=2}^{s} \norm{\fullCov^{(ij)}(x_1,x_k)}_{x_1,x_k} \leq h, \; (i,j) \in \ens{0,1}\times\ens{0,2} ,\; \{x_k\}_{k=1}^s \in\Ss_\Delta }
$$
where $\Ss_\Delta\eqdef \enscond{(x_k)_{k=1}^s\in\Xx^s}{\dsep(x_k,x_\ell)\geq \Delta,\; \forall k\neq \ell}$ is the set of $k$-tuples of $\Delta$-separated points.
We define $\inf \emptyset \eqdef +\infty$.
\end{defn}
Intuitively, these notions are similar to those appearing in the definition of non-degenerate dual certificates (and will ultimately serve in the proof of existence of such certificates): $r$ is a neighborhood size, $\constker_0$ represents the distance to $1$ of the kernel away from $x=x'$, and $\constker_2$ is the ``curvature'' of the kernel when $x \approx x'$. Finally, $\Delta$ is the ``minimal separation'' under which $s$ Diracs have minimal interference between them, or, in other words, the covariance kernel and its derivatives have low value. We formalize this in the following assumption.
\begin{ass}[Assumptions on the kernel.]\label{ass:kernel}
Suppose that $K$ is a real-valued kernel. For $i,j\leq 2$ and $i+j \leq 3$, assume that $B_{ij} \eqdef \sup_{x,x'\in\Xx}\norm{\fullCov^{(ij)}(x,x')}_{x,x'}< \infty$ and denote $B_i\eqdef B_{0i} + B_{1i}+1$. Assume that $K$ has positive curvature constants $\constker_0$ and $\constker_2$ at radius $0< \rnear< B_{02}^{-\frac12}$.  Let $s \in\NN$ be such that $\Delta \eqdef  \Delta(h,s)<\infty$ with $h \leq \frac{1}{64}\min\pa{\frac{\constker_0}{B_0},\frac{\constker_2}{B_2}}$.
\end{ass}

Under this assumption, the following theorem, which is the main result of this section, proves that a limit non-degenerate dual certificate can be constructed under minimal separation.
\begin{thm}\label{thm:admiss-kernel}
Under Assumption \ref{ass:kernel}, for all $\{x_k\}_{k=1}^s$ with $\min_{k\neq \ell}\dsep_\met(x_k,x_\ell) \geq \Delta$, there exists a unique function $\eta$ of the form \eqref{eq:certificate} which is $(\frac{\constker_0}{2},\frac{\constker_2}{4}, \rnear)$-nondegenerate. Moreover,
$$
\norm{\overline{\sign(a_j)}\diff{2}{\eta}(x) - \fullCov^{(02)}(x_j,x) }_x \leq \frac{\constker_2}{16} \qquad \forall  x\in\Bb_{\dsep_\met}(x_j;\rnear).
$$
\end{thm}
We delay the (slightly lengthy) proof of this result to the next subsection. Before that, we make a few comments.

\paragraph{Dependency on $s$.} As we have seen in the examples of Section \ref{sec:examples}, for a constant $h$ we generally let the minimal separation $\Delta = W(h,s)$ depend on $s$. Indeed, in dimension $d$, it is well known one can pack $C^d$ $\Delta$-separated points in a ball of radius $2\Delta$ for some constant $C$ (this is known as the kissing number). Hence, there exist $s$ $\Delta$-separated points such that
$$
 \sum_{k=2}^{s} \norm{\fullCov^{(ij)}(x_1,x_k)}_{x_1,x_k} \geq \min\pa{C^d, s} \sup_{\dsep(x,x')\geq \Delta}\norm{K^{(ij)}(x,x')}_{x,x'}\, .
$$
Therefore, while the kernel width can be independent of $s$ in low dimensions (and the trick is then to upper bound this by a constant bound $s\to \infty$, assuming the sum on the l.h.s. converges), as $d$ increases, the dependence on $s$ will become inevitable, otherwise $\Delta$ generally depends exponentially on $d$.

\paragraph{Babel function.}
The attentative reader might recognise the similarity of definition of kernel width $W(h,s)$ with the Babel function from compressed sensing \cite{tropp2004greed}, if we restrict the definition to $(i,j) = (0,0)$ and recall that $K(x,x') = \EE_\om[\overline{\phi_\om(x)}\phi_\om(x')]$. The Babel function of a $m\times N$ matrix $A$ with columns $\mathbf{a}_j$ is defined as
$$
\mu(s) = \max_{i\in [N]} \max \enscond{\sum_{j\in S}\abs{\dotp{\mathbf{a}_i}{\mathbf{a}_j}}}{S\subset [N], \abs{S} = s, i\neq S},
$$
and small value of $\mu(s)$ ensure that the sub-matrix $A_S^* A_S$, where $A_S$ is the matrix $A$ restricted to index set $S$ with $\abs{S}\leq s$, is well conditioned and invertible. Furthermore, recovery guarantees for Basis Pursuit and Orthogonal Matching Pursuit can be stated in terms of $\mu(s)$. In Theorem \ref{thm:admiss-kernel}, sufficient kernel width  also ensures that $\Phi_x^* \Phi_x$ is well conditioned and thereby provide performance guarantees for the BLASSO.

\subsection{Proof of Theorem \ref{thm:admiss-kernel}}

Before proving Theorem \ref{thm:admiss-kernel}, we illustrate the link between \emph{curvature of the kernel} as represented by $\constker_2$ in Def.~\ref{def:constker} and the \emph{quadratic decay} condition $\abs{\eta} \leq 1-\varepsilon \dsep_\met(x_i,\cdot)^2$ that we used in the definition of non-degenerate certificates (Def.~\ref{def:nondegen}). The resulting condition \eqref{eq:curv2quaddecay} is the one that we are actually going to prove in practice. The following Lemma is based on a generalized second-order Taylor expansion.
\begin{lem}\label{lem:curv2quaddecay}
Let $x_0 \in \Xx$ and $a \in \CC$ with $\abs{a} = 1$. Suppose that for some $\epsilon>0$, $B>0$ and $0<r\leq B^{-\frac12}$ we have: for all $x \in \Bb_{\dsep_\met}(x_0;r)$ and $v \in \CC^d$, it holds that $-K^{(02)}(x_0,x)[v,v] \geq \epsilon \norm{v}_{x}^2$ and $\norm{K^{(02)}(x_0,x)}_x \leq B$. Let $\eta:\Xx \to \CC$ be a smooth function.
\begin{enumerate}[label=(\roman*)]
\item If $\eta(x_0) = 0, \nabla \eta(x_0) = 0$ and 
\begin{equation}\label{eq:curvzero2quaddecay}
\norm{\diff{2}{\eta}(x)}_x \leq \delta \qquad \forall  x\in\Bb_{\dsep_\met}(x_0;r)
\end{equation}
then $\abs{\eta(x)} \leq \delta \dsep_\met(x_0,x)^2$ for all $x\in\Bb_{\dsep_\met}(x_0;r)$.
\item If $\eta(x_0) = a$, $\nabla \eta(x_0) = 0$ and
\begin{equation}\label{eq:curv2quaddecay}
\norm{\overline{a}\diff{2}{\eta}(x) - \fullCov^{(02)}(x_0,x) }_x \leq \delta \qquad \forall  x\in\Bb_{\dsep_\met}(x_0;r)
\end{equation}
for some $\delta < \frac{\varepsilon}{2}$, then, for all $x\in\Bb_{\dsep_\met}(x_0;r)$ we have $\abs{\eta(x)} \leq 1- \varepsilon' \dsep_\met(x_0,x)^2$ with $\varepsilon' = \frac{\varepsilon - 2\delta}{2}$.
\end{enumerate}
\end{lem}

\begin{proof}
We prove $(ii)$, the proof for $(i)$ is similar and simpler. Using \eqref{eq:curv2quaddecay} and the assumption on $\fullCov^{(02)}$, we can deduce that for all $v\in\RR^d$ we have
\begin{align*}
\rep{\overline{a} \diff{2}{\fulleta}(x)[v,v]} \leq -(\varepsilon-\delta)\norm{v}_x^2 \qandq \abs{\imp{\overline{a} \diff{2}{\fulleta}(x)[v,v]}} \leq \delta\norm{v}_x^2
\end{align*}
Given a geodesic $\gamma:[0,1] \to \RR^d$, it is a well-known property that
$
\ddot{\gamma} + \sum_{i,j} \Gamma_{ij}(\gamma) \dot{\gamma_i}  \dot{\gamma_j} = 0
$
where we recall that $\Gamma_{ij} \in \RR^d$ are the Christoffel symbols.
Therefore, we have
\begin{align*}
\frac{\mathrm{d}^2}{\mathrm{d}t^2}\eta(\gamma(t)) &= \dot{\gamma}(t)^\top\nabla^2 \eta(\gamma(t)) \dot{\gamma}(t) + \nabla \eta(t)^\top \ddot{\gamma}(t)\\
&=  \dot{\gamma}(t)^\top\nabla^2 \eta(\gamma(t)) \dot{\gamma}(t) - \nabla \eta(t)^\top \pa{\sum_{ij} \Gamma_{ij}(\gamma(t)) \dot{\gamma_j}(t)  \dot{\gamma_k}(t)}
\\
&= \dot{\gamma}(t)^\top \Hmtx \eta(\gamma(t)) \dot{\gamma}(t) = \diff{2}{\eta}(\gamma(t)) [\dot{\gamma}(t), \dot{\gamma}(t)]
\end{align*}
So, given any geodesic path with $\gamma(0) = x_0$ and $\gamma(1) = x$, since of course we have $\dsep_\met(x_0,\gamma(t))\leq\dsep_\met(x_0,x)\leq r$, by applying the inequalities above:
\begin{equation}\label{eq:taylor}
\begin{split}
\rep{\overline{a} \eta(x)} &= \rep{\overline{a} \pa{\eta(x_0) + \nabla \eta(x_0)^\top \dot{\gamma}(0) + \frac12 \int_0^1 (1-t) \frac{\mathrm{d}^2}{\mathrm{d}t^2}\eta(\gamma(t)) d t}} \\
&= 1 + \frac12 \int_0^1 (1-t) \rep{\overline{a} \diff{2}{\eta}(\gamma(t)) [\dot{\gamma}(t), \dot{\gamma}(t)]} \mathrm{d} t\\
&\leq 1- \pa{\varepsilon-\delta} \int_0^1 (1-t) \norm{\dot{\gamma}(t)}_{\gamma(t)}^2 \mathrm{d}t = 1- \frac{\pa{\varepsilon-\delta}}{2}\dsep_\met(x_0,x)^2.
\end{split}
\end{equation}
where the last line follows because $\norm{\dot{\gamma}(t)}_{\gamma(t)}$ is constant  for all $t\in[0,1]$. Similarly, we can show that $\rep{\overline{a} \eta(x)} \geq 1-\frac{B+\delta}{2} \dsep_\met(x_0,x)^2 \geq 0$ since $r\leq B^{-\frac12}$, and
$
\abs{\imp{\overline{a} \eta(x)}} \leq \frac{\delta}{2}\dsep_\met(x_0,x)^2
$,
from which we got $\abs{\eta(x)} \leq \rep{\overline{a} \eta(x)} + \abs{\imp{\overline{a} \eta(x)}} \leq 1- \frac{\varepsilon - 2\delta}{2} \dsep_\met(x_0,x)^2$.
\end{proof}

We can now proceed with the proof of Theorem \ref{thm:admiss-kernel}.

\begin{proof}[Proof of Theorem \ref{thm:admiss-kernel}]
Recall the block diagonal matric $D_\met$ from \eqref{eq:matrix-D-block}.
The system \eqref{eq:linsys1} is equivalent to
\begin{equation}\label{eq:eta_coeffs0}
\tilde \etaMat \binom{\tilde \alpha_1}{\tilde \alpha_2} = \SignVecPad_s.
\end{equation}
where $\tilde \etaMat = D_\met \etaMat D_\met$ and $\tilde \alpha = D_\met^{-1} \alpha$.
So, if $\tilde \etaMat$ is invertible, then we can write $\eta = \pa{\tilde \etaMat^{-1} \SignVecPad_s}^\top D_\met \etaFunc  = \pa{\etaMat^{-1} \SignVecPad_s}^\top  \etaFunc$. Therefore, we will proceed as follows: First, prove that $\tilde \etaMat$ is invertible. Second, bound the coefficients $\alpha_1$ and $\alpha_2$.  Third, prove that $\eta$ is nondegenerate.

We first prove that the matrix $\tilde \etaMat$ is invertible. To this end, we decompose it into blocks
\begin{equation}
\label{eq:Rdivide}
\tilde \etaMat =  \pa{\begin{matrix}
 \etaMat_0 &  \etaMat_1^\top \\
 \etaMat_1 &  \etaMat_2
\end{matrix}}
\end{equation}
where  $\etaMat_0\in \CC^{s\times s}$, $\etaMat_1  \in \CC^{sd\times s}$ and $\etaMat_2  \in \CC^{sd \times sd}$ are defined as
\begin{align*} 
&\etaMat_0 \eqdef (\fullCov(x_i,x_j))_{i,j=1}^s, \qquad
\etaMat_1 \eqdef(\met_{x_i}^{-\frac12} \nabla_1\fullCov(x_i,x_j))_{i,j=1}^s, \quad \etaMat_2 \eqdef (\met_{x_i}^{-\frac12} \nabla_1\nabla_2\fullCov(x_i,x_j) \met_{x_j}^{-\frac12})_{i,j=1}^s.
\end{align*}

To prove the invertibility of $\tilde \etaMat$, it suffices to prove that $\etaMat_2$ and its Schur complement  $\etaMat_S \eqdef \etaMat_0 - \etaMat_1 \etaMat_2^{-1} \etaMat_1^\top$ are both invertible. To show that $\etaMat_2$ is invertible, we define $A_{ij} = \met_{x_i}^{-\frac12}\nabla_1\nabla_2\fullCov(x_i,x_j) \met_{x_j}^{-\frac12}$, such that $\etaMat_2$ has the form:
\[
\etaMat_2 = \pa{\begin{matrix}
\Id & A_{12} & \ldots &  A_{1s}  \\
A_{21} & \Id & \ddots & \vdots \\
\vdots & \ddots & \ddots & \vdots \\
A_{s1}& \ldots & \ldots & \Id
\end{matrix}}
\]
and by Lemma \ref{lem:block_norm} in Appendix \ref{app:lin-alg}, Assumption \ref{ass:kernel} and \eqref{eq:matnorm_riemann2euclid}, we have
\begin{align*}
\normblock{\Id - \etaMat_2} \leq&~ \max_{i} \sum_{j\neq i} \norm{A_{ij}}_2  =\max_{i} \sum_{j\neq i} \norm{\fullCov^{(11)}(x_i,x_j)}_{x_i,x_j} \leq  h \leq 1/32.
\end{align*}
Since $\normblock{\Id - \etaMat_2} <1$, $\etaMat_2$ is invertible, and we have $\normblock{\etaMat_2^{-1}} \leq \frac{1}{1 - \normblock{\Id - \etaMat_2}} \leq \frac{4}{3}$. Next, again with Lemma \ref{lem:block_norm}, we can bound
\begin{align*}
\norm{\Id - \etaMat_0}_\infty =&~ \max_{i} \sum_{j\neq i}\abs{\fullCov(x_i,x_j)} \leq h 
\\
\norm{\etaMat_1}_{\infty \to \textup{block}} \leq&~\max_{i}\sum_j \norm{\met_{x_i}^{-\frac12}\nabla_1\fullCov(x_i,x_j)}_2= \max_{i}\sum_j \norm{\fullCov^{(10)}(x_i,x_j)}_{x_i} \leq h
\end{align*}
since $\fullCov^{(10)}(x,x)=0$. Hence, we have
\begin{equation}
\norm{\Id - \etaMat_S}_\infty \leq \norm{\Id-\etaMat_0}_\infty + \norm{\etaMat_1^\top}_{\textup{block}\to \infty}\normblock{\etaMat_2^{-1}} \norm{\etaMat_1}_{\infty \to \textup{block}} \leq h + \frac{4}{3}h^2 \leq 2h\eqdef h' <1. 
\end{equation}
Therefore the Schur complement of $\tilde \etaMat$ is invertible and so is $\tilde \etaMat$. Moreover, $\norm{\etaMat_S^{-1}}_\infty \leq \frac{1}{1-h'}$.

We can now define:
\[
\tilde \etaCoeff = \tilde \etaMat^{-1} \SignVecPad_s = \pa{\begin{matrix}\tilde  \etaCoeff_1 \\\tilde \etaCoeff_2 \end{matrix}}
\]
and, as described above, $\alpha = D_\met^{-1} \tilde \alpha$. The Schur's complement of $\tilde \etaMat$ allows us to express $\etaCoeff_1$ and $\etaCoeff_2$ as
\begin{equation}
\binom{\tilde \etaCoeff_1}{\tilde \etaCoeff_2} =\binom{ \etaMat_S^{-1} \sign(a)}{ -\etaMat_2^{-1} \etaMat_1 \etaMat_S^{-1} \sign(a)}
\end{equation}
and therefore we can bound
\begin{align*}
\norm{\etaCoeff_1}_\infty &\leq \norm{\etaMat_S^{-1}}_\infty \leq \frac{1}{1-h'} \\
\max_i \norm{ \etaCoeff_{2,i}}_{x_i} &= \normblock{\tilde \alpha_2} \leq\normblock{\etaMat_2^{-1}} \norm{\etaMat_1}_{\infty \to \textup{block}} \norm{\etaMat_S^{-1}}_\infty \leq 4h
\end{align*}
Moreover, we have
\begin{equation}\label{eq:alphaCloseToSign}
\norm{\etaCoeff_1 - \sign(a)}_\infty \leq \norm{\Id - \etaMat_S^{-1}}_\infty \leq \norm{\etaMat_S^{-1}}_\infty \norm{\Id-\etaMat_S}_\infty\leq \frac{h'}{1-h'}
\end{equation}

We can now prove that $\fulleta$ is non-degenerate.
For any $x$ such that $\dsep_\met(x, x_i)\geq \rnear$ for all $x_i$'s,  there exists at most one index $i$ such that $\dsep_\met(x,x_i) < \Delta/2$ and so, for all $j \neq i$, we have $\dsep_\met(x,x_j)\geq \Delta/2$. Therefore,
\begin{align*}
\abs{\fulleta(x)} =&~ \Bigg|\etaCoeff_{1,i} \fullCov(x_i,x) + \sum_{j \neq i} \etaCoeff_{1,j} \fullCov(x_j,x) + [\etaCoeff_{2,i}]\fullCov^{(10)}(x_i,x)  + \sum_{j \neq i}[\etaCoeff_{2,j}] \fullCov^{(10)}(x_j,x) \Bigg| \\
\leq&~ \norm{\etaCoeff_{1}}_\infty \pa{\abs{\fullCov(x_i,x)} + \sum_{j \neq i} \abs{\fullCov(x_j,x)}} + \max_i \norm{\etaCoeff_{2,i}}_{x_i} \pa{\norm{\fullCov^{(10)}(x_i,x)}_{x_i} + \sum_{j \neq i}\norm{\fullCov^{(10)}(x_j,x)}_{x_j}} \\
\leq&~ \frac{1}{1-h'}\pa{1-\constker_0 + h} + 4h\pa{B_{10} + h}
 \leq 1- \frac{\constker_0}{2}.
\end{align*}

Now, let $x$ be such that $\dsep_\met(x_i,x)\leq \rnear$.
Similarly, for all $j \neq i$ we have $\dsep_\met(x,x_j)\geq \Delta/2$. Observe that
\begin{align*}
\overline{\sign(a_i)} \diff{2}{\fulleta}(x) =&~ 
\fullCov^{(02)}(x_i,x) + \pa{\overline{\sign(a_i)} \etaCoeff_{1,i}-1}\fullCov^{(02)}(x_i,x)\\
&\quad + \overline{\sign(a_i)} \Bigg[  \sum_{j \neq i} \etaCoeff_{1,j} \fullCov^{(02)}(x_j,x) + [  \etaCoeff_{2,i}]\fullCov^{(12)}(x_i,x) + \sum_{j \neq i} [  \etaCoeff_{2,j}]\fullCov^{(12)}(x_j,x)\Bigg] 
\end{align*}
So,
\begin{align*}
&\norm{\overline{\sign(a_i)} \diff{2}{\fulleta}(x) - \fullCov^{(02)}(x_i,x)}_x\\
 &\leq \norm{
 \pa{\overline{\sign(a_i)} \etaCoeff_{1,i}-1}\fullCov^{(02)}(x_i,x) + \overline{\sign(a_i)} \Bigg[  \sum_{j \neq i} \etaCoeff_{1,j} \fullCov^{(02)}(x_j,x) + [ \etaCoeff_{2,i}] \fullCov^{(12)}(x_i,x) + \sum_{j \neq i} [  \etaCoeff_{2,j}]\fullCov^{(12)}(x_j,x)\Bigg] }_x\\
 &\leq \frac{h'}{1-h'} B_{02} + h \norm{ \alpha_1}_\infty + \max_i \norm{\etaCoeff_{2,i}}_{{x_i}}  \pa{B_{12} + h} \leq \frac{h'}{1-h'} B_{02} + \frac{h}{1-h'} + 4h B_{12} + 4h^2 \leq \frac{\constker_2}{16}
\end{align*}
We conclude using Lemma \ref{lem:curv2quaddecay} and $\frac{\constker_2 - 2 \constker_2/16}{2} \geq \constker_2/4$.
\end{proof}

\section{Sparse recovery}\label{sec:m-finite}

In this section, we formulate our main contribution, Theorem \ref{thm:main}, which is a detailed version of Theorem~\ref{thm:main-short}. In previous sections, we have shown that the existence of a non-degenerate dual certificates implies sparse recovery guarantees, and that in the limit case $m\to \infty$, a minimal separation assumption implies the existence of a dual certificate. Our main theorem is obtained by bounding the deviations from the limit case when $m$ is finite. We do so by extending the celebrated golfing scheme \cite{gross2011recovering} to the infinite-dimensional case.
We first begin by our assumptions on the feature functions $\phi_\om$.

\subsection{Almost bounded random features}

In order to bound the variation between $\fullCov$ and $\Cov$, we would ideally like the features $\varphi_\om$ and their derivatives to be uniformly bounded for all $\om$. However this may not be the case: think of $e^{i\om^\top x}$, which does not have a uniformly bounded gradient when the support of the distribution $\Lambda$ is not bounded. On the other hand, if $\Lambda(\om)$ has sufficient decay as $\norm{\om}$ increases, one could argue that the selected random features and their derivatives are uniformly bounded with high probability.
For $r\in \{0,1,2\}$, we define the random variables
\begin{equation}\label{eq:stochastic_grad_const}
L_r(\om) \eqdef \sup_{x\in \Xx} \norm{\diff{r}{\varphi_\om}(x)}_{x}.
\end{equation}
Note that $L_r(\omega)<\infty$ for each $\om$ since $\Xx$ is a bounded domain and $\phi_\om$ is smooth.

Since $\abs{\phi_\om(x) - \phi_\om(x')} = \abs{\int_{0}^1 \frac{d}{dt}\phi_\om(\gamma(t))\mathrm{d}t} = \abs{\int_{0}^1 \diff{1}{\phi_\om}(\gamma(t))[\dot{\gamma}(t)]\mathrm{d}t}$ for a smooth path from $x$ to $x'$, it is easy to see that
\begin{equation}\label{eq:feat_lip}
\abs{\phi_\om(x) - \phi_\om(x')} \leq L_1(\om) \dsep_\met(x,x')
\end{equation}
We will also require $\diff{2}{\phi_\om}(x)$ to be Lipschitz, to this end, we assume that for all $x,x'\in \Xx$, there exists $\tau_{x\to x'}:\CC^d \to \CC^d$ an isometric isomorphism with respect to $\met_x$, that is, such that $\dotp{u}{v}_{x} = \dotp{\tau_{x\to x'}u}{\tau_{x\to x'}v}_{x'}$, such that for all $\om$:
$$ 
L_3(\omega) \eqdef \inf \enscond{L>0}{ \sup_{\dsep_\met(x,x')\leq \rnear} \frac{\norm{ \diff{2}{\phi_\om}(x) - \diff{2}{\phi_\om}(x')[\tau_{x\to x'} \cdot, \tau_{x\to x'} \cdot] }_x}{\dsep_\met(x,x')} \leq L  } < \infty.
$$
where naturally
\[
\norm{ \diff{2}{\phi_\om}(x) - \diff{2}{\phi_\om}(x')[\tau_{x\to x'} \cdot, \tau_{x\to x'} \cdot] }_x =\sup_{\norm{u}_x \leq 1, \norm{v}_x \leq 1} \diff{2}{\phi_\om}(x)[u,v] - \diff{2}{\phi_\om}(x')[\tau_{x\to x'} u, \tau_{x\to x'} v] 
\]
and $\rnear$ comes from Assumption \ref{ass:kernel}.
One possible choice of $\tau_{x\to x'}$ is to choose the parallel transport along the unique geodesic connecting $x$ and $x'$. Another possible choice is to simply choose $\tau_{x\to x'}:v \mapsto \met_{x'}^{-\frac12} \met_x^{\frac12}v$. The latter choice  implies
\begin{equation}\label{eq:hessian_lipschitz_matrix_formulation}
\norm{ \diff{2}{\phi_\om}(x) - \diff{2}{\phi_\om}(x')[\tau_{x\to x'} \cdot, \tau_{x\to x'} \cdot] }_x  = \norm{\met_{x'}^{-\frac12} \Hmtx\phi_\om(x')\met_{x'}^{-\frac12} - \met_{x}^{-\frac12}   \Hmtx\phi_\om (x)\met_{x}^{-\frac12} }.
\end{equation}
which is a more convenient expression that we will use in the examples.

Finally, we let $F_r: [0,\infty) \to [0,1]$ be decaying tail functions such that
\begin{equation}\label{eq:tail_distr}
\PP_\om \pa{L_r(\om) > t} \leq F_r(t).
\end{equation}
Our sampling complexity will depend on the decay of these tail distributions so that the derivatives of the selected random features are bounded with high probability. A similar idea of stochastic incoherence was exploited in \cite{candes2011probabilistic} for deriving compressed sensing bounds.


\subsection{Main result}
Our main result is valid under the following assumption, which links the tail probabilities of the bounds on the feature functions and the final number of measurements $m$.
\begin{ass}[Assumption on the features and the sample complexity]\label{ass:feat}
For $\rho>0$, suppose that $m\in \NN$ and some constant $\{\Lu_i\}_{i=0}^3 \in \RR_+^4$ are chosen such that 
\begin{equation}\label{eq:stoc_lip_bd}
\begin{split}
\sum_{j=0}^3 F_j(\Lu_j) \leq \frac{\min(\constker_0,\constker_2,\rho)}{m} \qandq
\max_{j=0}^3 \pa{  \Lu_j^2 \sum_{i=0}^3 F_i({\Lu_i}) + 6\int_{\Lu_j}^\infty t F_j({t}) \mathrm{d}t } \leq \frac{\min\pa{\constker_0,\constker_2}}{m}
\end{split}
\end{equation}
and
\begin{equation}\label{eq:samp_ran}
m \gtrsim  s  \pa{C_1 \log(s) \log\pa{\frac{s}{\rho}} +C_2 \log\pa{\frac{(sN)^d}{\rho}}}
\end{equation}
where $N\eqdef  \frac{\Rr_\Xx \Lu_1}{\constker_0} +\frac{\rnear \Lu_{3}\Lu_0 + \Lu_2}{\constker_2}$, $C_1\eqdef (\Lu_{0}^2+\Lu_1^2) \sum_{r=0,2}\frac{B_r^2}{\constker_r^2}$, and $C_2\eqdef \frac{B_{22}\Lu_{01}^2}{B_2^2}  +  \sum_{r=0,2}\pa{\frac{\Lu_r^2}{\constker_r^2} + \frac{\Lu_{01}\Lu_r}{\constker_r}   }$ with $\Lu_{ij} = \sqrt{\Lu_i^2+\Lu_j^2}$.
\end{ass}

The constants $\Lu_r$ play the role of ``stochastic'' Lipschitz constant: for $r=0,1,2$, with high probability on $\om_j$, $\diff{r}{\phi_\om}(x)$ will be $\Lu_r$-bounded and $\Lu_{r+1}$-Lipschitz. The condition \eqref{eq:stoc_lip_bd} ensures that this is true with probability $1-\rho$, that is, with the same desired probability of failure. Then, the entire proof is done \emph{conditionally} on these bounds to hold.

Note also that, generally,  $\{\Lu_r\}$ depend on $m$, through \eqref{eq:stoc_lip_bd}. However, all our examples fall under two categories (see Sec.~\ref{sec:examples}):
\begin{enumerate}[label=(\roman*)]
\item either $\norm{\diff{r}{\phi_\om}(x)}_x$ is already uniformly bounded, in which case $\Lu_r$ can be chosen independently of $\rho$ and $m$, this is for instance the case of discrete Fourier sampling;
\item or the $F_r(t)$ are exponentially decaying, in which case we can show that $\Lu_r = \Oo\pa{\log\pa{\frac{m}{\rho}}^p}$ for some $p>0$, which only incurs additional logarithmic terms in the bound \eqref{eq:samp_ran}. This occurs in the case of sampling the Laplace transform or sampling the Fourier transform with respect to a Gaussian distribution.
\end{enumerate}

We are now ready to state the detailed version of Theorem~\ref{thm:main-short}, which is the main result of this paper.

\begin{thm}\label{thm:main}
Suppose that Assumptions \ref{ass:kernel} and \ref{ass:feat} hold. Let $y$ be as in \eqref{eq:meas} with $\min_{i\neq j}\dsep_\met(x_i,x_j)\geq \Delta$ and  $\norm{w}\leq \delta$. Then, with probability at least $1-\rho$,  any solution $\hat\mu$ of \eqref{eq:blasso} with $\la \sim \frac{\delta}{\sqrt{s}}$ satisfies
$$
\Tt_{\dsep_\met}^2 \pa{\abs{\hat\mu}, \sum_{j=1}^s \hat A_i \delta_{x_j} } \leq e
\qandq
\max_{i=1}^s \abs{\hat a_i - a_i} \leq e 
$$
where $\hat A_i = \abs{\hat\mu}\pa{\Bb_{\dsep_\met}(x_i;\rnear)}$, $\hat a_i = \hat\mu\pa{\Bb_{\dsep_\met}(x_i;\rnear)}$ and $e\lesssim \frac{1}{\min(\constker_0,\constker_2)}\pa{\abs{\tilde \mu_0}(\Xx) + \delta \cdot  \sqrt{s}  }$.
\end{thm}

The next section is dedicated to the proof of Theorem~\ref{thm:main} using an infinite-dimensional golfing scheme. Appendix~\ref{app:prelim} is dedicated to the proof of some technical Lemmas. Appendix~\ref{app:concentration} gathers all the concentration inequalities that we use in the golfing scheme, which are essentially many variants of Bernstein's inequality. Finally, Appendices \ref{app-discretefourier}, \ref{app-gaussian} and \ref{app-laplace} are dedicated to the computation of all the constants in Assumptions \ref{ass:kernel} and \ref{ass:feat} for the examples described in Section~\ref{sec:examples}, which can be quite verbose.

\newcommand{\rbe}{{\color{red} p_1 \constker_2}}
\newcommand{\rbee}{{\color{red}p_2\constker_2}}
\newcommand{\rbeo}{{\color{red} p_0 \constker_2}}
\newcommand{\rbeee}{{\color{red} p_3\constker_2}}
\newcommand{\rbeeee}{{\color{red} p_4\constker_2}}
\newcommand{\Cdel}{C_0}

\section{Proof of Theorem \ref{thm:main}}\label{sec:main-proof}
The main step towards proving Theorem \ref{thm:main} is to prove the existence of a dual certificate satisfying the properties described in Proposition \ref{prop:robustness}. More precisely, we are going to prove the following theorem.
\begin{thm}\label{thm:main-golf}
Suppose that Assumptions \ref{ass:kernel} and \ref{ass:feat} hold. Let $\ens{x_j}_{j=1}^s$ be such that $\min_{i\neq j}\dsep_\met(x_i,x_j)\geq\Delta$. Then, with probability at least $1-\rho$, there exists $p\in\CC^m$ with $\norm{p} \lesssim \sqrt{s}$ such that $\subeta = \Phi^* p$ is  $(\frac{\constker_0}{8}, \frac{3\constker_2}{8}, \rnear)$-nondegenerate.
\end{thm}
\paragraph{Outline of the proof.} The construction of the non-degenerate certificate includes several intermediate steps. As usual in this type of proof, we will first prove these properties on a finite $\varepsilon$-net that covers $\Xx$, then extend them to the whole space by regularity. Here we work with several nets $\xng_j\subset \xn_j$ and $\xfg \subset \xf$ whose precision will be adjusted later. The principle of the golfing scheme is  to work with an \emph{``approximate''} dual certificate $\etapp$ (which is actually not a dual certificate at all), then ``correct'' it to obtain the desired true certificate.
In details, we will go through the following steps:
\begin{enumerate}
\item First, show that with probability at least $1-\rho$, there is an approximate certificate $\etapp \in \Im(\Phi^*)$ such that, for some constant $c_0$ that will be adjusted later,
\begin{equation}\label{eq:prop_etapp}
\begin{dcases}
\sum_{j=1}^s \abs{\etapp(x_j) - \sign(a_j)}^2 + \norm{\diff{1}{\etapp}(x_j)}_{{x_j}}^2\leq c_0^2 &\text{for all $j=1,\ldots, s$} \\
\abs{\etapp(x)} \leq 1-\frac{\constker_0}{4} & \text{for all $x\in \xfg$} \\
\norm{ {\overline{\sign(a_j)}\diff{2}{\etapp}(x) - \fullCov^{(02)}(x_j,x)}  }_x \leq \frac{7\constker_2}{64} &\text{for all $j=1,\ldots, s$, $x\in \xng_j$}
\end{dcases}
\end{equation}
In other words, we relax the condition $\eta(x_j) = \sign(a_j)$, $\nabla \eta(x_j) = 0$, and replace it with the first equation above.
\item Second, correct the approximate certificate to obtain a function\footnote{Here we write $\subeta$ to distinguish from the ``limit'' certificate $\eta$ that we built in the case $m\to \infty$.} $\subeta \in \Im(\Phi^*)$ such that:
\begin{equation}\label{eq:prop_etagrid}
\begin{dcases}
\subeta(x_j) = \sign(a_j) \qandq \nabla \subeta(x_j) = 0 &\text{for all $j=1,\ldots, s$} \\
\abs{\subeta(x)} \leq 1-\frac{3\constker_0}{16} & \text{for all $x\in \xfg$} \\
\norm{ {\overline{\sign(a_j)}\diff{2}{\subeta}(x) - \fullCov^{(02)}(x_j,x)}  }_x \leq \frac{15\constker_2}{128} &\text{for all $j=1,\ldots, s$, $x\in \xng_j$}
\end{dcases}
\end{equation}
That is, $\subeta$ satisfy all the properties we want, but on the finite nets $\xfg, \xng_j$.
\item Third, bound the norm of the $p\in \CC^m$ corresponding to $\subeta = \Phi^* p$.
\item Then, use Assumption \ref{ass:feat} on the feature functions and the bound on $\norm{p}$ to show that actually, the $\subeta$ constructed above satisfy:
\begin{equation}\label{eq:prop_etafinal}
\begin{dcases}
\subeta(x_j) = \sign(a_j) \qandq \nabla \subeta(x_j) = 0 &\text{for all $j=1,\ldots, s$} \\
\abs{\subeta(x)} \leq 1-\frac{\constker_0}{8} & \text{for all $x\in \xf$} \\
\norm{ {\overline{\sign(a_j)}\diff{2}{\subeta}(x) - \fullCov^{(02)}(x_j,x)}  }_x \leq \frac{\constker_2}{8} &\text{for all $j=1,\ldots, s$, $x\in \xn_j$}
\end{dcases}
\end{equation}
which, by Lemma \ref{lem:curv2quaddecay}, will imply that $\subeta$ is non-degenerate with the desired constants and conclude the proof of Theorem \ref{thm:main-golf}.
\item In a fifth and final step, prove the existence of $s$ additional certificates $\subeta_j$ as appear in Prop.~\ref{prop:stab_near}. Combined with the existence of $\subeta$ and Prop.~\ref{prop:robustness} and \ref{prop:stab_near}, it concludes the proof of Theorem~\ref{thm:main}.
\end{enumerate}

We dedicate a subsection to each step of the proof. Before that, we start in the next subsection with some technical preliminaries and notations.

\subsection{Preliminaries}\label{sec:proofstep0}

Let us introduce some notations and show some technical bounds that will be handy. Recall the definitions of the sign vector $\SignVecPad_s$ from \eqref{eq:linsys1}, $\RFVec$, $\etaMat$ and $\etaFunc$ from \eqref{eq:gamma_vec}, \eqref{eq:fullmtx_func} and \eqref{eq:etaFunc}, and $D_\met$ from \eqref{eq:matrix-D-block}. We have the following additional bounds, whose proof, in Appendix \ref{app:proof_add}, follows similar arguments to that of Theorem \ref{thm:admiss-kernel}.
\begin{lem}\label{lem:additional_admissible}
Under Assumption \ref{ass:kernel}, $\etaMat$ and $\etaFunc$  defined as in \eqref{eq:fullmtx_func} and \eqref{eq:etaFunc} satisfy the following.
\begin{enumerate}[label=(\roman*)]
\item $\etaMat$ is invertible and  satisfies
\begin{equation}
\norm{\Id- D_\met \etaMat D_\met}_2\leq \frac12 \qandq \nB{\Id-D_\met \etaMat D_\met}\leq \frac12. \label{eq:bound_etaMat_specnorm}
\end{equation}
\item  For any vector $q \in \CC^{s(d+1)}$ and  any $x\in \Xx^\textup{far}$, we have
\begin{align}
\norm{D_\met \etaFunc(x)}_2 \leq B_0 \qandq  \abs{q^\top \etaFunc(x)} \leq B_0  \nB{D_\met^{-1} q} \label{eq:Bf0}
\end{align}
\item For any vector $q \in \CC^{s(d+1)}$ and any $x \in \Xx^\textup{near}$ we have the bound:
\begin{align}
\norm{\diff{2}{q^\top \etaFunc(.)}(x) }_x \leq\norm{D_\met^{-1}  q}B_2\qandq \norm{ \diff{2}{q^\top \etaFunc(.)}(x) }_x \leq \nB{D_\met^{-1} q}B_2 \label{eq:Bf2}
\end{align}
\end{enumerate}
\end{lem}

Now, for $\om_1,\ldots,\om_m$, denote the empirical versions of $\etaMat$ and $\etaFunc$ by:
\begin{equation}\label{eq:submtx_func}
\subetaMat \eqdef \frac{1}{m} \sum_{k=1}^m {\RFVec(\om_k)\RFVec(\om_k)^*} \qandq \subetaFunc(x) \eqdef \frac{1}{m} \sum_{k=1}^m \overline{\RFVec(\om_k)} \phi_{\om_k}(x).
\end{equation}
Recall the definition of $L_j(\om)$ and $\Lu_j$ in Assumption \ref{ass:feat}. Let the event $\Eve$ be defined by
\begin{equation}\label{eq:event}
\Eve \eqdef \bigcap_{k=1}^m E_{\omega_k} \qwhereq E_\om \eqdef \enscond{L_j(\om)\leq \Lu_j}{ j=0,1,2,3}.
\end{equation}
Since by Assumption \ref{ass:feat}, eq.~\eqref{eq:stoc_lip_bd}, we have $\PP(\Eve^c) \leq \rho$, a nondegenerate dual certificate can be constructed with probability at least $(1-\rho)^2 \geq 1-2\rho$ provided that, \emph{conditional on event $\Eve$}, a nondegenerate dual certificate can be constructed with probability at least $1-\rho$.

We therefore assume for the rest of this proof that event $\Eve$ holds and establish the probability conditional on $\Eve$ that a nondegenerate dual certificate exists. To control this probability, we will need to control the deviation of $\subetaFunc$ and $\subetaMat$ from their conditional expectations $\etaFunc_\Eve = \EE_\Eve[\subetaFunc]$ and $\etaMat_\Eve \eqdef \EE_\Eve[\subetaMat]$, where we denote $\EE_\Eve [\cdot] \eqdef \EE[\cdot | \Eve]$. The following Lemma, proved in Appendix \ref{app:proof_cdf_bd}, bounds the deviations between these.

\begin{lem}\label{lem:cdf_bd}
Under Assumption \ref{ass:kernel} and \ref{ass:feat}, we have:
\begin{enumerate}[label=(\roman*)]
\item $\norm{ D_\met(\etaMat - \etaMat_\Eve)D_\met}_2\leq 4\frac{(s+1)\min(\constker_0, \constker_2)}{m}$ and $\nB{ D_\met(\etaMat - \etaMat_\Eve)D_\met}\leq 8\frac{(s+1)\min(\constker_0, \constker_2)}{m}$
\item for all $x \in \xf$, $\norm{D_\met(\etaFunc(x) - \etaFunc_\Eve(x))}_2 \leq \frac{(B_0 + 2 \sqrt{s}) \min(\constker_0, \constker_2)}{m}$
\item for all $x \in \xn$, $\sup_{\norm{q}_2\leq 1}\norm{\diff{2}{(\etaFunc - \etaFunc_\Eve)^\top  D_\met q}(x)}_x \leq \frac{(B_2 + 2 \sqrt{s}) \min(\constker_0, \constker_2)}{m}$
\end{enumerate}
\end{lem}

\subsection{Step 1: construction of an approximate certificate with the golfing scheme}\label{sec:proofstep1}

The first step is to construct an approximate certificate $\etapp$ using the so-called ``golfing scheme''. The golfing scheme was introduced in \cite{gross2011recovering} and successfully used in compressed sensing for instance in \cite{candes2011probabilistic}. It can be intuitively explained as follows. Recall that the certificate constructed in Theorem \ref{thm:admiss-kernel} in the case $m\to \infty$ is of the form $\fulleta = (\etaMat^{-1} \SignVecPad)^{\top}\etaFunc$. It is therefore natural to try to show directly that $\subeta \eqdef (\subetaMat^{-1} \SignVecPad)^{\top}\subetaFunc$ is also nondegenerate by bounding the variation between $\fulleta$ and $\subeta$. This is the strategy adopted by Tang et al  \cite{tang2013compressed} and in our previous work \cite{2019-Poon-aistats}. However, as mentioned before, this proof technique requires the random signs assumption, otherwise a sub-optimal bound on $m$ is obtained.
To solve this, the golfing scheme starts by writing the following Neumann expansion: assuming that $\subetaMat$ is invertible, we have
\begin{equation}\label{eq:etainv}
\begin{split}
\subeta &= (\subetaMat^{-1} \SignVecPad)^{\top}\subetaFunc 
 = ( \etaMat^{-1} (\subetaMat \etaMat^{-1})^{-1} \SignVecPad)^{\top}\subetaFunc 
\\
&
=  \sum_{\ell=1}^\infty  \pa{\etaMat^{-1} \pa{\Id - \subetaMat \etaMat^{-1} }^{\ell-1} \SignVecPad }^\top \subetaFunc = \sum_{\ell=1}^\infty (\etaMat^{-1}  q_{\ell-1})^\top \subetaFunc
\end{split}
\end{equation}
where $q_{\ell} \eqdef \pa{\Id - \subetaMat \etaMat^{-1} } q_{\ell-1} $, $q_0 \eqdef \SignVecPad$. By cutting the sum above to a finite number of terms, one effectively obtains an approximate certificate that must be later corrected. However, there is an additional difficulty in analysing the sum, which comes from the fact that for each summand, $\subetaFunc$ and $\etaMat^{-1} q_{\ell-1}$ are  random variables which are not mutually independent. The idea of \cite{gross2011recovering,candes2011probabilistic} is to \emph{decouple} the random variables by partitioning the indices $\{1,\ldots, m\}$ into $\Lev$ disjoint blocks $\Bb_\ell$ of size $m_\ell$ with $\sum_{\ell=1}^J m_\ell = m$, for some $\Lev$ and $m_\ell$ that are adjusted below. Denote by $\subetaMat_\ell$ and $\subetaFunc_\ell$ the empirical versions of $\etaMat$ and $\etaFunc$ over the $m_\ell$ random variables included in $\Bb_\ell$, that is:
\begin{equation*}
\subetaMat_\ell \eqdef \frac{1}{m_\ell} \sum_{k\in \Bb_\ell} {\RFVec(\om_k)\RFVec(\om_k)^*} \qandq \subetaFunc_\ell(x) \eqdef \frac{1}{m_\ell} \sum_{k\in \Bb_\ell} \overline{\RFVec(\om_k)} \phi_{\om_k}(x).
\end{equation*}
Then, instead of \eqref{eq:etainv},  we consider
\begin{align*}
\etapp  =  \sum_{\ell=1}^{\Lev} (\etaMat^{-1} q_{\ell-1})^{\top}\subetaFunc_\ell 
\end{align*}
where $q_{\ell} \eqdef \pa{\Id - \subetaMat_\ell \etaMat^{-1}} q_{\ell-1}$, $q_0 \eqdef \SignVecPad$. Note that this can be rewritten as:
\begin{equation}\label{eq:q}
q_\ell = \SignVecPad_s - \sum_{p=1}^\ell \subetaMat_p \etaMat^{-1} q_{p-1}
\end{equation}
Now, the idea is  that one can control each term $q_{\ell-1}^{\top}\subetaFunc_\ell$ conditional on $q_{\ell-1}$ and  for appropriate choices of the blocksizes $m_\ell$, $\etapp$ can be shown to be approximately nondegenerate with high probability. Each additional term in the sum brings the certificate ``closer'' to its desired properties, hence the term ``golfing'' scheme.

\paragraph{Parameters and intermediate assumptions.} We set the error $c_0$ that appears in \eqref{eq:prop_etapp} as $$c_0 = \Cdel\min\pa{\frac{\constker_0}{B_0},\frac{\constker_2}{B_2}, 1}$$ for some universal constant $\Cdel$. We define the parameters of our golfing scheme as follows: 
\begin{align*}
&\Lev = \lceil \log(s) \rceil +2\, , \\
& c_1 = c_2 = \frac{c_0}{4\sqrt{\log(s)}} &&\qandq \forall \ell=3,\ldots,\Lev, \quad c_\ell = c_0\, , \\
&t_1 = 1- \frac{\constker_0}{2} + \frac{\constker_0}{8} \quad \t_2 = 4 B_0\sqrt{\log(s)}, &&\qandq \forall \ell=3,\ldots,\Lev, \quad t_\ell = 4 B_0 \log(s)\, , \\
&b_1 = \frac{3\constker_2}{32}, 
\quad b_2 = 4 B_2 \sqrt{\log(s)}, &&\qandq \forall \ell=3,\ldots,\Lev, \quad b_\ell = 4 B_2 \log(s)\, .
\end{align*}
We now formulate an intermediate set of assumptions, and proceed to show that: first, they imply the desired properties on $\etapp$, and second, they are valid with high probability. For $1\leq \ell\leq \Lev$, we define:
\begin{itemize}
\item[(I$_\ell$)] $\nB{D_\met q_\ell} \leq c_\ell \nB{D_\met q_{\ell-1}}$,
\item[(II$_\ell$)] For all $x\in \xfg$, $ \abs{(\etaMat^{-1} q_{\ell-1})^\top \subetaFunc_\ell(x)} \leq t_\ell \nB{D_\met q_{\ell-1}}$,
\item[(III$_\ell$)] If $\ell=1$: for all $j=1,\ldots,s$, $x\in \xng_j$, $\norm{\overline{ \sign(a_j)} \diff{2}{(\etaMat^{-1} \SignVecPad_s)^\top \subetaFunc_1}(x) - K^{(02)}(x_j, x)}_x \leq b_1$; and if $\ell\geq 2$: for all $x\in \xng$, $\norm{{ \diff{2}{(\etaMat^{-1} q_{\ell-1})^\top \subetaFunc_\ell}(x)}}_x \leq b_\ell \nB{D_\met q_{\ell-1}}$.
\end{itemize}

Let us now assume that (I$_\ell$), (II$_\ell$) and (III$_\ell$) are true for all $\ell$, and show that $\etapp$ satisfy the desired properties. We define $\Psi: \Cder{}(\Xx) \to \CC^{s(d+1)}$ by
\begin{equation}\label{eq:psi_func}
\Psi f \eqdef \left[f(x_1), \ldots, f(x_s), \nabla{f}(x_1)^\top, \ldots, \nabla{f}(x_s)^\top \right]^\top.
\end{equation}
In words, $\Psi$  evaluates a function and its first derivative at the points $\{x_j\}_{j=1}^s$. Note that for any vector $v \in \CC^{s(d+1)}$, by definition we have $\Psi(v^\top \subetaFunc_\ell) = \subetaMat_\ell v$. Using this, we have
\begin{align*}
&\sqrt{\sum_{j=1}^s \abs{\etapp(x_j) - \sign(a_j)}^2 + \norm{\diff{1}{\etapp}(x_j)}_{{x_j}}^2} \\
&\qquad = \norm{\SignVecPad_s - D_\met\Psi \etapp}  \leq \sqrt{2s} \nB{D_\met\pa{\SignVecPad_s - \Psi \etapp}} =\sqrt{2s} \nB{D_\met\pa{\SignVecPad_s - \Psi \pa{\sum_{\ell=1}^{\Lev} (\etaMat^{-1} q_{\ell-1})^{\top}\subetaFunc_\ell }}} \notag \\
&\qquad =\sqrt{2s} \nB{D_\met\pa{\SignVecPad_s - \sum_{\ell=1}^{\Lev} \subetaMat_\ell \etaMat^{-1} q_{\ell-1}}} \stackrel{\eqref{eq:q}}{=} \sqrt{2s}\nB{D_\met q_\Lev} \leq \sqrt{s} \prod_{\ell=1}^\Lev c_\ell \stackrel{\text{(I)}}{\leq} \frac{\sqrt{2s}c_0^{\Lev}}{16 \log(s)} \leq c_0\, ,
\end{align*}
since by adjusting $\Cdel$ we can have $c_0 \leq \pa{\frac{1}{\sqrt{6}}}^{\frac{1}{\log(3)-1}} \leq \pa{\frac{1}{\sqrt{2s}}}^{\frac{1}{\log(s)-1}}$ where the last inequality is valid for all $s$ and results from a simple function study. It proves the first part of \eqref{eq:prop_etapp}. Next, for all $x\in \xfg$,
\begin{align*}
\abs{\etapp(x)} &\leq \sum_{\ell=1}^\Lev \abs{(\etaMat^{-1} q_{\ell-1})^\top \subetaFunc_\ell(x)} \stackrel{\text{(II)}}{\leq} \sum_{\ell=1}^\Lev t_\ell \nB{D_\met q_{\ell-1}} \stackrel{\text{(I)}}{\leq} \sum_{\ell=1}^\Lev t_\ell \prod_{p=1}^{\ell-1} c_p\\
&\leq 1-\frac{\constker_0}{2} + \frac{\constker_0}{8} + B_0 c_0 + \frac{B_0}{4}  \sum_{\ell=2}^{\Lev-1} c_0^\ell  \leq 1- \frac{\constker_0}{2}+ \frac{\constker_0}{8} + B_0 c_0 + \frac{B_0c_0^2}{4(1-c_0)} \leq 1-\frac{\constker_0}{4}.
\end{align*}
since by our choice of $c_0$ and adjusting $\Cdel$, $ B_0 c_0 + \frac{B_0c_0^2}{4(1-c_0)}\leq\frac{\constker_0}{8}$. Similarly, for all $x\in \xng_j$,
\begin{align*}
&\norm{{\overline{\sign(a_j)} \diff{2}{\etapp}(x) - K^{(02)}(x_j,x)}}_x\\
&\leq   \norm{{\overline{\sign(a_j)} \diff{2}{(\etaMat^{-1} \SignVecPad_s)^\top \subetaFunc_1}(x)- K^{(02)}(x_j,x) }}_x + \sum_{\ell=1}^\Lev \norm{{\diff{2}{(\etaMat^{-1} q_{\ell-1})^\top \subetaFunc_\ell}(x)}}_x\\
& \leq \frac{3\constker_2}{32} + \sum_{\ell=2}^\Lev b_\ell \prod_{p=1}^{\ell-1} c_p = \frac{3\constker_2}{32} + B_2 c_0 +\frac{ B_2}{4}  \sum_{\ell=2}^{\Lev-1}c_0^\ell \leq \frac{3\constker_2}{32} + B_2 c_0 + \frac{B_2 c_0^2  }{4(1-c_0)} \leq \frac{7\constker_2}{64}
\end{align*}
since similarly, $ B_2 c_0 + \frac{B_2 c_0^2  }{4(1-c_0)} \leq \frac{\constker_2}{64}$. Hence (I$_\ell$), (II$_\ell$), (III$_\ell$) indeed implies \eqref{eq:prop_etapp}. Next we derive a condition on $m$ under which they are true with probability $1-\rho$ (conditional on event $\Eve$).

\paragraph{Probability of successful construction.} Let us now prove that (I$_\ell$), (II$_\ell$) and (III$_\ell$) are indeed valid with the desired probability. Let $p_1(\ell)$, $p_2(\ell)$ and $p_3(\ell)$ be the probabilities conditional on event $\Eve$ that (I$_\ell$), (II$_\ell$) and (III$_\ell$) fail, respectively. By a union bound, our goal is to derive a bound on $m$ such that $\sum_{k=1}^3 \sum_{\ell=1}^\Lev p_k(\ell) \leq \rho$. We do so by applying variants of Bernstein's concentration inequality, that are all detailed in Appendix \ref{app:concentration}. As we mentioned before, a crucial construction of the golfing scheme is that, at each step, $q_{\ell-1}$ and $\subetaFunc_\ell$ are mutually independent, such that we can reason conditionally on $q_{\ell-1}$ and treat it as a fixed vector when bounding the probabilities w.r.t. $\subetaFunc_\ell$ and $\subetaMat_\ell$.


We define $\bar q_\ell \eqdef D_\met^{-1} \etaMat^{-1}  q_\ell$ for short.
To bound $p_1(\ell)$, we first observe the recurrence relation $D_\met q_\ell = D_\met(\Id - \subetaMat_\ell \etaMat^{-1}) q_{\ell-1} = D_\met(\etaMat - \subetaMat_\ell) D_\met \bar q_{\ell-1}$. Moreover, by Lemma \ref{lem:additional_admissible} we have $\nB{D_\met^{-1}\etaMat^{-1}D_\met^{-1}} \leq \frac{1}{1- \nB{D_\met\etaMat D_\met}} \leq 2$, and therefore
$
\nB{D_\met q_{\ell-1}} \geq \frac{1}{\nB{D_\met^{-1}\etaMat^{-1}D_\met^{-1}}} \nB{ \bar q_{\ell-1}} \geq \frac{1}{2} \nB{\bar q_{\ell-1}}
$. Finally, by Lemma \ref{lem:cdf_bd} and our assumptions we have in particular that $\nB{D_\met(\etaMat_\Eve - \etaMat)D_\met} \leq \min_\ell c_\ell/4$. Therefore,
\begin{align*}
p_1(\ell) &= \PP_\Eve\pa{\nB{D_\met q_\ell} \geq c_\ell \nB{D_\met q_{\ell-1}}} \leq \PP_\Eve\pa{\nB{D_\met(\etaMat - \subetaMat_\ell) D_\met \bar q_{\ell-1}} \geq \frac{c_\ell}{2} \nB{\bar q_{\ell-1}}} \\
&\leq \PP_\Eve\pa{\nB{D_\met(\etaMat_\Eve - \subetaMat_\ell) D_\met \bar q_{\ell-1}} \geq \frac{c_\ell}{4} \nB{\bar q_{\ell-1}}}
\end{align*}
Finally, applying Lemma \ref{lem:bound_R_vec}, for some $\rho_\ell$ that we adjust later we obtain that
\begin{align*}
\PP_\Eve\pa{\nB{D_\met(\etaMat_\Eve - \subetaMat_\ell) D_\met \bar q_{\ell-1}} \geq \frac{c_\ell}{4} \nB{\bar q_{\ell-1}}} \leq \rho_\ell
\end{align*}
if $m_\ell \gtrsim \frac{s \Lu_{01}^2}{c_\ell^2} \log\pa{\frac{s}{\rho_\ell}}$.

For $p_2(\ell)$, we have
\begin{align*}
 \abs{(\etaMat^{-1} q_{\ell-1})^\top \subetaFunc_\ell(x)} = \abs{(\bar q_{\ell-1})^\top D_\met \subetaFunc_\ell(x)} &\leq \abs{(\bar  q_{\ell-1})^\top D_\met (\subetaFunc_\ell(x) - \etaFunc(x))} + \abs{(\bar  q_{\ell-1})^\top  D_\met \etaFunc(x)}\\
 &\leq \abs{(\bar q_{\ell-1})^\top D_\met (\subetaFunc_\ell(x) - \etaFunc(x))} +
 \begin{cases} B_0 \nB{ \bar q_{\ell-1}} &\ell\geq 2\\
 1-\frac{\constker_0}{2} &\ell=1
 \end{cases}
\end{align*}
 by Lemma \ref{lem:additional_admissible} for the case $\ell \geq 2$ and Theorem \ref{thm:admiss-kernel} for the case $\ell=1$. Hence,
\begin{align*}
p_2(\ell) &= \PP_\Eve \pa{ \exists x\in \xfg, \; \abs{(\etaMat^{-1} q_{\ell-1})^\top \subetaFunc_\ell(x)} > t_\ell \nB{ D_\met q_{\ell-1}}} \\
&\leq \PP_\Eve \pa{ \exists x\in \xfg, \; \abs{(\etaMat^{-1} q_{\ell-1})^\top \subetaFunc_\ell(x)} > \frac{t_\ell}{2} \nB{ \bar q_{\ell-1}}} \\
&\leq \PP_\Eve \pa{ \exists x\in \xfg, \; \abs{(\bar q_{\ell-1})^\top D_\met (\subetaFunc_\ell(x) - \etaFunc(x))} > \tilde t_\ell \nB{ \bar q_{\ell-1}}} \qwhereq \tilde t_\ell \eqdef
\begin{cases}
 \pa{\frac{t_\ell}{2} - B_0} &\ell\geq 2\\
 \frac{\constker_0}{16} & \ell=1
 \end{cases}.
\end{align*}
Since by Lemma \ref{lem:cdf_bd} we have in particular
\[
\abs{(\bar q_{\ell-1})^\top D_\met (\etaFunc_\Eve(x) - \etaFunc(x))} \leq \sqrt{2s}\nB{\bar q_{\ell-1}} \norm{D_\met (\etaFunc_\Eve(x) - \etaFunc(x))} \leq \frac{\tilde t_\ell}{2}\nB{\bar q_{\ell-1}}\, ,
\]
by Lemma \ref{lem:bound_f} and a union bound we have
\[
p_2(\ell)\leq \PP_\Eve\pa{ \exists x\in \xfg, \; \abs{(\bar q_{\ell-1})^\top D_\met (\subetaFunc_\ell(x) - \etaFunc_\Eve(x))} > \frac{\tilde t_\ell}{2} \nB{ \bar q_{\ell-1}}} \leq \rho_\ell
\]
provided that $m_\ell \gtrsim s \pa{\frac{\Lu_0^2}{\tilde t_\ell^{2}} + \frac{\Lu_{01} \Lu_0}{\tilde t_\ell}} \log\pa{\frac{\abs{\xfg}}{\rho_\ell}}$.

For $p_3(\ell)$, fix $j$, for any $x \in \xng_j$: in the case $\ell\geq 2$, by Lemma \ref{lem:additional_admissible}, 
\begin{align*}
 \norm{\diff{2}{(D_\met \bar q_{\ell-1})^\top  \subetaFunc_\ell}(x)}_x &\leq \norm{\pa{\diff{2}{(D_\met \bar  q_{\ell-1})^\top (\subetaFunc_\ell- \etaFunc)}(x)}}_x + \norm{\diff{2}{(D_\met \bar  q_{\ell-1})^\top  \etaFunc}(x) }_x\\
 &\leq \norm{\diff{2}{(D_\met \bar q_{\ell-1})^\top (\subetaFunc_\ell - \etaFunc)}(x) }_x + B_2 \nB{\bar q_{\ell-1}} 
\end{align*}
and for $\ell=1$, by Theorem \ref{thm:admiss-kernel},
\begin{align*}
&\norm{{\overline{\sign(a_j)} \diff{2}{(D_\met \bar q_{0})^\top  \subetaFunc_1}(x) - \fullCov^{(02)}(x_j,x)}}_x\\
&\leq \norm{{\overline{\sign(a_j)} \diff{2}{(D_\met \bar q_{0})^\top  \etaFunc}(x) - \fullCov^{(02)}(x_j,x)}}_x + \norm{\diff{2}{(D_\met \bar q_{0})^\top (\subetaFunc_1 - \etaFunc)}(x) }_x  \\
&\leq \frac{\constker_2}{16} + \norm{\diff{2}{(D_\met \bar q_{0})^\top (\subetaFunc_1 - \etaFunc)}(x) }_x.
\end{align*}
Therefore, by the same computation as before,
\begin{align*}
p_3(\ell) \leq \PP_\Eve \pa{\exists x\in \xng, \; \norm{\diff{2}{(D_\met \bar q_{\ell-1})^\top (\subetaFunc_\ell - \etaFunc)}(x) }_x > \tilde b_\ell \nB{\bar q_{\ell-1}}}, \text{ where } \tilde b_\ell\eqdef \begin{cases} \pa{\frac{b_\ell}{2} - B_2} &\ell\geq 2 \\
\frac{\constker_2}{64} &\ell=1.
\end{cases}
\end{align*}
Again using Lemma \ref{lem:cdf_bd} we bound $\norm{\diff{2}{(D_\met \bar q_{\ell-1})^\top (\etaFunc_\Eve - \etaFunc)}(x) }_x \leq \frac{\tilde b_\ell}{2} \nB{\bar q_{\ell-1}}$ and
\begin{align*}
p_3(\ell) \leq \PP_\Eve \pa{\exists x\in \xng, \; \norm{\diff{2}{(D_\met \bar q_{\ell-1})^\top (\subetaFunc_\ell - \etaFunc_\Eve)}(x) }_x > \frac{\tilde b_\ell}{2} \nB{\bar q_{\ell-1}}} \leq \rho_\ell
\end{align*}
by Lemma \ref{lem:bound_f_hess} and a union bound, provided that $m_\ell \gtrsim s\pa{\frac{\Lu_2^2}{\tilde b_\ell^{2}} + \frac{\Lu_2\Lu_{01}}{\tilde b_\ell}} \log\pa{\frac{\abs{\xng}}{\rho_\ell}}$.

Choosing $\rho_1 = \rho_2 = \rho/9$ and $\rho_\ell = \rho/(9J)$ for $\ell\geq 3$, recalling that obviously $\constker_r \leq B_r$ for $r=1,2$ and denoting $N_0 = \abs{\xfg}$ and $N_2 = \abs{\xng}$ for short, we have 
$
\sum_{k=1}^3 \sum_{\ell=1}^\Lev p_k(\ell) \leq \rho
$ 
provided that
\begin{align*}
m_1 = m_2 \gtrsim&~ s \sum_{r=0,2} \pa{\Lu_{01}^2 \frac{B_r^2}{\constker_r^2} \log(s) \log\pa{\frac{s}{\rho}}+ \pa{\frac{\Lu_r^2}{\constker_r^2} + \frac{\Lu_{01}\Lu_r}{\constker_r}} \log\pa{\frac{N_r}{\rho}}}
\end{align*}
and for $\ell\geq 3$,
\[
m_\ell \gtrsim s \sum_{r=0,2} \pa{\Lu_{01}^2 \frac{B_r^2}{\constker_r^2} \log\pa{\frac{s\log(s)}{\rho}}+ \pa{\frac{\Lu_r^2}{B_r^2 \log^2(s)} + \frac{\Lu_{01}\Lu_r}{B_r \log(s)}} \log\pa{\frac{N_r\log(s)}{\rho}}}
\]
Therefore, conditionally on $\Eve$, $\etapp$ can be constructed with probability at least $1-\rho$ if $m \gtrsim m_1+m_2+ \Lev m_3$,
for which it is sufficient that
\begin{align}
m &\gtrsim s \sum_{r=0,2} \pa{\Lu_{01}^2 \frac{B_r^2}{\constker_r^2} \log(s) \log\pa{\frac{s}{\rho}}+ \pa{\frac{\Lu_r^2}{\constker_r^2} + \frac{\Lu_{01}\Lu_r}{\constker_r}} \log\pa{\frac{N_r \log(s)}{\rho}}} \label{eq:mbound1}
\end{align}

\subsection{Step 2: correcting the approximate certificate} \label{sec:proofstep2}

The second step of our proof is to ``correct'' the previously constructed approximate certificate $\etapp$ to obtain a certificate $\eta \in \Im(\Phi^*)$ satisfying \eqref{eq:prop_etagrid}. Recalling the definition \eqref{eq:psi_func} of $\Psi$, let $e\eqdef \Psi \etapp - \SignVecPad_s$ be the error made by $\etapp$ and define $$
\subeta \eqdef \etapp - \etae, \qwhereq 
\etae \eqdef (\subetaMat^{-1} e)^\top \subetaFunc.
$$
Then,
$$
\Psi \subeta = \Psi \etapp - e = \SignVecPad_s\, ,
$$
and we have indeed that $\subeta(x_i) = \sign(a_i)$ and $\nabla \subeta(x_i) = 0$. We will now bound the deviations of $\subeta$ on the grids $\xfg$ and $\xng$, using the fact that $e$ has a small norm. Note that there is a subtlety here: $e$ itself is random, and not independent of $\subetaFunc$ or $\subetaMat$. So we must use ``uniform'' concentration bounds.

Using Lemma \ref{lem:additional_admissible} in combination with Lemma \ref{lem:cdf_bd} and Lemma \ref{lem:bound_etaMat}, we have that with probability at least $1-\rho$:
\begin{equation}
\label{eq:bound_subetaMat}
\norm{\Id - D_\met \subetaMat D_\met} \leq \norm{\Id - D_\met \etaMat D_\met} + \norm{D_\met (\etaMat - \etaMat_\Eve) D_\met} + \norm{D_\met (\etaMat_\Eve - \subetaMat) D_\met} \leq \frac12 + \frac{1}{8} + \frac{1}{8} = \frac{3}{4}
\end{equation}
and therefore
\begin{equation}
\label{eq:bound_subetaMatinv}
\norm{D_\met^{-1} \subetaMat^{-1} D_\met^{-1} } \leq 4\, .
\end{equation}
By Lemma \ref{lem:additional_admissible}, \ref{lem:cdf_bd}, \ref{lem:bound_f_alone} and a union bound to respectively bound each term in the following triangular inequality, with probability $1-\rho$ we have
\[
\forall x \in \xfg,\quad \norm{D_\met \subetaFunc(x)} \leq \norm{D_\met \etaFunc(x)} + \norm{D_\met (\etaFunc_\Eve(x)-\etaFunc(x))} + \norm{D_\met (\subetaFunc(x)-\etaFunc_\Eve(x))} \leq 2 B_0
\]
if $m \gtrsim B_0^{-2} \log\pa{\frac{\abs{\xfg}}{\rho}}(s \Lu_{01}^2 + \sqrt{s} \Lu_{01} \Lu_0)$. Then, for all $x\in \xfg$, since by adjusting $\Cdel$ we can have in particular
$\norm{D_\met e}\leq \frac{c_0}{\sqrt{s}} \leq c_0 \leq \frac{1}{128} \min\pa{ \frac{\constker_2}{B_2} , \frac{\constker_0}{ B_0}}$, we have
\begin{align*}
\abs{\subeta(x)} \leq \abs{\etapp(x)} +  \norm{D_\met \subetaFunc(x)} \norm{D_\met^{-1} \subetaMat^{-1} D_\met^{-1} } \norm{D_\met e} \leq 1-\frac{3\constker_0}{16},
\end{align*}

Similarly, by Lemma \ref{lem:additional_admissible}, \ref{lem:cdf_bd}, with probability $1-\rho$ we have for all $x \in \xng$ and $q \in \CC^{s(d+1)}$,
\begin{align*}
\norm{\diff{2}{\subetaFunc^\top D_\met q}(x)}_x &\leq \norm{\diff{2}{\etaFunc^\top D_\met q}(x)}_x + \norm{\diff{2}{(\etaFunc_\Eve - \etaFunc)^\top D_\met q}(x)}_x + \norm{\diff{2}{(\etaFunc_\Eve - \subetaFunc)^\top D_\met q}(x)}_x \\
&\leq (B_2 + B_2/2) \norm{q}  +  \norm{q}\sup_{\norm{v}_x \leq 1} \norm{\frac{1}{m} \sum_{k=1}^m D_\met\overline{\RFVec(\om_k)} g_{\om_k}(v) - \EE_\Eve D_\met  \overline{\RFVec(\om)} g_{\om}(v)} 
\end{align*}
where $g_\om(v) \eqdef \diff{2}{\phi_\om}(x)[v,v]$.
By Lemma \ref{lem:bound_f_hess_uniform} and and a union bound, for all $x\in\xng$, $$\sup_{\norm{v}_x \leq 1} \norm{\frac{1}{m} \sum_{k=1}^m D_\met\overline{\RFVec(\om_k)} g_{\om_k}(v) - \EE_\Eve D_\met  \overline{\RFVec(\om)} g_{\om}(v)}  \leq B_2$$
if $
m \gtrsim \frac{s B_{22} \Lu_{01}^2 + \sqrt{s} \Lu_{01} \Lu_2 B_2}{B_2^2} \pa{\log\pa{\frac{\abs{\xng}}{\rho}} + d \log\pa{\frac{s \Lu_{01} \Lu_{2}}{B_2}}}$. Using this property with $q \eqdef D_\met^{-1} \subetaMat^{-1} e$ such that $\norm{q} \leq 4 c_0$, and by adjusting $\Cdel$, we obtain: for all $x\in\xng_j$,
\begin{align*}
\norm{\overline{\sign(a_j)} \diff{2}{\subeta}(x)  - \fullCov^{(02)}(x_j,x)}_x &\leq \norm{\overline{\sign(a_j)} \diff{2}{\etapp}(x)  - \fullCov^{(02)}(x_j,x)}_x + \norm{\diff{2}{\subetaFunc^\top D_\met q}(x)}_x \\
&\leq \frac{7\constker_2}{64} + \frac{\constker_2}{128} = \frac{15\constker_2}{128}
\end{align*}
which concludes the second step of our proof. By combining the bounds on $m$ that we obtained with \eqref{eq:mbound1}, after simplification we still obtain
\begin{equation}
m \gtrsim s \sum_{r=0,2} \pa{\Lu_{01}^2 \frac{B_r^2}{\constker_r^2} \log(s) \log\pa{\frac{s}{\rho}}+ \pa{\frac{\Lu_r^2}{\constker_r^2} + \frac{\Lu_{01}\Lu_r}{\constker_r} + \frac{B_{22}}{B_2^2}\Lu_{01}^2} \log\pa{\frac{N'_r \log(s)}{\rho}}} \label{eq:mbound2}
\end{equation}
with $N'_0 = N_0 = \abs{\xfg}$ but $N'_2 = \abs{\xng} + (s\Lu_{01} \Lu_2/B_2)^d$.
%
%
%

\subsection{Step 3: Bounding the norm $\norm{p}$} \label{sec:proofstep4}
\newcommand{\papp}{p^{\mathrm{app}}}
\newcommand{\pe}{p^\mathrm{e}}
In this section we upper bound $\norm{p}$ where $\Phi^* p = \subeta$, for the $\subeta$ that we have constructed in the previous section.
We recall that $\Phi^* p = \frac{1}{\sqrt{m}} \sum_{k=1}^m p_k \phi_{\om_k}(\cdot)$, and
\begin{align*}
\etapp = \sum_{\ell=1}^\Lev (\etaMat^{-1}q_{\ell-1})^\top \subetaFunc_\ell = \frac{1}{\sqrt{m}}\sum_\ell \frac{\sqrt{m}}{m_\ell} \sum_{k\in \Bb_\ell} (\etaMat^{-1}q_{\ell-1})^\top \overline{\RFVec(\om_k)} \phi_{\om_k} = \Phi^* \papp,
\end{align*}
where $\papp \eqdef (p_\ell)_{\ell=1}^\Lev\in \CC^m$ and $p_\ell \eqdef \frac{\sqrt{m}}{m_\ell} \pa{\RFVec(\om_k)^* \etaMat^{-1} q_{\ell-1}}_{k\in \Bb_j} \in\CC^{m_\ell}$.
So, $\norm{\papp}^2 = \sum_{\ell=1}^\Lev \norm{p_\ell}^2_2$. To upper bound this, for each $\ell=1,\ldots, \Lev$,
\begin{align*}
\frac{m_\ell}{m}\norm{p_\ell}^2_2 &=  \frac{1}{m_\ell} \sum_{k\in \Bb_\ell} q_{\ell-1}^* \etaMat^{-1} \RFVec(\om_k) \RFVec(\om_k)^* \etaMat^{-1} q_{\ell-1} = q_{\ell-1}^* \etaMat^{-1} \subetaMat_\ell \etaMat^{-1} q_{\ell-1} \\
& = q_{\ell-1}^* \etaMat^{-1} (\subetaMat_\ell \etaMat^{-1} -\Id) q_{\ell-1} + q_{\ell-1}^* \etaMat^{-1} q_{\ell-1}  = q_{\ell-1}^* \etaMat^{-1} q_{\ell} + q_{\ell-1}^* \etaMat^{-1} q_{\ell-1} \\
& \leq \norm{D_\met^{-1} \etaMat^{-1} D_\met^{-1}} \norm{D_\met q_{\ell-1}} \pa{ \norm{D_\met q_{\ell-1}} + \norm{D_\met q_{\ell}} } \\
&\leq 4s \nB{D_\met q_{\ell-1}} \pa{\nB{D_\met q_\ell} + \nB{D_\met q_{\ell-1}}} \leq 4s \pa{ c_\ell + 1} \prod_{i=1}^{\ell-1} c_i^2.
\end{align*}
where we have used $\norm{D_\met^{-1} \etaMat^{-1} D_\met^{-1}} \leq 2$ by Lemma \ref{lem:additional_admissible}, $\norm{\cdot} \leq \sqrt{2s}\nB{\cdot}$, and the computation that precedes for $\nB{D_\met q_\ell}$. For $\ell=1,2$
$
\frac{m}{m_\ell} = \Oo(1)
$ and
$\frac{m}{m_3} = \Oo( \log(s))$.
Also, for $\ell\geq 3$, 
\begin{equation*}
\pa{ c_\ell + 1}\prod_{i=1}^{\ell-1} c_i^2 = 
\pa{1+c_0}\frac{c_0^{\ell-1}}{16\log(s)}
\end{equation*}
Therefore,
$$
\norm{\papp}^2 \lesssim 4s \pa{ 1 + \frac{c_0}{4\sqrt{\log(s)}} +  \frac{c_0^2}{16 \log(s)} + (1+c_0) \frac{c_0^2}{16(1-c_0)} } \lesssim s.
$$
On the other hand, $\etae = \Phi^* \pe$ where $\pe=\pa{ \RFVec(\om_k)^* \etaMat^{-1} e}_{k=1}^m$.  
So,
$$
\norm{\pe}^2 = e^*  \etaMat^{-1} \subetaMat  \etaMat^{-1} e \leq 8\norm{D_\met e}^2 \lesssim 1.
$$
Therefore, $\subeta = \Phi^* p $ with $\norm{p}^2 \lesssim s$.

\subsection{Step 4: Nondegeneracy on the entire domain} \label{sec:proofstep3}

We conclude by showing that the $\subeta$ constructed in the previous sections is indeed nondegenerate on the entire domain. For this we simply need to control the Lipschitz constants of $\subeta$ and its Hessian, which are in fact directly related to $\norm{p}$.
%
%
Let any $x \in \xf$, and $x'\in \xfg$ be the point in the grid closest to it. Under $\Eve$, we have
\begin{align*}
\abs{\subeta(x)} &\leq 1-\frac{3\constker_0}{16} + \abs{\subeta(x) -\subeta(x')} = 1-\frac{3\constker_0}{16} + \abs{(\Phi^* p)(x) -(\Phi^*p)(x')} \\
&\leq 1-\frac{3\constker_0}{16} + \norm{p}\sqrt{\frac{1}{m} \sum_{k=1}^m \abs{\phi_{\om_k}(x) - \phi_{\om_k}(x')}^2} \leq 1-\frac{3\constker_0}{16} + \Lu_{1} \norm{p} \dsep_\met(x,x')
\end{align*}

Hence we prove the first part of \eqref{eq:prop_etafinal} by choosing $\xfg$ such that $\dsep_\met(x,x') \leq \frac{\constker_0}{16\Lu_{1} \norm{p}}$, which results in
\begin{align*}
\abs{\xfg} = \pa{\frac{C\Rr_\Xx \Lu_1 \norm{p}}{\constker_0}}^d
\end{align*}
for an appropriate constant $C$.

Now, for any $x \in \xn_j$, and $\x'\in \xng_j$ closest to it, we write
\begin{equation}\label{eq:proofstep3hessianbound}
\begin{split}
\norm{\overline{\sign(a_j)} \diff{2}{\subeta}(x)  - \fullCov^{(02)}(x_j,x)}_x &\leq \norm{\diff{2}{\subeta}(x)  - \diff{2}{\subeta}(x')[\tau_{\x \to x'} \cdot,\tau_{\x \to x'} \cdot]}_x \\
&\quad+ \norm{\overline{\sign(a_j)}\diff{2}{\subeta}(x')[\tau_{\x \to x'} \cdot,\tau_{\x \to x'} \cdot] -\fullCov^{(02)}(x_j,x')[\tau_{\x \to x'} \cdot,\tau_{\x \to x'} \cdot]}_x \\
&\quad+ \norm{\fullCov^{(02)}(x_j,x')[\tau_{\x \to x'} \cdot,\tau_{\x \to x'} \cdot] - \fullCov^{(02)}(x_j,x)}_x
\end{split}
\end{equation}
We bound each of these terms. For the first, under $\Eve$ we have
\begin{align*}
&\norm{\diff{2}{\subeta}(x)  - \diff{2}{\subeta}(x')[\tau_{\x \to x'} \cdot,\tau_{\x \to x'} \cdot]}_x \\
&\quad\leq \norm{p}\sqrt{\frac{1}{m} \sum_{k=1}^m \norm{\diff{2}{\phi_{\om_k}}(x) - \diff{2}{\phi_{\om_k}}(x')[\tau_{\x \to x'} \cdot,\tau_{\x \to x'} \cdot]}_x^2} \leq \Lu_3 \norm{p}\dsep_\met(x,x')
\end{align*}
For the second term in \eqref{eq:proofstep3hessianbound}, we have
\begin{align*}
&\norm{\overline{\sign(a_j)}\diff{2}{\subeta}(x')[\tau_{\x \to x'} \cdot,\tau_{\x \to x'} \cdot] -\fullCov^{(02)}(x_j,x')[\tau_{\x \to x'} \cdot,\tau_{\x \to x'} \cdot]}_x \\
&\quad = \norm{\overline{\sign(a_j)}\diff{2}{\subeta}(x') -\fullCov^{(02)}(x_j,x')}_{x'} \leq \frac{15\constker_2}{128}
\end{align*}
from what we have proved in the previous section.

Finally, for the third term in \eqref{eq:proofstep3hessianbound} we naturally introduce $\fullCov_\Eve^{(ij)}$ defined as $\fullCov^{(ij)}$ in \eqref{eq:defKij}, but by replacing $\EE$ with the conditional $\EE_\Eve$. From Lemma \ref{lem:cdf_bd} the deviation between $\fullCov^{(02)}$ and $\fullCov_\Eve^{(02)}$ can be bounded by
\[
\forall x\in \xn,~\norm{\fullCov_\Eve^{(02)}(x_j,x) - \fullCov^{(02)}(x_j,x)}_x = \norm{\diff{2}{(\etaFunc_\Eve - \etaFunc)^\top D_\met u_j}(x)}_x \leq \frac{\constker_2}{512}
\]
where $u_j$ is the j$th$ canonical vector of $\CC^{s(d+1)}$. Moreover, by Assumption \ref{ass:feat} it is easy to see that
\[
\norm{\fullCov_\Eve^{(02)}(x_j,x')[\tau_{\x \to x'} \cdot,\tau_{\x \to x'} \cdot] - \fullCov_\Eve^{(02)}(x_j,x)}_x \leq \Lu_0 \Lu_3 \dsep_\met(x,x')
\]
Hence by a triangular inequality we have
\begin{align*}
\norm{\fullCov^{(02)}(x_j,x')[\tau_{\x \to x'} \cdot,\tau_{\x \to x'} \cdot] - \fullCov^{(02)}(x_j,x)}_x \leq&\norm{\fullCov^{(02)}(x_j,x')[\tau_{\x \to x'} \cdot,\tau_{\x \to x'} \cdot] - \fullCov_\Eve^{(02)}(x_j,x')[\tau_{\x \to x'} \cdot,\tau_{\x \to x'} \cdot]}_x \\
&+\norm{\fullCov_\Eve^{(02)}(x_j,x')[\tau_{\x \to x'} \cdot,\tau_{\x \to x'} \cdot] - \fullCov_\Eve^{(02)}(x_j,x)}_x\\
&+\norm{\fullCov_\Eve^{(02)}(x_j,x) - \fullCov^{(02)}(x_j,x)}_x \leq \frac{\constker_2}{256} + \Lu_0 \Lu_3 \dsep_\met(x,x')
\end{align*}
Therefore, \eqref{eq:proofstep3hessianbound} becomes
\begin{equation}\label{eq:proofstep3hessianboundfinal}
\norm{\overline{\sign(a_j)} \diff{2}{\subeta}(x)  - \fullCov^{(02)}(x_j,x)}_x \leq \Lu_3(\Lu_0 + \norm{p})\dsep_\met(x,x') + \frac{15\constker_2}{128} + \frac{\constker_2}{256}
\end{equation}
We prove the desired property on $\diff{2}{\subeta}$ by choosing $\dsep_\met(x,x') \leq \frac{\constker_2}{256\Lu_3(\Lu_0 + \norm{p})}$, which yields
\begin{align*}
\abs{\xng} = s \abs{\xng_j} = s\pa{\frac{C \rnear \Lu_0\Lu_3 \norm{p}}{\constker_2}}^d
\end{align*}
for an appropriate constant $C$.
Gathering everything with \eqref{eq:mbound2}, we finally obtain
\begin{equation}
m \gtrsim s \sum_{r=0,2} \pa{\Lu_{01}^2 \frac{B_r^2}{\constker_r^2} \log(s) \log\pa{\frac{s}{\rho}}+ \pa{\frac{\Lu_r^2}{\constker_r^2} + \frac{\Lu_{01}\Lu_r}{\constker_r}} \log\pa{\frac{\bar N_r^d}{\rho}}} \label{eq:mbound3}
\end{equation}
with $\bar N_0 = \frac{s \Rr_\Xx \Lu_1}{\constker_0}$, $\bar N_2 = \frac{s (\rnear \Lu_0\Lu_3 + \Lu_2)}{\constker_2}$.

\subsection{Step 5: additional certificates}\label{sec:proofstep5}
Nondegeneracy of $\subeta$ directly allows us to apply Proposition \ref{prop:robustness} to deduce stability away from the sparse support in the reconstructed measure. In order to apply Proposition \ref{prop:stab_near}, we need to construct an additional $s$ certificates $\eta_j$, which are however ``simpler'' to construct since they need to interpolate a ``sign vector'' that has only one non-zero coordinate, and do not require the golfing scheme to do so. 

For each $j=1,\ldots,s$, let $u_j$ be the vector of length $s(d+1)$ whose $j^{th}$ entry is one and all other entries are zero. Define the functions
 $$\eta_j^+ \eqdef \pa{ \etaMat^{-1} \binom{1_s}{0_{sd}} }^\top \etaFunc(x)\qandq \eta_j^- \eqdef  \pa{ \etaMat^{-1} \pa{2 u_j- \binom{1_s}{0_{sd}}}}^\top \etaFunc(x),$$ and 
\[
\eta_j\eqdef \frac{1}{2}(\eta_j^+ + \eta_j^-) =  \pa{ \etaMat^{-1} u_j}^\top \etaFunc(x).
\]
 By Theorem \ref{thm:admiss-kernel}, $\eta_j^+$ and $\eta_j^-$ are nondegenerate (limit) dual certificates with respect to signs $1_s$ and $-1_s + 2 u_j$ respectively, and $\eta_j$ satisfies, for all $\ell \neq j$:
\begin{equation}\label{eq:prop_etaj}
\begin{split}
&\eta_j(x_j) = 1,\quad \nabla \eta_j(x_j) = 0 \qandq \eta_j(x_\ell) = 0,\quad \nabla \eta_j(x_\ell) = 0 \\ 
&\abs{\eta_j(x)} \leq \frac12\pa{\abs{\eta_j^+(x)} + \abs{\eta_j^-(x)}} \leq 1-\frac{\constker_0}{4},\quad \forall x \in \xf \\
&\norm{\diff{2}{\eta_j}(x) - \fullCov^{(02)}(x_j,x)}_x \leq \frac{\constker_2}{16},\quad \forall x \in \xn_j \\
&\norm{\diff{2}{\eta_j}(x)}_x \\
&\qquad \leq \frac12\pa{\norm{\diff{2}{\eta_j^+}(x) - \fullCov^{(02)}(x_\ell,x)}_x + \norm{-\diff{2}{\eta_j^-}(x) - \fullCov^{(02)}(x_\ell,x)}_x} \leq \frac{\constker_2}{16},\quad \forall x \in \xn_\ell
\end{split}
\end{equation} 
Thus, using Lemma \ref{lem:curv2quaddecay} to translate the last two conditions into quadratic decay, we conclude that $\eta_j$ satisfies the conditions of Proposition \ref{prop:stab_near}.

To conclude, we will show that $$\subeta_j\eqdef  \pa{ \subetaMat^{-1} u_j }^\top \subetaFunc \in \Im{\Phi^*}$$
does not deviate too much from $\eta_j$ and satisfies the conditions of Proposition \ref{prop:stab_near}. Note that by construction, $\subeta_j(x_j) = 1$, $\subeta_j(x_\ell) = 0$ for all $\ell\neq j$, and $\nabla \subeta_j(x_\ell) = 0$ for all $\ell$. It therefore remains to control the deviation of $\subeta_j$ from $\eta_j$ on $\xf$ and $\diff{2}{\subeta_j}$ from $\diff{2}{\eta_j}$ on $\xn$.

\begin{prop}\label{prop:eta_j}
Under Assumption \ref{ass:kernel} and \ref{ass:feat}, suppose that $\min_{i\neq j}\dsep_\met(x_i,x_j)\geq \Delta$. Then, with probability at least $1-\rho$, for all $j=1,\ldots, s$, there exists
$\subeta_j = \Phi^* p_j$ where $\norm{p_j} \leq 4$ which satisfies, for all $\ell \neq j$:
\begin{equation}\label{eq:prop_subetaj}
\begin{split}
&\subeta_j(x_j) = 1,\quad \nabla \subeta_j(x_j) = 0 \qandq \subeta_j(x_\ell) = 0,\quad \nabla \subeta_j(x_\ell) = 0 \\ 
&\abs{\subeta_j(x)} \leq 1-\frac{\constker_0}{8},\quad \forall x \in \xf \\
&\norm{\diff{2}{\eta_j}(x) - \fullCov^{(02)}(x_j,x)}_x \leq \frac{\constker_2}{8},~ \forall x \in \xn_j,\quad\norm{\diff{2}{\eta_j}(x)}_x \leq \frac{\constker_2}{8},~ \forall x \in \xn_\ell
\end{split}
\end{equation} 
\end{prop}
The proof controls the deviation between $\subeta_j$ and $\eta_j$ on a fine grid using Bernstein's concentration inequalities, and extend the bound to the entire domain using Lipschitz properties of $\subeta_j$. As we mentioned above, the proof of this result is conceptually simpler than the deviation bounds on $\etapp$ since $\norm{u_j} = 1$. We therefore defer its proof to Appendix~\ref{app:proof_prop_subetaj}. Using Lemma \ref{lem:curv2quaddecay}, we have therefore constructed the additional certificates to apply Proposition~\ref{prop:stab_near} and conclude the proof of Theorem~\ref{thm:main}.


\section{Conclusion and outlooks}

In this paper, we have presented an unifying geometric view on the problem of sparse measures recovery from random measurements. This theoretical analysis highlights the key role played by the invariant Fisher metric to define a precise notion of Rayleigh limit in the case of possibly non-translation invariant measurement kernels. We analyzed several examples including Laplace measurements in imaging, and left partially open some other important examples such as one-hidden-layer neural networks. Analyzing the super-resolution regime (going below the Rayleigh limit) requires stringent assumptions, such as positivity of the measures. Beyond the 1-D case, this is still mostly an open question, and we refer to~\cite{2019-Poon-MultiDim} for some partial results.

\bibliographystyle{siam}
\bibliography{biblio}

\clearpage

\appendix
\section{Preliminaries}\label{app:prelim}

In this Appendix, we provide the proofs to some technical lemmas in the paper, and give useful tools.

\subsection{Linear algebra tools}\label{app:lin-alg}

We give the following simple lemma.
\begin{lem}\label{lem:block_norm}
For $1\leq i,j \leq s$, take any scalars $a_{ij} \in \CC$, vectors $Q_{ij}, R_{ij} \in \CC^d$ and square matrices $A_{ij} \in \CC^{d\times d}$.

\begin{enumerate}[label=(\roman*)]
\item For $q \in \CC^{sd}$ and $M \in \CC^{sd \times sd}$, we have $\normblock{q} \leq \norm{q} \leq \sqrt{s} \normblock{q}$, and as a consequence $\norm{M} \leq \sqrt{s} \normblock{M}$ and $\normblock{M} \leq \sqrt{s} \norm{M}$. Similarly, for $q \in \CC^{s(d+1)}$ and $M \in \CC^{s(d+1) \times s(d+1)}$, we have $\nB{q} \leq \norm{q} \leq \sqrt{2s} \nB{q}$, and as a consequence $\norm{M} \leq \sqrt{2s} \nB{M}$ and $\nB{M} \leq \sqrt{2s} \norm{M}$.
\item Let $M \in \CC^{sd\times sd}$ be a matrix formed by blocks :
\[
M = \left( \begin{matrix}
A_{11} & \ldots & A_{1s} \\
\vdots & \ddots & \vdots \\
A_{s1} & \ldots & A_{ss}
\end{matrix}
\right)
\]
Then we have
\begin{equation}
\label{eq:block_norm}
\normblock{M} = \sup_{\normblock{x} = 1}\normblock{Mx} \leq \max_{1\leq i\leq s} \sum_{j=1}^s \norm{A_{ij}}
\end{equation}
Now, let $M \in \RR^{sd\times s}$ be a rectangular matrix formed by stacking vectors $Q_{ij} \in \RR^{d}$:
\[
M = \left( \begin{matrix}
Q_{11} & \ldots & Q_{1s} \\
\vdots & \ddots & \vdots \\
Q_{s1} & \ldots & Q_{ss}
\end{matrix}
\right)
\]
Then,
\begin{equation}
\label{eq:block_norm_vec}
\norm{M}_{\infty \to \textup{block}}  \leq \max_{1\leq i\leq s} \sum_{j=1}^s \norm{Q_{ij}}_2, \quad \norm{M^\top}_{\textup{block} \to \infty}  \leq \max_{1\leq i\leq s} \sum_{j=1}^s \norm{Q_{ji}}_2
\end{equation}
\item Consider $M \in \CC^{s(d+1)\times s(d+1)}$ decomposed as
\[
M = \left(\begin{matrix}
a_{11} & \ldots & a_{1s} & Q^\top_{11} & \ldots & Q^\top_{1s} \\
\vdots & \ddots & \vdots &\vdots & \ddots & \vdots \\
a_{s1} & \ldots & a_{ss} & Q^\top_{s1} & \ldots & Q^\top_{ss} \\
R_{11} & \ldots & R_{1s} & A_{11} & \ldots & A_{1s} \\
\vdots & \ddots & \vdots &\vdots & \ddots & \vdots \\
R_{s1} & \ldots & R_{ss} & A_{s1} & \ldots & A_{ss}
\end{matrix}\right).
\]
Then,
\begin{align*}
\norm{M}^2 &\leq 
\max_i \pa{\sum_{j=1}^s \abs{a_{ij}} + \norm{Q_{ij}}}\cdot    \max_j \pa{\sum_{i=1}^s \abs{a_{ij}} + \norm{Q_{ij}}  } \\
&\qquad+ \max_i \pa{\sum_{j=1}^s \norm{R_{ij}} + \norm{A_{ij}} } \cdot \max_{j=1}^s \pa{ \sum_{i=1}^s\norm{R_{ij}}  + \norm{A_{ij}} }.
\end{align*}
and
\begin{align*}
\nB{M} &\leq \max_i \{\sum_{j}{ \abs{a_{ij}} + \norm{Q_{ij}}}, \;  \sum_{j}{ \norm{R_{ij}} + \norm{A_{ij}}}\}
\end{align*}

\end{enumerate}
\end{lem}
\begin{proof}
The proof is simple linear algebra.
\begin{enumerate}[label=(\roman*)]
\item This is immediate by writing the definitions.
\item Let $x$ be a vector with $\normblock{x} \leq 1$ decomposed into blocks $x = [x_1,\ldots,x_s]$ with $x_i \in \CC^d$, we have
\begin{align*}
\normblock{Mx} &= \max_{1\leq i\leq s} \norm{\sum_{j=1}^s A_{ij} x_j} \leq \max_{i}\sum_j \norm{A_{ij}} \norm{x_j} \leq \max_{i}\sum_j \norm{A_{ij}}
\end{align*}
Similarly,
\[
\norm{M^\top x}_\infty= \max_{1\leq i\leq s} \norm{\sum_{j=1}^s Q_{ji}^\top x_j} \leq \max_{i}\sum_j \norm{Q_{ji}} \norm{x_j} \leq \max_{i}\sum_j \norm{Q_{ji}}
\]
Then, taking $x\in \CC^s$ such that $\norm{x}_\infty \leq 1$, we have
\[
\normblock{M x} = \max_{1\leq i\leq s} \norm{\sum_{j=1}^s x_j Q_{ij}} \leq \max_{i}\sum_j \norm{Q_{ij}}
\]
\item Taking $x=[x_1,\ldots,x_s,X_1,\ldots,X_s] \in \CC^{s(d+1)}$ with $\norm{x} = 1$, we have
\begin{align*}
\norm{Mx}^2 &= \sum_{i=1}^s \pa{\sum_{j=1}^s a_{ij}x_j + Q_{ij}^\top X_j}^2 + \norm{\sum_{j=1}^s R_{ij}x_j + A_{ij} X_j}^2 \\
&\leq \sum_{i=1}^s \pa{\sum_{j=1}^s \abs{a_{ij}}x_j^2 + \norm{Q_{ij}}\norm{ X_j}^2} \pa{\sum_{j=1}^s \abs{a_{ij}} + \norm{Q_{ij}}}  \\
&\qquad+\sum_{i=1}^s \pa{\sum_{j=1}^s \norm{R_{ij}} x_j^2 + \norm{A_{ij}} \norm{X_j}^2}\pa{\sum_{j=1}^s \norm{R_{ij}} + \norm{A_{ij}} }\\
&=  \max_i \pa{\sum_{j=1}^s \abs{a_{ij}} + \norm{Q_{ij}}}\cdot  \max\pa{ \max_j \sum_{i=1}^s \abs{a_{ij}},  \max_j\sum_{i=1}^s\norm{Q_{ij}}} \norm{x}^2   \\
&\qquad+ \max_i \pa{\sum_{j=1}^s \norm{R_{ij}} + \norm{A_{ij}} } \cdot \max\pa{\max_{j=1}^s \sum_{i=1}^s\norm{R_{ij}}  , \max_j \sum_{i=1}^s \norm{A_{ij}} }\norm{x}^2.
\end{align*}

Now, if $\nB{x} = 1$, we have
\begin{align*}
\nB{Mx} &= \max_i \pa{\abs{\sum_{j=1}^s a_{ij}x_j + Q_{ij}^\top X_j} , \; \norm{\sum_{j=1}^s R_{ij}x_j + A_{ij} X_j} }\\
&\leq \max_i \pa{\sum_{j=1}^s \abs{a_{ij}} +\norm{ Q_{ij}} ,\; \sum_{j=1}^s \norm{R_{ij}} + \norm{A_{ij} } } 
\end{align*}
\end{enumerate}
\end{proof}

\subsection{Proof of Lemma \ref{lem:additional_admissible}}\label{app:proof_add}
The proof is similar to that of Theorem \ref{thm:admiss-kernel}.
\begin{enumerate}[label=(\roman*)]
\item 
We bound the spectral norm of $\Id - D_\met \etaMat D_\met$.
By Lemma \ref{lem:block_norm},
\begin{align*}
\norm{(\Id - D_\met \etaMat D_\met)}^2 &\leq 
\max_i \pa{\sum_{\substack{j=1\\ j\neq i}}^s \abs{K(x_i,x_j)} + \sum_{j=1}^s\norm{K^{(10)}(x_i,x_j)}_{x_i}}^2  \\
&\qquad+ \max_i \pa{\sum_{\substack{j=1}{j\neq i}}^s \norm{ K^{(10)}(x_j,x_i)}_{x_j} + \sum_{j=1}^s  \norm{ K^{(11)}(x_i,x_j) }_{x_i,x_j} }^2 \leq 8 h^2
\end{align*}
by assumption on the kernel widths. Hence $\etaMat$ is invertible. Similarly, by again applying Lemma \ref{lem:block_norm}, $\nB{D_\met \etaMat D_\met- \Id} \leq 2h$.
\item 
Let $x\in \xf$, then we have
\begin{align*}
\norm{D_\met \etaFunc(x)} &\leq \pa{\sum_{i=1}^s \abs{\fullCov(x_i,x)}^2 + \norm{\fullCov^{(10)}(x_i,x)}_{x_i}^2}^\frac12 \leq B_{00} + B_{10} + 2h \leq B_0
\end{align*}
for which, similar to the proof above, we have used the fact that $x$ is $\Delta/2$-separated from at least $s-1$ points $x_i$. 
Similarly, for any vector $q = [q_1,\ldots,q_s,Q_1,\ldots,Q_s]\in \CC^{s(d+1)}$  and any $x\in \xf$, we have
\begin{align*}
\norm{q^\top \etaFunc(x)} &\leq \sum_{i=1}^s \abs{q_i} \abs{\fullCov(x_i,x)} + \norm{ Q_i}_{x_i}\norm{\fullCov^{(10)}(x_i,x)}_{x_i} \notag \\
&\leq \nB{D_\met^{-1} q}\pa{ B_{00}  +B_{10} + 2h }\leq B_0 \nB{D_\met^{-1} q}.
\end{align*}

\item
For any $x \in \xn$ we have the bound:
\begin{align*}
\norm{\diff{2}{q^\top \etaFunc}(x)  }_x &= \norm{\sum_{i=1}^s q_i  \fullCov^{(02)}(x_i,x)  + [Q_i]\fullCov^{(12)}(x_i,x) }_x \notag\\
&\leq \norm{D_\met^{-1} q}\pa{\sum_{i=1}^s \norm{\fullCov^{(02)}(x_i,x) }_x^2 + \norm{ \fullCov^{(12)}(x_i,x)}_{x_i,x}^2 }^\frac12 \leq\norm{D_\met^{-1}  q}B_2
\end{align*}
and
\begin{align*}
\norm{\diff{2}{q^\top \etaFunc}(x)}_x &= \norm{\sum_{i=1}^s q_i \fullCov^{(02)}(x_i,x) + [Q_i]\fullCov^{(12)}(x_i,x)}_x \notag\\
&\leq \nB{D_\met^{-1} q}\pa{\sum_{i=1}^s \norm{\fullCov^{(02)}(x_i,x)}_x + \norm{\fullCov^{(12)}(x_i,x)}_{x_i,x}} \leq \nB{D_\met^{-1}  q}B_2
\end{align*}
which concludes the proof.
\end{enumerate}

\subsection{Proof of Lemma \ref{lem:cdf_bd}}\label{app:proof_cdf_bd}

First note that, for $X = n^{-1} \sum_{k=1}^m f(\om_k)$ any empirical average, since the $\om_k$ are iid, we have $\EE_\Eve[X] = \EE_{E_\om}[f(\om)]$, and therefore $\etaFunc_\Eve = \EE_{E_\om} [\gamma(\om) \gamma(\om)^*]$, and so on.

We now prove a general bound, that we then implement for each item. Let $A = A_\om$ be a random matrix that depends on $\om$, such that $\norm{\EE[A]} \leq B$ and $\norm{A} \leq L(\om)$, for any matrix norm $\norm{\cdot}$. We have
\begin{align*}
\EE[A] = \EE[A \bun_{E_\om}] + \EE[A \bun_{E_\om^c}] = \EE_\Eve[A] \PP(E_\om) + \EE[A \bun_{E_\om^c}] 
\end{align*}
by Bayes' rule, and therefore,
\begin{equation}\label{eq:bound-EA-E_EA}
\norm{\EE[A] - \EE_\Eve[A]} \leq \frac{\norm{\EE[A]} \PP(E_\om^c) + \EE[\norm{A} \bun_{E_\om^c}]}{\PP(E_\om)} \leq \frac{ B \PP(E_\om^c) + \EE[L(\om)\bun_{E_\om^c}]}{\PP(E_\om)}
\end{equation}
Then, if we let $E_{\om,q}$ be the event that  $L_q(\om) \leq \Lu_q$, so $E_\om = \cap_{q=0}^3 E_{\om,q}$, by the union bound we get $\PP(E_{\om}^c) \leq \sum_q \PP(E_{\om,q}^c) \leq \sum_q F_q(\Lu_q) \leq \frac{\min(\constker_0,\constker_2)}{m\max_j(\Lu_j^2)} \leq \frac12$, and in particular $\PP(E_\om) \geq \frac12$. In the following $L(\om)$ will be a sum of some of the $L_q(\om)^2$, so we bound $\EE[ L_q(\om)^2 1_{E_\om^c}] \leq \sum_j \EE[ L_q(\om)^2  1_{E_{\om,j}^c}]$ and we have
\begin{align*}
\EE[ L_q(\om)^2 1_{E_{\om,j}^c}]  &= \int_0^\infty \PP(L_q(\om)^2 1_{E_{\om,j}^c} \geq t) \mathrm{d}t = \int_0^\infty \PP\pa{(L_q(\om)^2 \geq t) \cap (L_j(\om) \geq \Lu_j)} \mathrm{d}t \\
&\leq \Lu_q^2 F_j(\Lu_j) + \int_{\Lu_q^2}^\infty F_q(\sqrt{t}) \mathrm{d}t = \Lu_q^2 F_j(\Lu_j) + 2\int_{\Lu_q}^\infty tF_q(t) \mathrm{d}t
\end{align*}
where we have bounded $\PP\pa{(L_q(\om)^2 \geq t) \cap (L_j(\om) \geq \Lu_j)}$ by respectively $\PP(L_j(\om) \geq \Lu_j) \leq F_j(\Lu_j)$ in the first term and by $\PP(L_q(\om)^2 \geq t) \leq F_q(\sqrt{t})$ in the second term. Hence by Assumption \ref{ass:feat} we have
\begin{equation}\label{eq:bound-intL2_1_E}
\EE[ L_q(\om)^2 1_{E_{\om}^c}] \leq \frac{\min(\constker_0, \constker_2)}{m}
\end{equation}

We can now obtain the desired results by combining \eqref{eq:bound-EA-E_EA} and \eqref{eq:bound-intL2_1_E} each time:
\begin{enumerate}[label=(\roman*)]
\item we let $A =D_\met \RFVec(\om) \RFVec(\om)^* D_\met$. We have $\norm{\EE[A]} \leq 2$ by Lemma \ref{lem:additional_admissible}, and $\norm{\RFVec(\om) \RFVec(\om)^*} \leq s L_{01}^2(\om)$. When applied with the norm $\nB{\cdot}$, we get $\nB{\EE[A]} \leq 2$, and $\nB{\RFVec(\om) \RFVec(\om)^*} \leq 2s L_{01}^2(\om)$ by Lemma \ref{lem:block_norm} $(iii)$.
\item we let $A = D_\met \RFVec(\om) \phi_\om(x)$. We have $\norm{\EE[A]} \leq B_0$ by Lemma \ref{lem:additional_admissible}, and $\norm{A} \leq \sqrt{s} L_{01}(\om) L_0(\om) \leq \frac12\sqrt{s} (L_{01}(\om)^1+ L_0(\om)^2)$.
\item we let $A = (\tilde \RFVec(\om)^\top q) \met_x^{-\frac12}(\Hmtx
\phi_\om)(x) \met_x^{-\frac12}$. We have $\norm{\EE[A]} \leq B_2 \norm{q}$ by Lemma \ref{lem:additional_admissible}, and $\norm{A} \leq \sqrt{s} L_{01}(\om) L_2(\om)$.
\end{enumerate}

\section{Concentration bounds} \label{app:concentration}

In this section, we detail the various Berstein concentration inequalities that we used in the golfing scheme. More precisely, we present some probabilistic bounds on deviation of $\subetaFunc$ and $\subetaMat$ from their deterministic counterparts $\etaFunc$ and $\etaMat$, conditional on event $\Eve$ (recall their definitions in \eqref{eq:fullmtx_func}, \eqref{eq:submtx_func} and \eqref{eq:event}). Define the shorthands
$$
L_{ij}(\om) \eqdef  \sqrt{L_i(\om)^2+L_j(\om)^2} \qandq
\Lu_{ij}\eqdef \sqrt{\Lu_i^2+\Lu_j^2}.
$$
Observe that conditional on $\Eve$, we have
\begin{equation}\label{eq:gamma-bound}
\norm{D_\met \RFVec(\om)} \leq \sqrt{s(\Lu_0^2+\Lu_1^2)} = \sqrt{s}\Lu_{01}
\end{equation}

All this section is done under the assumptions of Theorem \ref{thm:main}, and we will use several times the following from Lemma \ref{lem:additional_admissible} and \ref{lem:cdf_bd}:
\begin{equation}
\label{eq:norm_D_etaMat_D}
\norm{D_\met \etaMat_\Eve D_\met} \leq1+  \norm{\Id - D_\met \etaMat D_\met} + \norm{D_\met(\etaMat - \etaMat_\Eve) D_\met} \leq 2
\end{equation}

\subsection{Elementary concentration inequalities}
To begin, we first recall some elementary concentration inequalities.

\begin{lem}[Matrix Bernstein for complex matrices]\label{lem:bernstein_matrix}
Let $Y_1,\ldots, Y_M$ be a sequence of $d_1\times d_2$ complex random matrices with $\EE[Y_\ell] = 0$, $\norm{Y_\ell}_{2\to 2} \leq K$ for all $\ell =1,\ldots, M$ and set
$$
\sigma^2 \eqdef \max\left\{ \norm{\sum_{\ell=1}^M \EE(Y_\ell Y_\ell^*)}_{2\to 2}, \norm{\sum_{\ell=1}^M \EE(Y_\ell^* Y_\ell) }_{2\to 2}\right \}.
$$
Then,
$$
\PP\left(  \norm{ \frac{1}{M} \sum_{\ell=1}^M Y_\ell}_{2\to 2} \geq t \right) \leq 2(d_1+d_2) \exp \left( -\frac{M t^2/2}{\sigma^2/M + Kt/3} \right).
$$
\end{lem}

\begin{lem}[Vector Bernstein for complex vectors \citep{minsker2017some}] \label{lem:vec-bernstein} 
Let $Y_1,\ldots, Y_M\in \CC^d$ be a sequence of independent random vectors such that $\EE[Y_i] = 0$, $\norm{Y_i}_2\leq K$ for $i=1,\ldots, M$ and set
$$
\sigma^2 \eqdef \sum_{i=1}^M \EE\norm{Y_i}_2^2.
$$
Then, for all $t\geq (2K + 6\sigma)/M$, 
$$
\PP\pa{\norm{\frac{1}{M} \sum_{i=1}^M Y_i}_2 \geq t} \leq 28 \exp \pa{ - \frac{M t^2/2}{\sigma^2/M + t K /3}}
$$
\end{lem}

\subsection{Deviation between $\etaFunc_\Eve$ and $\subetaFunc$}\label{app:bound_f}

\begin{lem}[Bound against a fixed vector]\label{lem:bound_f}
Let $q \in \CC^{s(d+1)}$ and $x \in \Xx$. For all $u>0$ we have
$$
\PP_\Eve\pa{\abs{(\etaFunc_\Eve(x) - \subetaFunc(x))^\top D_\met q} \geq u \norm{q} } \leq 4\exp\pa{\frac{-m u^2}{2\Lu_0^2 + 2\sqrt{s} \Lu_{01} \Lu_0 u/3)}}.
$$
As a corollary,
$$
\PP_\Eve\pa{\abs{(\etaFunc_\Eve(x) - \subetaFunc(x))^\top D_\met q} \geq u \nB{q} } \leq 4\exp\pa{\frac{-m u^2}{4s( 2 \Lu_0^2 + \sqrt{2} \Lu_{01} \Lu_0 u/3)}}.
$$
\end{lem}

\begin{proof}
Assume $\norm{q}_2 = 1$ without lost of generality. We apply the classical (scalar) Bernstein inequality. By defining $Y_k\eqdef \varphi_{\om_k}(x) \RFVec(\om_k)^* D_\met q - \EE_E[\varphi_{\om}(x) \RFVec(\om)^\top D_\met q]$, we have
$(\subetaFunc(x)-\etaFunc_\Eve(x))^\top D_\met q =  \frac{1}{m} \sum_{k=1}^m Y_k$. To apply Bernstein's inequality, observe that for each $k=1,\ldots, m$, $\EE_\Eve[Y_k] = 0$, and conditional on event $\Eve$, we have $\abs{Y_k} \leq  2\sqrt{s} \Lu_{01} \Lu_0$ and $\EE_E \abs{Y_k}^2 = \EE_E \abs{\varphi_{\om_k}(x)}^2 \abs{\RFVec(\om_k)^* D_\met q}^2 \leq \Lu_0^2 \norm{D_\met \etaMat_\Eve D_\met} \leq 2 \Lu_0^2$ by \eqref{eq:norm_D_etaMat_D}.
Therefore,
$$
\PP\pa{\abs{\frac{1}{m} \sum_{k=1}^m Y_k} \geq u} \leq 4 \exp\pa{\frac{-m u^2}{2\Lu_0^2 + 2\sqrt{s} \Lu_{01} \Lu_0 u/3)}}.
$$
The last inequality follows because $\nB{q} \geq \norm{q}_2/\sqrt{2s}$.
\end{proof}

\begin{lem}[Uniform bound]\label{lem:bound_f_alone}
Fix $x\in \Xx$. For all $u>\frac{4\sqrt{s}\Lu_{01} \Lu_0}{m} + \frac{6\sqrt{s}\Lu_{01}}{\sqrt{m}}$ we have
$$
\PP_\Eve\pa{\norm{D_\met(\etaFunc_\Eve(x) - \subetaFunc(x))} \geq u} \leq 4\exp\pa{\frac{-m u^2}{s\Lu_{01}^2 + 2\sqrt{s} \Lu_{01} \Lu_0 u/3)}}.
$$
\end{lem}
\begin{proof}
We apply the vector Bernstein inequality (Lemma \ref{lem:vec-bernstein}). By defining $Y_k\eqdef D_\met \overline{\RFVec(\om_k)} \varphi_{\om_k}(x) - \EE_E[ D_\met \overline{\RFVec(\om_k)} \varphi_{\om_k}(x) ]$, we have
$D_\met(\subetaFunc(x)-\etaFunc_\Eve(x)) = \frac{1}{m} \sum_{k=1}^m Y_k$. Observe that for each $k=1,\ldots, m$, $\EE_\Eve[Y_k] = 0$, and conditional on event $\Eve$, we have $\abs{Y_k} \leq  2\sqrt{s} \Lu_{01} \Lu_0$ and $\EE_E \norm{Y_k}^2 = \EE_E \abs{\varphi_{\om_k}(x)}^2 \norm{D_\met \RFVec(\om_k)}^2 \leq s\Lu_{01}^2$.
Therefore, for all $u \geq \frac{4\sqrt{s}\Lu_{01} \Lu_0}{m} + \frac{6\sqrt{s}\Lu_{01}}{\sqrt{m}}$,
$$
\PP\pa{\norm{\frac{1}{m} \sum_{k=1}^m Y_k} \geq u} \leq 4 \exp\pa{\frac{-m u^2}{s\Lu_{01}^2 + 2 \sqrt{s}\Lu_{01} \Lu_0 u/3)}}.
$$
The last inequality follows because $\nB{q} \geq \norm{q}_2/\sqrt{2s}$.
\end{proof}

\subsection{Deviation between $\diff{2}{\etaFunc_\Eve}$ and $\diff{2}{\subetaFunc}$}\label{app:bound_f_hess}

\begin{lem}[Bound against a fixed vector]\label{lem:bound_f_hess}
Let $q \in \CC^{s(d+1)}$ and $x \in \Xx$. For all $u>0$ we have
\begin{equation}
\PP_\Eve\pa{\norm{ \diff{2}{(\etaFunc_\Eve - \subetaFunc)^\top D_\met q}(x)}_x \geq u \norm{q}}
 \leq 4d \exp \pa{ \frac{-m u^2}{2\Lu_2^2 + 2\sqrt{s}\Lu_{01} \Lu_2 u/3 )}}.
\end{equation}
as a corollary
\begin{equation}
\PP_\Eve\pa{\norm{ \diff{2}{(\etaFunc_\Eve - \subetaFunc)^\top D_\met q}(x)}_x \geq u \nB{q}}
 \leq 4d \exp \pa{ \frac{-m u^2}{4s( 2\Lu_2^2 + \sqrt{2} \Lu_{01} \Lu_2 u/3 )}}.
\end{equation}
\end{lem}

\begin{proof}
Assume $\norm{q}=1$ without lost of generality. Recalling the definitions of Sec. \ref{sec:diff_geo}, we have
\begin{align*}
\norm{ \diff{2}{(\etaFunc_\Eve - \subetaFunc)^\top D_\met q}(x)}_x = \norm{\met_x^{-\frac12} \Hmtx\pa{(\etaFunc_\Eve - \subetaFunc)^\top D_\met q}(x)\met_x^{-\frac12}}
\end{align*}
We now apply Lemma \ref{lem:bernstein_matrix}. Define
\[
Y_k = (q^\top D_\met \RFVec(\om_k)) \met_x^{-\frac12}  \Hmtx\pa{ \phi_{\om_k}}(x) \met_x^{-\frac12}   - \EE_\Eve (q^\top D_\met \RFVec(\om)) \met_x^{-\frac12}  \Hmtx\pa{ \phi_{\om}}(x) \met_x^{-\frac12}.
\]
which are indeed symmetric matrices.
We have $\EE_E Y_k = 0$ and conditional on event $E$,
\begin{align*}
\norm{Y_k} &\leq 2 \sqrt{s} \Lu_{01} \Lu_2.
\end{align*}
Furthermore, defining $A = (q^\top D_\met \RFVec(\om_k)) \met_x^{-\frac12}  \Hmtx\pa{ \phi_{\om_k}}(x) \met_x^{-\frac12} $ (which is symmetric), we have
\begin{align*}
0 \preceq \EE_\Eve [Y_j Y_j^*] &\preceq \EE_\Eve \pa{A A^*} - \EE_\Eve A\EE_\Eve A^*  \preceq \EE_\Eve \pa{A A^*}  \preceq \Lu_2^2 \EE_\Eve \abs{q^\top D_\met \RFVec(\om)}^2 \Id \preceq \Lu_2^2  \norm{D_\met \etaMat_\Eve D_\met} \Id \preceq 2\Lu_2^2 \Id
\end{align*}
and thus $\norm{ \EE_\Eve [Y_j Y_j^*]} \leq  2\Lu_2^2$.
Therefore, the matrix Bernstein's inequality yields
\[
\PP\left(  \norm{ \frac{1}{m} \sum_{\ell=1}^m Y_\ell}_2 \geq u \right) \leq 4d \exp \left( \frac{-m u^2}{2\Lu_2^2 + 2\sqrt{s}  \Lu_{01} \Lu_2 u/3 } \right).
\]
The last inequality follows because $\nB{q} \geq \norm{q}_2/\sqrt{2s}$.
\end{proof}

\begin{lem}[Uniform bound]\label{lem:bound_f_hess_uniform}
Let $x \in \Xx$.  
Let $\Bb_x \eqdef \enscond{v \in \CC^d}{\norm{v}_x \leq 1}$ and  given $v\in\Bb_x$, let $g_\om(v) \eqdef \diff{2}{\phi_\om}(x)[v,v] \in \CC$. Then, for all $u>\frac{4\sqrt{s} \Lu_{01} \Lu_2}{m} + \frac{6\sqrt{2} \Lu_2^2}{\sqrt{m}}$,
$$
\PP\pa{
 \sup_{v\in \Bb_x} \norm{\frac{1}{m} \sum_{k=1}^m D_\met\overline{\RFVec(\om_k)} g_{\om_k}(v) - \EE_\Eve D_\met  \overline{\RFVec(\om)} g_{\om}(v)} 
\geq u } \leq  \exp\pa{C d \log \pa{\frac{s \Lu_{01} \Lu_2}{u}} - \frac{m u^2}{s \Lu_{01}^2 B_{22} + 2 \sqrt{s}\Lu_{01} \Lu_2 u/3)}}
$$
for some constant $C$.

\end{lem}

\begin{proof}

We use a covering net strategy: let $\Vv =\{ v_1,\ldots, v_N\}$ be a covering $\varepsilon$-net of $\Bb_x $, for $\varepsilon>0$ that we will adjust later. Fix $v \in \Vv$, and define $Y_k = D_\met\overline{\RFVec(\om_k)} g_{\om_k}(v) - \EE_\Eve D_\met\overline{\RFVec(\om_k)} g_{\om_k}(v) \in \CC^{s(d+1)}$ centered $i.i.d.$ variables. We have $\EE_\Eve Y_k = 0$, $\abs{Y_k} \leq 2 \sqrt{s} \Lu_{01}\Lu_2$ and $\EE_\Eve\norm{Y_k}^2 \leq \EE_\Eve \abs{g_{\om}(v)}^2 \norm{D_\met \RFVec(\om)}^2 \leq s \Lu_{01}^2 B_{22}$. Hence applying Lemma \ref{lem:vec-bernstein}: for all $u \geq \frac{4\sqrt{s} \Lu_{01} \Lu_2}{m} + \frac{6\sqrt{B_{22}s} \Lu_{01} }{\sqrt{m}}$,
$$
\PP_\Eve\pa{\norm{\frac{1}{m} \sum_{k=1}^m Y_k} \geq u} \leq 4 \exp\pa{\frac{-m u^2}{s \Lu_{01}^2 B_{22} + 2 \sqrt{s}\Lu_{01} \Lu_2 u/3)}}.
$$
Next, we use the fact that for all $\om$
\[
\abs{D_\met\overline{\RFVec(\om)} g_{\om}(v) - D_\met\overline{\RFVec(\om)} g_{\om}(v')} \leq 2\sqrt{s}\Lu_{01} \Lu_2 \norm{v-v'}_x
\]
Hence by choosing
\[
\abs{\Vv} \sim \pa{\frac{\sqrt{s} \Lu_{01} \Lu_2}{u}}^d
\]
and using a union bound on $\abs{\Vv}$ we conclude the proof.
\end{proof}

\subsection{Deviation between $\etaMat_\Eve$ and $\subetaMat$}\label{app:bound_etaMat}

\begin{lem}[Bound in spectral norm]\label{lem:bound_etaMat}
For all $u>0$, it holds that,
\begin{equation}\label{eq:bound_proba_etaMatR}
\begin{split}
\mathbb{P}_\Eve \pa{\norm{D_\met(\etaMat_\Eve-\subetaMat) D_\met}\geq u}  \leq 
4(d+1)s \exp\left(- \frac{mu^2}{ 2 s \Lu_{01}^2  + 2 s \Lu_{01}^2 u/3} \right).
\end{split}
\end{equation}
\end{lem}
\begin{proof}
To bound this probability, we apply Lemma \ref{lem:bernstein_matrix} with $Y_k \eqdef (D_\met\RFVec(\om_k))(D_\met\RFVec(\om_k))^* - \etaMat_\Eve$. We have, conditional on event $E$:
\[
\EE_\Eve[ Y_k]= 0,\quad \norm{Y_k} \overset{\eqref{eq:gamma-bound}}{\leq} 2s\Lu_{01}^2.
\]
Also,
\begin{align*}
0 \preceq \EE_\Eve[Y_k Y_k^*] = \EE_\Eve[Y_k^* Y_k] &= \EE_\Eve[\norm{D_\met \RFVec(\om_k)}^2 (D_\met \RFVec(\om_k))(D_\met \RFVec(\om_k))^*] - (D_\met \etaMat_\Eve D_\met)^2\\
&
\preceq \EE_\Eve[\norm{D_\met \RFVec(\om_k)}^2 (D_\met \RFVec(\om_k))(D_\met \RFVec(\om_k))^*]  \preceq s \Lu_{01}^2 \norm{D_\met\etaMat_\Eve D_\met}\Id
\end{align*}
So, $\norm{\EE[Y_k^* Y_k]} = \norm{\EE[Y_k Y_k^*]} \leq s \Lu_{01}^2 \norm{D_\met \etaMat_\Eve D_\met} \leq 2s \Lu_{01}^2$ by \eqref{eq:norm_D_etaMat_D}.
By choosing $K=2 s\Lu_{01}^2$ and  $\sigma^2 =  m  s \Lu_{01}^2 \norm{D_\met \etaMat_\Eve D_\met}$ in  Lemma \ref{lem:bernstein_matrix}, we obtain
$$\PP_\Eve \pa{\norm{D_\met(\etaMat_\Eve - \subetaMat)D_\met} \geq u}
\leq 4(d+1)s \exp\left(- \frac{mu^2}{2 s \Lu_{01}^2 + 2 s\Lu_{01}^2 t/3} \right).
$$
\end{proof}

\begin{lem}\label{lem:appl-vec-bern}
For $i=1,\ldots, s$, let $S_i = \{s+(i-1)d +1, \ldots, s+id\}$,  $q\in \CC^{s(d+1)}$. 
Then, for all 
$u\geq \frac{4\sqrt{s}\Lu_{01} \Lu_1}{m} + \frac{6\sqrt{2}\Lu_1}{\sqrt{m}}$,
$$
\PP_\Eve\pa{\norm{ (D_\met(\etaMat_\Eve - \subetaMat)D_\met q)_{S_i}  }_2 > u\norm{q}_2} \leq 28 \exp\pa{ \frac{- m u^2/2}{2\Lu_1^2 +2u\sqrt{s} \Lu_{01} \Lu_1/3 }}.
$$
As a corollary, for all 
$
u\geq \frac{4\sqrt{2}s\Lu_{01} \Lu_1}{m} + \frac{12\sqrt{s}\Lu_1}{\sqrt{m}}
$, we have
$$
\PP_\Eve\pa{\norm{ (D_\met (\etaMat_\Eve - \subetaMat) D_\met q)_{S_i}  }_2 > u\nB{q}} \leq 28 \exp\pa{ \frac{- m u^2}{4 s \pa{2 \Lu_1^2 +\sqrt{2} u { \Lu_{01} \Lu_1}/3  } }} .
$$
\end{lem}
\begin{proof}
Fix $i\in \{1,\ldots, s\}$. Without loss of generality, assume that $\norm{q}_2 = 1$.
The claim of this lemma follows by applying Lemma \ref{lem:vec-bernstein}.
Let $$Y_k = \met_{x_i}^{-\frac12} \overline{\nabla \varphi_{\om_k}(x_i)} (\RFVec(\om_k)^*D_\met q) -  \EE_\Eve \pa{\met_{x_i}^{-\frac12} \overline{\nabla \varphi_{\om_k}(x_i)} (\RFVec(\om_k)^*D_\met q)} \in \CC^d,
 $$
and observe that $(D_\met(\subetaMat - \etaMat_\Eve)D_\met q)_{S_i} = \frac{1}{m}\sum_k Y_k$.
We apply Lemma \ref{lem:vec-bernstein}. Observe that conditional on event $\Eve$, we have
\[
\norm{Y_k}_2 \leq 2\norm{q}_2 \norm{D_\met \RFVec(\om_k)}_2 \norm{\diff{1}{\varphi_{\om_k}}(x_i)}_{x_i} \leq 2 \sqrt{s} \Lu_{01} \Lu_1
\]
and
\begin{align*}
\EE_\Eve \norm{Y_k}^2 \leq \EE_\Eve \abs{ \RFVec(\om_k)^* D_\met q}^2 \norm{\met_{x_i}^{-\frac12}\nabla \varphi_{\om_k}(x_i)}_2^2  \leq \Lu_1^2 q^* D_\met \etaMat_\Eve D_\met q \leq \Lu_1^2 \norm{D_\met \etaMat_\Eve D_\met}_2 \leq 2\Lu_1^2
\end{align*}
by \eqref{eq:norm_D_etaMat_D}.
Therefore, for all $u\geq \frac{4\sqrt{s}\Lu_{01} \Lu_1}{m} + \frac{6\sqrt{2}\Lu_1}{\sqrt{m}}$,
$$
\PP_\Eve\pa{\norm{\frac{1}{m} \sum_{i=1}^m Y_i }_2 \geq u} \leq 28 \exp\pa{- \frac{m u^2/2}{2\Lu_1^2+ 2u\sqrt{s} \Lu_{01} \Lu_1/3 }}
$$
The last inequality follows because $\nB{q} \geq \norm{q}_2/\sqrt{2s}$.
\end{proof}

\begin{lem}[Bound in block norm]\label{lem:bound_R_vec}
Let $q\in \CC^{s(d+1)}$ be any vector. For all
$
u\geq \frac{4\sqrt{2}s\Lu_{01} \Lu_1}{m} + \frac{12\sqrt{s}\Lu_1}{\sqrt{m}}
$ 
we have
\begin{equation}
\PP_\Eve \pa{\nB{D_\met(\etaMat_\Eve - \subetaMat) D_\met q} \geq u \nB{q}} \leq  32 s \exp\pa{ \frac{- m u^2}{4s \pa{2 \Lu_{01}^2 +\sqrt{2} u \Lu_{01}\Lu_1 /3  } }}
\end{equation}
\end{lem}
\begin{proof}
Let $S_0 \eqdef \{1,\ldots, s\}$ and $S_j \eqdef \{ s+(j-1)d+1, \ldots, s+jd\}$ for $j=1,\ldots, s$.
By the union bound
\begin{align}
\PP_\Eve&\pa{\nB{D_\met (\etaMat_\Eve - \subetaMat) D_\met q} \geq u \nB{q}}
\notag \\
&\leq \sum_{j=1}^s \PP_\Eve\pa{\abs{(D_\met (\etaMat_\Eve - \subetaMat)  D_\met q)_{j}} \geq u \nB{q}} + \sum_{j=1}^s \PP_\Eve\pa{\norm{(D_\met (\etaMat_\Eve - \subetaMat) D_\met q)_{S_j}}_2 \geq u \nB{q}}
\label{eq:tobound1}
\end{align}
To bound the first sum, observe that for $j=1,\ldots, s$,
$( D_\met (\etaMat - \subetaMat) D_\met q)_{j} = (D_\met(\etaFunc(x_j) - \subetaFunc(x_j)))^\top q$  and apply Lemma \ref{lem:bound_f}. The second sum can be bounded by applying Lemma \ref{lem:appl-vec-bern}.

\end{proof}

\subsection{Proof of Proposition \ref{prop:eta_j}}\label{app:proof_prop_subetaj}

We fix a particular $j=1,\ldots,s$, do the proof for $\subeta_j$, then use a union bound to conclude. As before, it is enough to establish the probability that $\subeta_j$ satisfies the properties of Proposition \ref{prop:eta_j} conditional on event $\Eve$. We proceed in the same way as in the main proof of the golfing scheme: first we show that $\subeta_j$ satisfies the desired property on a finite grid, then we bound $\norm{p_j}$, and finally we use the latter to extend the non-degeneracy to the whole space. As mentioned in the paper, the first step is considerably simpler and more direct than the golfing scheme, since the ``sign'' vector $u_j$ is of norm $1$.


\paragraph{Deviation bounds on a grid.} Similar to our previous argument, we will bound the deviation between $\subeta_j$ and $\eta_j$ on a finite grid $\xfg\subset \xf$ whose precision we will later adjust, and between $\diff{2}{\subeta_j}$ and $\diff{2}{\eta_j}$ on $\xng \subset \xn$.
We will show that
\begin{align*}
&\forall x \in \xfg,~ \abs{\eta_j(x) - \subeta_j(x)} \leq \frac{\constker_0}{16} \\
&\forall x \in \xng,~ \norm{\diff{2}{\eta_j}(x) - \diff{2}{\subeta_j}(x)}_x \leq \frac{\constker_2}{32}.
\end{align*}

Let $\hat q_j \eqdef D_\met^{-1} \subetaMat^{-1} u_j$ and $ q_j \eqdef D_\met^{-1} \etaMat^{-1} u_j$.  Note that $q_j$ is deterministic and $\norm{q_j}\leq 2$ for all $j$. Recall also that $\eta_j = q_j^\top D_\met \etaFunc(x)$ and $\subeta_j = \hat q_j^\top D_\met \subetaFunc(x)$. For $x\in\xfg$,
\begin{align*}
&\abs{\eta_j(x) - \subeta_j(x) } \leq \abs{q_j^\top D_\met (\etaFunc(x) - \subetaFunc(x))} + \norm{D_\met^{-1}( \etaMat^{-1} - \subetaMat^{-1} )D_\met^{-1}} \norm{D_\met \subetaFunc(x)} 
\\
&\leq  \abs{q_j^\top D_\met (\etaFunc(x) - \subetaFunc(x))} + 8 \norm{D_\met (\etaMat - \subetaMat)}  \norm{D_\met \subetaFunc(x)} 
\end{align*}
where the last line is valid with probability $1-\rho$ by Lemma \ref{lem:additional_admissible} and \eqref{eq:bound_subetaMatinv}.
Similarly,
\begin{align*}
&\norm{\diff{2}{\eta_j}(x) - \diff{2}{\subeta_j}(x) }_x\\
& \leq
\norm{\diff{2}{q_j^\top D_\met (\etaFunc - \subetaFunc)}(x)}_x + \norm{D_\met^{-1}( \etaMat^{-1} - \subetaMat^{-1} )D_\met^{-1}}  \sup_{\norm{v}_x\leq 1}\norm{\frac{1}{m} \sum_{k=1}^mD_\met \overline{\RFVec(\om_k)} \diff{2}{\phi_{\om_k}}(x)[v,v] }\\
&\leq 
\norm{\diff{2}{q_j^\top D_\met (\etaFunc - \subetaFunc)}(x)}_x +8  \norm{D_\met (\etaMat - \subetaMat)}   \sup_{\norm{v}_x\leq 1}\norm{\frac{1}{m} \sum_{k=1}^mD_\met \overline{\RFVec(\om_k)} \diff{2}{\phi_{\om_k}}(x)[v,v] }
\end{align*}
where again,  the last line is valid with probability $1-\rho$ by Lemma \ref{lem:additional_admissible} and \eqref{eq:bound_subetaMatinv}. 
Therefore, we simply have to show that with probability at least $1-\rho$,
\begin{itemize}
\item[(i)] For $j=1,\ldots,s$,  $\abs{q_j^\top D_\met (\etaFunc_\Eve(x) - \subetaFunc(x))} \leq \constker_0/32$ for all $x\in \xfg$.
\item[(ii)] For $j=1,\ldots, s$,  $\norm{\diff{2}{q_j^\top D_\met (\etaFunc - \subetaFunc)}(x)}_x \leq \constker_2/64$ for all $x\in \xng$.
\item[(iii)] $ \norm{D_\met \subetaFunc(x)} \leq 2 B_0$ for all $x\in \xfg$.
\item[(iv)] $  \sup_{\norm{v}_x\leq 1}\norm{\frac{1}{m} \sum_{k=1}^mD_\met \overline{\RFVec(\om_k)} \diff{2}{\phi_{\om_k}}(x)[v,v] } \leq 2 B_2$ for all $x\in \xng$.
\item[(v)] $ \norm{D_\met( \etaMat - \subetaMat )D_\met} \leq \min\pa{\frac{\constker_0}{512 B_0}, \frac{\constker_2}{1024 B_2}}$.
\end{itemize}
By applying  Lemma \ref{lem:cdf_bd} and recalling our choice of $m$,
(i) follows by Lemma \ref{lem:bound_f}, (ii) follows by Lemma \ref{lem:bound_f_hess},
(iii) follows by Lemma \ref{lem:bound_f_alone},
(iv) follows by Lemma \ref{lem:bound_f_hess_uniform},
and (v) follows by Lemma \ref{lem:bound_etaMat}.

\paragraph{Bound on $p_j$.} By the same computations as in Section \ref{sec:proofstep3}, we have $\subeta_j(x) = (\subetaMat^{-1} u_j)^\top \subetaFunc(x) = \Phi^* p_j$ with $p_j = \frac{1}{\sqrt{m}}\pa{\RFVec(\om_i)^* \subetaMat^{-1} u_j}_{i=1}^m$. Therefore,
\begin{align*}
\norm{p_j}_2^2 &= \frac{1}{m} \sum_{i=1}^m u_j^* \subetaMat^{-1}  \RFVec(\om_i) \RFVec(\om_i)^* \subetaMat^{-1} u_j = u_j^* \subetaMat^{-1} u_j \\
&\leq \norm{D_{\met} \subetaMat^{-1} D_{\met}^{-1}} \leq 4
\end{align*}
with probability $1-\rho$, by \eqref{eq:bound_subetaMatinv}.

\paragraph{Extension to the whole domain.}
We proceed as in Section \ref{sec:proofstep4}. By the same computations, we obtain: for any $x \in \xf$ and $x' \in \xfg$,
\[
\abs{\subeta_j(x)} \leq \abs{\eta_j(x')} + \abs{\subeta_j(x') - \eta_j(x')} + \abs{\subeta_j(x) -\subeta_j(x')} \leq 1-\frac{\constker_0}{4} + \frac{\constker_0}{16} + \Lu_{1} \norm{p_j} \delta_\met(x,x')
\]
and therefore, we choose
\begin{align*}
\abs{\xfg} \sim \pa{\frac{\Rr_\Xx \Lu_1}{\constker_0}}^d
\end{align*}

For the second covariante derivative, as in Section \ref{sec:proofstep4} we get: for all $x \in \xn_j$ and $x' \in \xng_j$,
\begin{equation*}
\begin{split}
&\norm{\overline{\sign(a_j)} \diff{2}{\subeta_j}(x)  - \fullCov^{(02)}(x_j,x)}_x \\
&\quad\leq \norm{\diff{2}{\subeta_j}(x)  - \diff{2}{\subeta_j}(x')[\tau_{\x \to x'} \cdot,\tau_{\x \to x'} \cdot]}_x \\
&\qquad+ \norm{\diff{2}{\subeta_j}(x')[\tau_{\x \to x'} \cdot,\tau_{\x \to x'} \cdot] -\diff{2}{\eta_j}(x')[\tau_{\x \to x'} \cdot,\tau_{\x \to x'} \cdot]}_x \\
&\qquad+ \norm{\overline{\sign(a_j)}\diff{2}{\eta_j}(x')[\tau_{\x \to x'} \cdot,\tau_{\x \to x'} \cdot] -\fullCov^{(02)}(x_j,x')[\tau_{\x \to x'} \cdot,\tau_{\x \to x'} \cdot]}_x \\
&\qquad+ \norm{\fullCov^{(02)}(x_j,x')[\tau_{\x \to x'} \cdot,\tau_{\x \to x'} \cdot] - \fullCov^{(02)}(x_j,x)}_x \\
&\quad \leq \frac{\constker_2}{32} + \frac{\constker_2}{16} + \frac{\constker_2}{64} + \Lu_3(\Lu_0 + \norm{p_j})\dsep_\met(x,x')
\end{split}
\end{equation*}
and similarly for $\ell\neq j$, for all $x \in \xn_\ell$ and $x' \in \xng_\ell$,
\begin{equation*}
\begin{split}
\norm{\diff{2}{\subeta_j}(x)}_x &\leq \norm{\diff{2}{\subeta_j}(x)  - \diff{2}{\subeta_j}(x')[\tau_{\x \to x'} \cdot,\tau_{\x \to x'} \cdot]}_x \\
&\quad+ \norm{\diff{2}{\subeta_j}(x')[\tau_{\x \to x'} \cdot,\tau_{\x \to x'} \cdot] -\diff{2}{\eta_j}(x')[\tau_{\x \to x'} \cdot,\tau_{\x \to x'} \cdot]}_x + \norm{\diff{2}{\eta_j}(x')[\tau_{\x \to x'} \cdot,\tau_{\x \to x'} \cdot]}_x \\
&\quad \leq \frac{\constker_2}{32} + \frac{\constker_2}{16} + \Lu_3\norm{p_j}\dsep_\met(x,x')
\end{split}
\end{equation*}
and therefore we conclude by setting
\begin{align*}
\abs{\xng} \sim s\pa{\frac{\rnear \Lu_3\Lu_{0}}{\constker_2}}^d
\end{align*}

The final bound on $m$ is satisfied with the one we obtained previously \eqref{eq:mbound3}.

\section{Application: Discrete Fourier sampling}\label{app-discretefourier}

\newcommand{\fq}{f}
\newcommand{\cf}{C_f}
\newcommand{\sig}{x}
\newcommand{\sigg}{x'}
\newcommand{\hm}{H^\infty}
\newcommand{\km}{\kappa^\infty}

In this section, we consider the case of sampling Fourier coefficients as described in \cite{candes-towards2013}.
Let $\fq \in \NN$ and $\Xx \in \TT^d$ the $d$-dimensional torus.
Let   $\Omega = \enscond{\om\in \ZZ^d}{\norm{\om}_\infty\leq \fq }$, $\phi_\om(x) \eqdef  e^{\mathrm{i} 2\pi \om^\top x}$, and  $\Lambda(\om) = \prod_{j=1}^d g(\omega_j)$ where $g(j) = \frac{1}{\fq } \sum_{k=\max(j-\fq ,-\fq )}^{\min(j+\fq ,\fq )}(1-\abs{k/\fq })(1-\abs{(j-k)/\fq })$.

\paragraph{The kernel and Fisher metric}
The associated kernel is the multivariate Jackson kernel
$
\fullCov(x,x') = \prod_{i=1}^d \kappa(x_i-x_i'),
$
where $$\kappa(x)\eqdef \pa{ \frac{\sin\pa{\pa{\tfrac{\fq }{2}+1} \pi x}}{\pa{\tfrac{\fq }{2}+1} \sin(\pi x)} }^4,
$$
with constant metric tensor \[
\met_x =  \cf \Id\qandq \dsep_\met(x,x') = \cf^{\frac12} \norm{x-x'}_2.
\]
where $\cf \eqdef -\kappa''(0) = \frac{\pi^2}{3}\fq(\fq+4) \sim f^2$.
 Note that $\fullCov^{(ij)} = \nabla_1^i \nabla_2^j \fullCov$ and $\norm{\fullCov^{(ij)}}_{x,x'}=   \cf^{-(i+j)/2}\norm{ \nabla_1^i \nabla_2^j \fullCov}$. Moreover, since the metric is constant,  we have $\norm{\cdot}_x = \cf^{\frac12}\norm{\cdot}$ for all $x$. The domain diameter is $\Rr_\Xx = \cf^{\frac12} d^{\frac12}$.

\paragraph{Sampling bounds}
Suppose that $\fq\geq 128$.  The rest of this section consists of Lemmas which bound the parameters in Theorem \ref{thm:main}: We show in Lemma \ref{lem:Jackson_hessian} that by choosing $\rnear = \frac{1}{8\sqrt{2}}$, for all $\dsep_\met(x,x') \leq \rnear$, we can set $\constker_2 = (1-6\rnear^2)/(1-\rnear^2/(2-\rnear^2) - \rnear^2) \geq 0.941$. In Lemma \ref{lem:near_bound_Jackson}, we show that for all $\dsep_\met(x,x') \geq \rnear$, $\abs{\fullCov(x,x') } \leq 1-1/(8^3\cdot 2)$, so we can set $\constker_0 \eqdef 0.00097$. Moreover, the uniform bounds given in Lemma \ref{lem:Jackson_upbd} imply that
$$
\frac{\min(\constker_0,\constker_2)}{32\max_{i,j} B_{ij}} = \Oo(d^{-\frac12}).
$$
So, for $h = \Oo(d^{-\frac12})$,  by Lemma \ref{lem:far_bounds_Jackson}, we have
$
W(h,s) = \Oo(s^{\frac{1}{4}} d^{\frac12}).
$
Gradient bounds are computed in Section \ref{sec:jackson-grads}.

To summarise, Theorem \ref{thm:main} is applicable with:
\begin{itemize}
\item[(i)] $B_{00} = B_{02} = B_{12} = \Oo(1)$, $B_{01} = \Oo(d^{\frac12})$,  $B_{22} = \Oo(d)$ and $C_\met =0$.
\item[(ii)] $\rnear = 1/(8\sqrt{2})$, $\constker_0 = 0.00097$, $\constker_2 = 0.941$.
\item[(iii)] $\Delta = \Oo(d^{\frac12} s_{\max}^{\frac{1}{4}})$.
\item[(iv)] $\Lu_i = \Oo(d^{i/2})$.
\end{itemize}
and
$$
m\gtrsim d^2 s  \pa{ \log(s) \log\pa{\frac{s}{\rho}} + \log\pa{\frac{(fd)^d}{\rho}}}.
$$

\subsection{Preliminaries: properties of the univariate kernel}\label{sec:Jackson-props}
We first summarise  in Section \ref{sec:Jackson-props}  some key properties of the univariate Jackson kernel $\kappa$ when $\fq \geq 128$ which were derived in \cite{candes-towards2013}.

From \cite[Equations (2.20)-(2.24) and (2.29)]{candes-towards2013},
for all $t \in [-1/2,1/2]$  and $\ell=0,1,2,3$:
\begin{equation}\label{eq:upp_bd_Jackson}
\begin{split}
 1- \frac{\cf}{2}t^2 &\leq \kappa(t) \leq 1-  \frac{\cf}{2}t^2 + 8 \pa{\frac{1+2/\fq}{1+2/(2+\fq)}}^2  \cf^2 t^4 \leq  1-  \frac{\cf}{2}t^2 + 8  \cf^2 t^4 \\
 \abs{\kappa'(t)} &\leq \cf t,\quad \abs{\kappa''(t)} \leq \cf, \quad \abs{\kappa'''(t)} \leq 3 \pa{\frac{1+2/\fq}{1+2/(2+\fq)}}^2 \cf^2 t \leq 12 \cf^2 t \\
\kappa'' &\leq -\cf + \frac{3}{2} \pa{\frac{1+2/\fq}{1+2/(2+\fq)}}^2 \cf^2 t^2  \leq -\cf + 6 \cf^2 t^2.
\end{split}
\end{equation}

By \cite[Lemma 2.6]{candes-towards2013},
\begin{align*}
\abs{\kappa^{(\ell)}(t)} \leq \begin{cases}
 \frac{\pi^\ell H_\ell(t) }{(\fq+2)^{4-\ell} t^4}, &t\in [\frac{1}{2\fq}, \frac{\sqrt{2}}{\pi}]\\
 \frac{\pi^\ell H_\ell^\infty}{(\fq+2)^{4-\ell} t^4}, &t\in [\frac{\sqrt{2}}{\pi}, \frac{1}{2}),
\end{cases}
\end{align*}
where $H_0^\infty \eqdef 1$, $H_1^\infty \eqdef 4$, $H_2^\infty \eqdef 18$ and $H_3^\infty \eqdef 77$, and $H_\ell(t) \eqdef \alpha^4(t) \beta_\ell(t)$, with
$$
\alpha(t) \eqdef \frac{2}{\pi(1-\frac{\pi^2 t^2}{6})}, \quad \bar \beta(t) \eqdef  \frac{\alpha(t)}{\fq t} = \frac{2}{\fq t \pi(1-\pi^2t^2/6)}
$$
and
$\beta_0(t) \eqdef 1$, $\beta_1(t) \eqdef 2+2\bar \beta(t)$, $\beta_2 \eqdef 4+7\bar \beta(t) + 6\bar \beta(t)^2$ and $\beta_3(t) \eqdef 8+24\bar \beta + 30\bar \beta(t)^2 + 15 \bar \beta(t)^3$.
Let us first remark that $\bar \beta$ is decreasing on $I \eqdef [\frac{1}{2\fq}, \frac{\sqrt{2}}{\pi}]$, so $\abs{\bar \beta(t)} \leq \abs{\bar \beta(1/(2\fq))} \approx 1.2733$, and $a(t) \leq a(\sqrt{2}/\pi) = \frac{3}{\pi}$ on $I$. Therefore, on $I$, $H_0(t) \leq \frac{3}{\pi}$, $H_1(t) \leq 3.79$, $H_2(t) \leq 18.83$ and $H_3(t) \leq 98.26$, and we can conclude that on $[\frac{1}{2\fq}, \frac{1}{2})$, we have
$$
\abs{\kappa^{(\ell)}(t)} \leq \frac{\pi^\ell \hm_\ell}{(\fq+2)^{4-\ell} t^4}
$$
where $\hm_0 = 1$, $\hm_1 \eqdef 4$, $\hm_2 \eqdef 19$, $\hm_3 \eqdef 99$. 
Combining with \eqref{eq:upp_bd_Jackson}, we have
\begin{equation}\label{eq:uniform_upp_bd_Jackson}
\norm{\kappa^{(\ell)}}_\infty \leq \km_\ell
\end{equation}
 where $\km_0 \eqdef 1$, $\km_2 \eqdef \cf$,
\begin{align*}
\km_1 \eqdef \sqrt{\cf} \max\pa{\frac{2\pi^4 }{(\frac12+\frac{1}{\fq})^3} \frac{f}{\sqrt{\cf}}, \frac{\sqrt{\cf}}{2f} } = \Oo(\sqrt{\cf}) \\
 \km_3 \eqdef (\cf)^{3/2} \max\pa{\frac{99\pi^3}{(\frac12+\frac{1}{\fq})} \pa{\frac{2\fq}{\sqrt{\cf}}}^4, \frac{6\sqrt{\cf}}{\fq}  } = \Oo((\cf)^{3/2}).
\end{align*}

Finally, given $p\in (0,1)$,
\begin{align*}
(\fq+2)^4 t^4 \geq (1+ p(\fq+2)^2 t^2)^2, \qquad \forall \; t \geq \frac{1}{\sqrt{(1-p)}(\fq+2) }.
\end{align*}
Choosing $p=\frac{1}{2}$ and using $(\fq+2)^2 =( \frac{3}{\pi^2}\cf + 4) \geq \frac{3}{\pi^2} \cf$, we have
\begin{equation}\label{eq:kappa_decay}
\abs{\kappa^{(\ell)}(t)} \leq  \frac{\km_\ell}{(1+ \frac{3}{2\pi^2} \cf t^2)^2}, \qquad \forall\; t^2\geq \frac{2\pi^2}{3 \cf },
\end{equation}

In the following sections, we will repeatedly make use of \eqref{eq:upp_bd_Jackson}, \eqref{eq:uniform_upp_bd_Jackson} and \eqref{eq:kappa_decay}.

\subsection{Notation}

For notational convenience, write $t_i\eqdef x_i-x_i'$, $\kappa_i \eqdef \kappa(t_i)$, $\kappa_i'\ \eqdef \kappa'(t_i)$, and so on. Let
\[
\fullCov_i \eqdef \prod_{\substack{k=1\\k\neq i}}^d \kappa_k,\quad \fullCov_{ij} \eqdef \prod_{\substack{k=1 \\ k\neq i,j}}^d \kappa_k \qandq \quad \fullCov_{ij\ell} \eqdef \prod_{\substack{ k=1 \\ k\neq i,j,\ell}}^d \kappa_k.
\]
 With this, we have:
\begin{align*}
\partial_{1,i} \fullCov(\sig,\sigg) =&~ \kappa'_i \fullCov_i \\
\partial_{1,i}\partial_{2,i} \fullCov(\sig,\sigg) =&~ -\kappa''_i \fullCov_i, \qandq\forall i\neq j, \; \partial_{1,i} \partial_{2,j} \fullCov(\sig,\sigg) = -\kappa_i' \kappa_j' \fullCov_{ij}.
\end{align*}
Where convenient, we sometimes write $\fullCov(t)  = \fullCov(x-x') \eqdef \fullCov(x,x')$.
Given a symmetric matrix $M$, we write $\lambda_{\min}(M)$ to denote the smallest eigenvalue of $M$.

\subsection{Bounds when $\norm{t}$ is small}\label{sec:Jackson_neigh}

\begin{lem}\label{lem:Jackson_hessian}
Suppose that $\cf \norm{x-x'}_2^2 \leq c$ with $c>0$ such that
$$
\epsilon \eqdef   \pa{1-  6 c}\pa{1- \frac{c}{2-{c}}} - c > 0
 $$
 Then, 
$-\dotp{\fullCov^{(02)}(x-x') q}{q}\geq  \epsilon \norm{q}_{x}$.
\end{lem}
\begin{proof}


Let  $q\in \RR^d$,
and note that
\begin{equation}\label{eq:Jackson2nd}
\begin{split}
- \dotp{\nabla_2^2 K q}{q}
&= - \sum_i \pa{ q_i \kappa_i'' \fullCov_i - \kappa_i' \sum_{j\neq i} q_j \kappa_j' \fullCov_{ij}} q_i\\
&=  - \pa{\sum_i q_i^2 \kappa_i'' \fullCov_i - \sum_i q_i \kappa_i \sum_{j\neq i} q_j \kappa_j \fullCov_{ij}}
\\
&\geq  \norm{q}_x^2  \frac{1}{\cf}  \pa{ - \max_i \ens{\kappa_i'' \fullCov_i} - \sum_j \abs{\kappa_j'}^2 }.
\end{split}
\end{equation}
We first consider $ \kappa_i'' \fullCov_i$: By applying \eqref{eq:upp_bd_Jackson}, we obtain
\begin{align*}
 \kappa_i'' &\leq -\cf + 6\cf^2 t_i^2,
\\
\fullCov_i &\geq \prod_{j\neq i} \pa{1-\frac{\cf}{2}t_i^2} \geq 1- \frac{\cf}{2} \norm{t}_2^2 - \pa{\frac{\cf}{2} \norm{t}_2^2}^3 - \pa{\frac{\cf}{2} \norm{t}_2^2}^5-\cdots\\
&\qquad \geq 1- \frac{\cf\norm{t}_2^2}{2(1-\frac{\cf}{2}\norm{t}_2^2)}.
\end{align*}
and hence, 
\[
  \kappa_i''\fullCov_i  \leq \pa{-\cf +  6\cf^2 \norm{t}_2^2}\pa{1- \frac{\cf\norm{t}_2^2}{2(1-\frac{\cf}{2}\norm{t}_2^2)}}
 \]
For the second term in \eqref{eq:Jackson2nd}, again, by applying \eqref{eq:upp_bd_Jackson}, we obtain
\[
\sum_j \abs{\kappa_j'}^2 \leq \cf^2\norm{t}_2^2 .
\]
Therefore, for $\norm{q}_x =1$, we have
$$
-\dotp{\fullCov^{(02)}(x-x') q}{q} \geq  \pa{1-  6\cf \norm{t}_2^2}\pa{1- \frac{\cf\norm{t}_2^2}{2(1-\frac{\cf}{2}\norm{t}_2^2)}} - \cf \norm{t}_2^2
$$

\end{proof}

\begin{lem}\label{lem:near_bound_Jackson}
Assume that $\frac{1}{8 \sqrt{\cf}} \geq   \norm{t}_2 $
Then,
\begin{align*}
K(t) \leq   1 -  \frac{\cf}{4} \norm{t}_2^2 +  16\cf^2\norm{t}_2^4.
\end{align*}
Consequently, for all
\begin{equation*}
0< c  \leq  \frac{1}{8\sqrt{2\cf} },
\end{equation*} and all $t$ such that $\norm{t}_2 \geq c$, 
\begin{align*}
\abs{K(t)} \leq   1 - \frac{\cf}{8} c^2.
\end{align*}
\end{lem}
\begin{proof}

First note that by \eqref{eq:upp_bd_Jackson},
$$
\abs{\kappa(u)} \leq  1-  \frac{\cf}{2} u^2 + 32  \cf^2 u^4 = 1- u^2 g(u)
$$
where
$$
g(u) \eqdef  \cf\pa{ \frac{1}{2} - 32  \cf u^2 },
$$
and note that $g(u) \in (0, \tfrac{\cf}{2})$ for $u\in (0,1/(8\sqrt{\cf})$.
So, writing $t = (t_i)_{i=1}^d$ and $g_j \eqdef g(t_j)$, we have
\begin{align*}
&K(t) = \prod_{j=1}^d \kappa(t_i ) \leq \prod_{j=1}^d \pa{1- t_j^2\cdot  g(t_j)} 
\\
&= 1 -  \sum_{j=1}^d t_j^2 g_j + \sum_{j\neq k} t_j^2  t_k^2  g_j g_k - \sum_{j\neq k \neq \ell } t_j^2 t_k^2 t_\ell^2 g_j g_k  g_\ell + \cdots\\
&= 1+ \sum_{\ell=1}^d \sum_{j\in \Jj_\ell} (-1)^\ell \prod_{i=1}^\ell (t_{j_i}^2  g_{j_i}),
\end{align*}
where $\Jj_\ell \eqdef \enscond{j\in \NN^d}{ j\leq d, \text{ all entries of } j \text{ are distinct}}$.
Note that for odd integers $\ell$,
\begin{align*}
- \sum_{j\in \Jj_\ell} \prod_{i=1}^\ell (t_{j_i}^2  g_{j_i}^2)  + \sum_{j\in \Jj_{\ell+1}} \prod_{i=1}^{\ell+1} (t_{j_i}^2  g_{j_i}) 
&\leq  - \sum_{j\in \Jj_\ell} \prod_{i=1}^\ell (t_{j_i}^2  g_{j_i})  + \pa{\sum_{j\in \Jj_{\ell}} \prod_{i=1}^{\ell} (t_{j_i}^2  g_{j_i}^2)} \pa{\sum_{k=1}^d t_k^2 g_k}\\
&\leq - \pa{\sum_{j\in \Jj_{\ell}} \prod_{i=1}^{\ell} (t_{j_i}^2  g_{j_i}^2)}\pa{ 1-  \frac{\cf}{2}\norm{t}_2^2} <0
\end{align*}
since $\pa{ 1-  \frac{\cf}{2}\norm{t}_2^2} >0$.
Also,
$$
 \sum_{j=1}^d t_j^2 g_j 
\leq  \frac{\cf}{2} \sum_{j=1}^d t_j^2 <1,
$$
by assumption.
So,
\begin{align*}
&K(t) \leq  1 -  \sum_{j=1}^d t_j^2 g_j + \sum_{j\neq k} t_j^2  t_k^2  g_j g_k \\
&\leq  1 -  \sum_{j=1}^d t_j^2 g_j + \frac{1}{2} \pa{\sum_{j} t_j^2    g_j}^2
\leq   1 - \frac{1}{2}  \sum_{j=1}^d t_j^2 g_j\\
&\leq 1 - \frac{\cf}{2}   \pa{ \frac{1}{2} \sum_{j=1}^d t_j^2 - 32  \cf  \sum_{j=1}^dt_j^4 }
\leq  1 -  \frac{\cf}{4} \norm{t}_2^2 +  16\cf^2\norm{t}_2^4.
\end{align*}
Finally, observe that the function
$$
q(z) \eqdef  \frac{\cf}{4} z^2 - 16 \cf^2  z^4
$$
is positive and increasing on the interval $[0, \frac{1}{8\sqrt{2\cf} }]$. So, for $t$ satisfing \begin{equation}
c \leq \norm{t}_2 \leq  \frac{1}{8\sqrt{2\cf} } ,
\end{equation} we have
$
\abs{K(t)} \leq 1- q(c) \leq 1- \frac{\cf}{8} c^2 .
$
Finally, since $\abs{K(t)}$ is decreasing as $t$ increases, we trivially have that $
\abs{K(t)} \leq 1- q(c )
$ for all $t$ with $\norm{t}_2\geq c$.

\end{proof}

\subsection{Bounds when $\norm{t}$ is large}\label{sec:Jackson_sep}

\begin{lem}\label{lem:far_bounds_Jackson}
Let $i,j\in \{0,1,2\}$ with $i+j\leq 3$.
Let $\bar A \geq \sqrt{\tfrac{4\pi^2}{3 }}$ and   $\norm{t}_2 \geq \bar A \sqrt{d} s_{\max}^{1/4}/\sqrt{\cf}$. Then, we have $\norm{\fullCov^{(ij)}(t) }_{x,x'} \leq d^{\frac{i+j-4}{2}} (\bar A^4  s_{\max})^{-1}$.
\end{lem}
\begin{proof}
Write $t = (t_j)_{j=1}^d$.
To bound $K(t) = \prod_{j=1}^d \kappa(t_j)$, we want to make use of the bounds on $\km_j$ from \eqref{eq:kappa_decay}. We can do this for each $t_j$ such that $\abs{t_j} \geq \sqrt{\tfrac{2\pi^2}{3 \cf }}$. Note that there exists at least one such $t_j$ since $\norm{t}_\infty \geq \norm{t}_2/\sqrt{d}  \geq \bar A s_{\max}^{1/4}/\sqrt{\cf} \geq \sqrt{\tfrac{2\pi^2}{3 \cf }}$.
 If $\{\abs{t_j}\}_{j=1}^k \subset [0,  \sqrt{\tfrac{2\pi^2}{3 \cf }})$ for $k\leq d-1$, then
 $$
k {\frac{2\pi^2}{3 \cf }} + \sum_{j=k+1}^d t_j^2  \geq  \norm{t}_2^2   \geq \frac{\bar A^2 d s_{\max}^{1/2}}{\cf},
$$ 
which implies that $\sum_{j=k+1}^d t_j^2  \geq \frac{1}{\cf}\pa{ \bar A^2 d s_{\max}^{1/2} - \frac{2\pi^2(d-1)}{3} } \geq \frac{\bar A^2 d s_{\max}^{1/2}}{2\cf} $,
by our assumptions on $\bar A$. Therefore, we may assume that we have some $d\geq p\geq 1$ such that $\{b_j\}_{j=1}^p \subseteq \{t_j\}$ with $\abs{b_j} \geq \sqrt{\tfrac{2\pi^2}{3 \cf }}$ and $\norm{b}_2 \geq \frac{\bar A \sqrt{d} \sqrt[4]{ s_{\max}}}{\sqrt{2\cf} }$. 
Observe that
$$
\prod_{j=1}^p (1+ \frac{3\cf}{2\pi^2} b_j^2) \geq 1+ \frac{3\cf}{2\pi^2} \sum_{j=1}^p b_j^2 = 1+ \frac{3\cf}{2\pi^2} \norm{b}^2_2 \geq 1+  \frac{3}{4\pi^2} {\bar A^2 d \sqrt{ s_{\max}}}.
$$
So, by applying the fact that $\abs{\kappa} \leq 1$, $\km_0 = 1$ and \eqref{eq:kappa_decay}, we have
\begin{align*}
\abs{K(t)} &\leq \prod_{j=1}^p \abs{\kappa(b_j)} \leq  \prod_{j=1}^p \frac{1}{ \pa{1 + \frac{3\cf}{2\pi^2} b_j^2}^2} \leq \frac{1}{\pa{1+  \frac{3}{4\pi^2} {\bar A^2 d \sqrt{ s_{\max}}} }^2}.
\end{align*}

For $\abs{\kappa_i' \fullCov_i}$, if $i\not\in \enscond{j}{ \abs{t_j}> \sqrt{\tfrac{2\pi^2}{3 \cf }} }$, then
$$
\abs{\kappa_i' \fullCov_i} \leq  \norm{\kappa_i'}_\infty \prod_{j=1}^p \abs{\kappa(b_j)} \leq \frac{ \norm{\kappa_i'}_\infty}{\pa{1+  \frac{3}{4\pi^2} \bar A^2 d \sqrt{s_{\max}}}^2},
$$
and otherwise, we have 
$
\abs{\kappa_i' \fullCov_i} \leq  \abs{\kappa'(t_i)} \prod_{j\neq i} \abs{\kappa(b_j)} \leq \frac{\km_1}{\pa{1+  \frac{3}{4\pi^2} \bar A^2 d \sqrt{s_{\max}}}^2},
$
In a similar manner, writing $V\eqdef \pa{1+  \frac{3}{4\pi^2} \bar A^2 d \sqrt{s_{\max}}}^{-2}$, we can deduce that
\begin{align*}
&\abs{\kappa_i' \fullCov_i}  \leq \kappa_1^{\max} V,\qquad \abs{\kappa_i'' \fullCov_i} \leq \km_2 V, \qquad\abs{\kappa_i' \kappa_j' \fullCov_{ij}} ^2 \leq (\km_1)^2 V\\
&\abs{\kappa_i''' \fullCov_i} ^3 \leq \km_3 V, \qquad\abs{\kappa_i'' \kappa_j' \fullCov_{ij}} ^3 \leq \km_2 \kappa^{\max}_1 V, \qquad \abs{\kappa_i'\kappa_j'\kappa_\ell' \fullCov_{ij\ell}}  \leq (\kappa^{\max}_1)^3 V.
\end{align*}

Therefore, 
$$
\norm{\fullCov^{(10)}}_{x,x'} = \frac{1}{\sqrt{\cf}}
\norm{\nabla_1 K} \leq  \frac{1}{\sqrt{\cf}} \sqrt{\sum_{j=1}^d \abs{\kappa_j' \fullCov_j}^2} \leq \frac{\km_1}{\sqrt{\cf}}  V \sqrt{d} \lesssim \frac{1}{\bar A^4 d^{3/2} s_{\max}}.
$$

Using Gershgorin theorem, we have
\begin{align*}
\norm{\nabla^2_2 \fullCov(\sig,\sigg)} \leq&~ \max_{1\leq i\leq d} \ens{\abs{\kappa''_i \fullCov_i} + \abs{\kappa'_i}\sum_{j\neq i} \abs{\kappa'_j}\abs{\fullCov_{ij}}}
\end{align*}
and hence,
\begin{align*}
\norm{\fullCov^{(02)}}_{x'} &=  \frac{1}{\cf}\norm{\nabla_2^2 K} \leq \frac{1}{\cf} \max_{i=1}^d \ens{ \abs{\kappa_i'' \fullCov_i} + \abs{\kappa_i'} \sum_{j\neq i} \abs{\kappa_j' \fullCov_{ij}}} \\
&\leq \frac{1}{\cf} V\pa{\km_2  + (\km_1)^2 (d-1)  } \leq  \frac{\max\{\km_2, (\km_1)^2\} }{\cf} V d \lesssim \frac{1}{\bar A^4 d s_{\max}}.
\end{align*}
Note also that $\norm{\fullCov^{(11)}}_{x,x'} = \norm{\fullCov^{(02)}}_{x'}$.
Finally, since
\begin{align*}
\norm{\partial_{1,i} \nabla^2_2 \fullCov(\sig,\sigg)} \leq&~ \max\Bigg\lbrace \abs{\kappa'''_i \fullCov_i} + \abs{\kappa''_i}\sum_{j\neq i} \abs{\kappa'_j}\abs{\fullCov_{ij}},\\
&\qquad \max_{j\neq i}\ens{\abs{\kappa''_j\kappa'_i \fullCov_{ij}} + \abs{\kappa'_j\kappa''_i\fullCov_{ij}} + \abs{\kappa'_i}\abs{\kappa'_j} \sum_{l\neq i,j} \abs{\kappa'_l}\abs{\fullCov_{ij\ell}}}\Bigg\rbrace,
\end{align*}
we have
\begin{align*}
\norm{\fullCov^{(12)}}_{x,x'} &= \frac{1}{\cf^{3/2}}\norm{\nabla_1 \nabla_2^2 K} \\
&\leq \frac{1}{\cf^{3/2}} \sqrt{d} V \max\pa{ \km_3  + \km_2 \km_1 (d-1), 2\km_2 \km_1  + (d-1) (\km_1)^3  }
\\
&\leq d^{3/2} \max\{\km_3, \km_1\km_2, (\km_1)^3\} \frac{1}{\cf^{3/2}} V \lesssim \frac{1}{\bar A^4 d^{1/2} s_{\max}}
\end{align*}
\end{proof}

\subsection{Uniform bounds}

\begin{lem}\label{lem:Jackson_upbd}
If $\rnear \sim 1/\sqrt{\cf}$, then $B_0=\Oo(1)$, $B_{01} = \Oo(\sqrt{d})$, $B_{02} = B_{12} = \Oo(1)$ and $B_{22} = \Oo(d)$.
\end{lem}
\begin{proof}
We have $\abs{K} \leq 1$, and 
$$
\norm{\nabla \fullCov}^2 \leq \sum_i \abs{\kappa_i}^2\abs{K_i}^2 \leq d (\km_1)^2 \lesssim {\cf} d,
$$
so $B_{01} = \Oo(\sqrt{d})$.

From \eqref{eq:Jackson2nd},  for all $\norm{q}=1$, 
$$
\dotp{\nabla_2^2 K(t) q}{q} \leq \max_i \abs{\kappa_i''} \norm{q}_2^2 + \norm{q}_2^2 \sum_i \abs{\kappa_i}^2
\leq \cf + \cf^2 \norm{t}^2 = \Oo(\cf),
$$
for $\norm{t}\lesssim 1/\sqrt{\cf}$.
So, since $\rnear\leq 2/\sqrt{\cf}$, $\norm{\fullCov^{02}(t)}\leq 2\eqdef B_{02}$. 
For the bound on $B_{12}$:
\begin{align*}
\norm{\fullCov^{(12)}}_{x,x'} &= \sup_{\norm{q}=\norm{p}=1}\frac{1}{\cf^{3/2}} \Bigg( \sum_{k} \sum_{k\neq i} \partial_{1,i} \pa{\partial_{2,k}^2 K p_i q_k^2 
+ \partial_{1,i} \partial_{2,i}\partial_{2,k} K p_i q_i q_k}\\ 
&+ \sum_i \sum_k \sum_j \partial_{1,i} \partial_{2,j}\partial_{2,k} p_i p_j p_k + \sum_i \sum_{j\neq i} \partial_{1,i}\partial_{2,i} \partial_{2,j} K p_i q_i q_j + \sum_i \partial_{1,i}\partial_{2,j}^2 K p_i q_i^2 \Bigg)\\
=&\sup_{\norm{q}=\norm{p}=1}\frac{1}{\cf^{3/2}} \Bigg( \sum_{k} \sum_{k\neq i} \kappa_i' \kappa_k'' K_{ik} p_i q_k^2 
+ \kappa_i'' \kappa_k'  K_{ik} p_i q_i q_k\\ 
&+ \sum_i \sum_k \sum_j \kappa_i'\kappa_k'\kappa_j'  K_{ijk} p_i p_j p_k + \sum_i \sum_{j\neq i} \kappa_i'' \kappa_j'  K_{ij} p_i q_i q_j + \sum_i \kappa_i' \kappa_j''  K_{ij} p_i q_i^2 \Bigg)\\
&\leq \frac{1}{\cf^{3/2}}\Bigg( 3 \norm{\kappa''}_\infty \sqrt{ \sum_i \abs{\kappa_k'}^2} + \pa{{ \sum_i \abs{\kappa_k'}^2}}^{3/2} + \norm{\kappa'}_\infty \norm{\kappa''}_\infty\Bigg)\\
&\leq \frac{1}{\cf^{3/2}}\pa{ 3\cf^2 \norm{t} + \cf^3 \norm{t}^3 + \Oo( \cf^{3/2}) } = \Oo(1)
\end{align*}
for $\norm{t} \leq 1/\cf^{1/2}$.

\end{proof}

\subsection{Gradient bounds}\label{sec:jackson-grads}
The derivatives of the  random features  are uniformly bounded with
\begin{equation}\label{eq:fourier_grad_bd}
\norm{\nabla^j \phi_\om(x)}= \norm{\om}^j \leq \fq^j d^{j/2} \sim \cf^{j/2} d^{j/2}
\end{equation}
  So, we can set  $\Lu_i = \Oo(d^{i/2})$ for $i=0,1,2$.
For $\Lu_3$,  the condition \eqref{eq:hessian_lipschitz_matrix_formulation}  is simply
$$
\cf^{-1} \norm{\nabla^2 \phi_\om(x) - \nabla^2 \phi_\om(x')} \leq \Lu_3 \cf^{\frac12} \norm{x-x'},
$$
so $\Lu_3 = \Oo(d^{3/2})$ by \eqref{eq:fourier_grad_bd}.

\section{Application: Continuous Fourier sampling with the Gaussian kernel}\label{app-gaussian}

\newcommand{\normmah}[2]{\norm{#2}_{#1}}
\def\Cdcov{C_{\dsep}}
\def\Cmetrictensor{C_H}
\newcommand{\order}[1]{\Oo\left(#1\right)}


In this section, we consider the case of continuous Fourier sampling with Gaussian frequencies, which may appear for instance in sketched Gaussian mixture learning \cite{gribonval2017compressive}. Let $\Xx \subset \RR^d$ be any bounded subset of $\RR^d$.
Let $\Omega = \RR^d$, $\phi_\om(x) \eqdef  e^{\mathrm{i} \om^\top x}$, and  $\Lambda(\om) =\mathcal{N}(0,\Sigma^{-1})$, for a known covariance matrix $\Sigma$.

\paragraph{The kernel and Fisher metric} The associated kernel is the Gaussian kernel
\[
\fullCov(\sig,\sigg) = \exp\pa{-\frac12 \normmah{\Sigma^{-1}}{\sig-\sigg}^2}
\]
with constant metric tensor
\[
\met_x = \Sigma^{-1} \qandq \dsep(x,x') = \normmah{\Sigma^{-1}}{x-x'} = \norm{\Sigma^{-\frac12}(x-x')}
\]


\paragraph{Sampling bounds} The rest of this section consists of Lemmas which bound the parameters in Assumptions \ref{ass:kernel} and \ref{ass:feat}. We show that by choosing $\rnear = \frac{1}{\sqrt{2}}$, we obtain $\constker_2=\frac12 e^{-\frac{1}{4}}$ and $\constker_0 \eqdef 1-e^{-\frac{1}{4}}$. Moreover, Lemma \ref{lem:gaussian_upbd} gives uniform bounds in
$
B_{ij} = \Oo(1)
$ 
and, for $h = \Oo(1)$, 
$
W(h,s) = \Oo(\sqrt{\log s} + 1).
$
Gradient bounds are computed in Section \ref{sec:gaussian_grads}.

\subsection{Properties of the kernel}

\paragraph{Notations.} For simplicity define $t = \sig-\sigg$, b an abuse of notations $\fullCov_\Sigma(t) = \exp\pa{-\frac12 \normmah{\Sigma^{-1}}{t}^2}$ and for $u\in \RR$, $\kappa(u) = \exp\pa{-\frac12 u^2}$. Denote by $\ens{e_i}$ the canonical basis of $\RR^d$, and by $f_i = \Sigma^{-1} e_i$ the $i^{th}$ row of $\Sigma^{-1}$.

\paragraph{Gradients of the kernel.} We have the following:
\begin{align*}
\nabla \fullCov_\Sigma(t)=&~-\Sigma^{-1} t \fullCov_ \Sigma(t) \\
\nabla^2\fullCov_\Sigma(t)=&~\pa{-\Sigma^{-1} + \Sigma^{-1}tt^\top\Sigma^{-1}}\fullCov_\Sigma(t) \\
\partial_{i}\nabla^2\fullCov_\Sigma(t)=&~\pa{\Sigma^{-1} t f_i^\top + f_i t^\top \Sigma^{-1}}\fullCov_\Sigma(t) - (t^\top f_i)\nabla^2 \fullCov_\Sigma(t) \\
\partial_{ij}\nabla^2\fullCov_\Sigma(t)=&~\pa{-\Sigma^{-1} ((t^\top f_j)t f_i^\top + (t^\top f_i)t f_j^\top) + (f_i f_j^\top + f_j f_i^\top)}\fullCov_\Sigma(t) - f_{ij}\nabla^2 \fullCov_\Sigma(t) - (t^\top f_i) \partial_j \nabla^2 \fullCov_\Sigma(t)
\end{align*}

Then we observe that for any $q\geq 1$ the function $f_q(r) = r^q e^{-\frac12 {r^2}}$ defined on $\RR_+$ is increasing on $[0,\sqrt{q}]$ and decreasing after, and its maximum value is $f_q(\sqrt{q}) = \pa{\frac{q}{e}}^{q/2}$. Furthermore, it is easy to see that we have $f_q(r) = r^q e^{-r^2/2} \leq \pa{\frac{2q}{2}}^{\frac{q}{2}} e^{-r^2/4}$ and therefore $f(r) \leq \varepsilon$ if $r \geq 2\pa{\log\pa{\frac{1}{\varepsilon}} + \frac{q}{2}\log\pa{\frac{2q}{e}}}$.

\subsection{Bounds when $\norm{t}$ is small}
\begin{lem}\label{lem:gaussian_constker2}
For all $\dsep_\met(x,x') \leq \rnear \eqdef \frac{1}{\sqrt{2}}$ and all $v \in T_x \Mm$, we have $-\fullCov^{(02)}(x,x')[v,v] \geq \constker_2 \norm{v}_x^2$ where $\constker_2=\frac12 e^{-\frac{1}{4}}$.
\end{lem}
\begin{proof}
From the derivations above we have $\fullCov^{(02)}(x,x')[v,v] = v^\top \nabla_2^2 \fullCov_\Sigma(t) v = (-1+\dsep_\met(x,x')^2)\kappa(\dsep_\met(x,x')) \norm{v}^2_x \leq (\rnear^2 -1)\kappa(\rnear) \norm{v}_x$.
\end{proof}

\subsection{Bounds when $\norm{t}$ is large}
\begin{lem}\label{lem:gaussian_constker0}
For all $\dsep_\met(x,x') \geq \rnear$ we have $\abs{\fullCov(x,x') } \leq 1-\constker_0$, where $\constker_0 \eqdef 1-e^{-\frac{1}{4}}$, and for $h= \Oo(1)$, $W(h,s) = \Oo(\sqrt{\log s} + 1)$.
\end{lem}
\begin{proof}
For the first inequality we have $\abs{\fullCov} \leq \kappa(\rnear) = 1- (1-e^{-\frac{1}{4}})$.

Then, from \eqref{eq:matnorm_riemann2euclid}, the fact that the metric tensor is constant, and the expressions for the derivatives of the kernel above, it is immediate that
\begin{align*}
\norm{\fullCov^{(10)}(x,x')}_{x,x'} &= \norm{\fullCov^{(01)}(x,x')}_{x,x'} = \norm{\Sigma^\frac12 \nabla \fullCov_\Sigma(t)}_1 \\
&= \dsep(x,x') \kappa(\dsep(x,x')) = f_1(\dsep(x,x')) \\
\norm{\fullCov^{(02)}(x,x')}_{x,x'} &= \norm{\fullCov^{(11)}(x,x')}_{x,x'} = \norm{\Sigma^\frac12 \nabla^2 \fullCov_\Sigma(t) \Sigma^\frac12}_2 \\
&= (\dsep(x,x')^2+1)\kappa(\dsep(x,x')) = f_2(\dsep(x,x')) + f_0(\dsep(x,x'))
\end{align*}

For $\fullCov^{(12)}$, again since the metric tensor $\met$ is constant, we observe that
\begin{align*}
[q]\fullCov^{(12)}(x,x')[v_1,v_2] &= v_1^\top\left(\sum_{i=1}^d q_i \pa{\partial_{i}\nabla^2\fullCov_\Sigma(t)} \right)v_2
\end{align*}
and
\begin{align*}
\norm{\fullCov^{(12)}(x,x')} &= \sup_{\norm{\Sigma^{-\frac12}q}_2 \leq 1, \norm{\Sigma^{-\frac12}v_i}_2 \leq 1} \abs{v_1^\top\left(\sum_{i=1}^d q_i \pa{\partial_{i}\nabla^2\fullCov_\Sigma(t)} \right)v_2} \\
&= \sup_{\norm{q}_2 \leq 1, \norm{v_i}_2 \leq 1} \abs{v_1^\top\left(\sum_{i=1}^d (\Sigma^\frac12 q)_i \Sigma^\frac12 \pa{\partial_{i}\nabla^2\fullCov_\Sigma(t)}\Sigma^\frac12 \right)v_2} \\
&= \sup_{\norm{q} = 1} \norm{\Sigma^{\frac12}\pa{\sum_{i=1}^d (\Sigma^\frac12 q)_i \partial_{i}\nabla^2\fullCov_\Sigma(t)}\Sigma^\frac12}_2 \, .
\end{align*}

Using, $\sum_i (\Sigma^{\frac12}q)_i f_i = \Sigma^{-\frac12} q$, we observe that
\begin{align*}
&\Sigma^\frac12 \pa{\sum_i (\Sigma^\frac12 q)_i \Sigma^{-1} t f_i^\top}\Sigma^\frac12 = \Sigma^{-\frac12} t \pa{\sum_i q^\top \Sigma^\frac12 e_i e_i^\top\Sigma^{-\frac12}} = \Sigma^{-\frac12} t q^\top \\
& \Sigma^\frac12 \sum_i (\Sigma^\frac12 q)_i (t^\top f_i) \nabla^2 \fullCov_\Sigma(t)\Sigma^\frac12 = (q^\top \Sigma^{-\frac12} t) \left(\Sigma^\frac12\nabla_2^2 \fullCov(x,x')\Sigma^\frac12\right)
\end{align*}
Hence at the end of the day
\[
\norm{\fullCov^{(12)}(x,x')} \leq (3\dsep(x,x') + \dsep(x,x')^3)\kappa(\dsep(x,x')) = 3f_1(\dsep(x,x')) + f_3(\dsep(x,x'))
\]
Therefore, for $h=\Oo(1)$, using the properties of the functions $f_q$ it is immediate that $W(h,s) = \Oo(\sqrt{\log s} + 1)$.
\end{proof}

\subsection{Uniform bounds}

\begin{lem}\label{lem:gaussian_upbd}
For $(i,j) \in \{0,1,2\}$, we have $B_{ij} = \Oo(1)$.
\end{lem}
\begin{proof}
The bounds for $i+j\leq 3$ are immediate using the identities in the proof of Lemma \ref{lem:gaussian_constker0} and the properties of the functions $f_q$.

By the same reasoning we have
\[
\norm{\fullCov^{(22)}(x,x')} = \sup_{\norm{q_1}=1, \norm{q_2}=1} \norm{\Sigma^\frac12 \sum_{ij} (\Sigma^\frac12 q_1)_i (\Sigma^\frac12 q_2)_j \partial_{ij} \nabla^2\fullCov_\Sigma(t)\Sigma^\frac12}
\]
and we have
\begin{align*}
& \Sigma^\frac12 \pa{\sum_{ij} (\Sigma^\frac12 q_1)_i (\Sigma^\frac12 q_2)_j \Sigma^{-1} (t^\top f_j) (t f_i^\top) }\Sigma^{\frac12} = (q_2^\top \Sigma^{-\frac12} t) \Sigma^{-\frac12} t q_1^\top \\
& \Sigma^\frac12 \pa{\sum_{ij} (\Sigma^\frac12 q_1)_i (\Sigma^\frac12 q_2)_j f_i f_j^\top }\Sigma^{\frac12} = q_1 q_2^\top\\
& \Sigma^\frac12 \pa{\sum_{ij} (\Sigma^\frac12 q_1)_i (\Sigma^\frac12 q_2)_j f_{ij} }\Sigma^{\frac12} = q_1 q_2^\top\\
& v_1\Sigma^\frac12 \pa{\sum_{ij} (\Sigma^\frac12 q_1)_i (\Sigma^\frac12 q_2)_j (t^\top f_i)\partial_j \nabla^2 \fullCov_\Sigma(t)}\Sigma^{\frac12} v_2 = (q_1^\top \Sigma^{-\frac12} t) [q_2] K^{(12)}(x,x') [v_1,v_2]
\end{align*}
Hence
\[
\norm{\fullCov^{(22)}(x,x')} \leq [3f_0 + 6f_2 + f_4](\dsep(x,x'))
\]
and $B_{22} = \Oo(1)$.
\end{proof}

\subsection{Gradient bounds}\label{sec:gaussian_grads}

For $j = \{0,1,2\}$, we have $\diff{j}{\phi_\om}(x)[q_1,\ldots,q_j] = \pa{\prod_i \om^\top q_i}\phi_\om(x)$ and therefore
\begin{align*}
\norm{\diff{j}{\phi_\om}(x)}_x \leq \norm{\Sigma^\frac12 \om}^j_2
\end{align*}
And then, from \eqref{eq:hessian_lipschitz_matrix_formulation}, using $\tau_{x\to x'} = \Id$,
\begin{align*}
\norm{ \diff{2}{\phi_\om}(x) - \diff{2}{\phi_\om}(x')[\tau_{x\to x'} \cdot, \tau_{x\to x'} \cdot] }_x  &= \norm{\Sigma^{\frac12} \pa{\nabla_2^2\phi(x') -   \nabla_2^2\phi(x)}  \Sigma^\frac12}_2 \\
&= \norm{\Sigma^\frac12 \om}_2^2 \abs{\phi_\om(x) - \phi_\om(x')} \\
&= \norm{\Sigma^\frac12 \om}_2^2 \abs{\om^\top(x-x')} \leq \norm{\Sigma^\frac12 \om}^3_2 \dsep_\met(x,x')
\end{align*}

Since $\om \sim \Nn(0,\Sigma^{-1})$, $\norm{\Sigma^\frac12 \om}^j = W^{\frac{j}{2}}$ where $W$ is a $\chi^2$ variable with $d$ degrees of freedom. 
Then, we use the following Chernoff bound \citep{Dasgupta2003}: for $x\geq d$, we have
\[
\PP(W\geq x) \leq \pa{\frac{ex}{d}e^{-\frac{x}{d}}}^{\frac{d}{2}} \leq \pa{e \pa{\sqrt{\frac{x}{d}}}^2 e^{-\frac12 \cdot \pa{\sqrt{\frac{x}{d}}}^2} e^{-\frac{x}{2d}}}^{\frac{d}{2}}\leq 2^{\frac{d}{2}}e^{-\frac{x}{4}}
\]
by using $x^2e^{-\frac{x^2}{2}} \leq \frac{2}{e}$.

Hence we can define the $F_j$ such that, for all $t \geq d^{j/2}$, $\PP(L_j(\om) \geq t) \leq F_j(t) = 2^{\frac{d}{2}} \exp\pa{-\frac{t^\frac{2}{j}}{4}}$, and $F_j(\bar L_j)$ is smaller than some $\delta$ if $\bar L_j \propto \pa{d + \log\frac{1}{\delta}}^{\frac{j}{2}}$.
Then we must choose the $L_j$ such that $\int_{\bar L_j} t F_j(t) d t$ is bounded by some $\delta$. Taking $\bar L_j \geq d^{j/2}$ in any case, we have
\begin{align*}
\int_{\bar L_j} tF_j(t) d t &= 2^{\frac{d}{2}} \int_{\bar L_j} t\exp\pa{-\frac{t^{\frac{2}{j}}}{4}} d t = 2^{\frac{d}{2}} \int_{\bar L_j^{\frac{2}{j}}} (j/2) t^{j-1} \exp\pa{-\frac{t}{4}} d t \\
& = 2^{\frac{d}{2}} (j/2) \int_{\bar L_j^{\frac{2}{j}}} \pa{t^{j-1}\exp\pa{-\frac{t}{8}}} \exp\pa{-\frac{t}{8}} d t \leq 2^{\frac{d}{2}} (j/2) \pa{\frac{8(j-1)}{e}}^{j-1} \int_{\bar L_j^{\frac{2}{j}}} \exp\pa{-\frac{t}{8}} d t \\
&= 2^{\frac{d}{2}} j \pa{\frac{8(j-1)}{e}}^{j-1} 8 \exp\pa{-\bar L_j^{\frac{2}{j}}/8}
\end{align*}
Hence this quantity is bounded by $\delta$ if
$
\bar L_j \propto \pa{d + \log\pa{\frac{1}{\delta}}}^{\frac{j}{2}}
$. 
Then we have $\bar L_j^2 F_i(\bar L_i) = \bar L_j^2 2^{\frac{d}{2}} \exp\pa{-\frac{\bar L_i^{\frac{2}{i}}}{4}}$ which is also bounded by $\delta$ if
$
\bar L_j \propto \pa{d + \pa{\log\frac{d}{\delta}}^2}^\frac{j}{2}
$. 
At the end of the day, our assumptions are satisfied for
\[
\bar L_j \propto \pa{d + \pa{\log\frac{d m}{\rho}}^{2}}^\frac{j}{2}
\]
\subsubsection{Gaussian mixture model learning}

We apply the mixture model framework with the base distribution:
\[
P_\theta = \mathcal{N}(\theta,\Sigma)
\]
The random features on the data space are $\phi'_\om(x) = C e^{i\om^\top x}$ with Gaussian distribution $\om \sim \Lambda = \mathcal{N}(0,A)$ for some constant $C$ and matrix $A$ that we will choose later. Then, the features on the parameter space are $\phi_\om(\theta) = \EE_{x\sim P_\theta} \phi'_\om(x) = Ce^{i\om^\top \theta} e^{-\frac12 \normmah{\Sigma}{\om}^2}$ (that is, the characteristic function of Gaussians). Then, it is possible to show \citep{gribonval2017compressive} that the kernel is
\[
\fullCov(\theta,\theta') = C^2 \frac{\abs{A^{-1}}^\frac12}{\abs{2\Sigma + A^{-1}}^\frac12}e^{-\frac12 \normmah{(2\Sigma + A^{-1})^{-1}}{\theta-\theta'}^2}
\]
Hence we choose $A = c \Sigma^{-1}$, $C = (1+2c)^{\frac{d}{4}}$, and we come back to the previous case $\fullCov(\theta,\theta') = e^{-\frac12 \normmah{\tilde \Sigma^{-1}}{\theta-\theta'}^2}$ with covariance $\tilde \Sigma = (2+1/c)\Sigma$. Hence $\constker_i = \order{1}$, $B_{ij} = \order{1}$, $\dsep(\theta,\theta') = \normmah{\tilde \Sigma^{-1}}{\theta - \theta'} = \frac{1}{\sqrt{2+1/c}}\normmah{\Sigma^{-1}}{\theta - \theta'}$.

\paragraph{Admissible features.} Unlike the previous case, the features are directly bounded and Lipschitz. We have
\begin{align*}
\abs{\phi_\om(\theta)} &\leq C \eqdef L_0,\\
\norm{\diff{j}{\phi_\om(\theta)}} &= C\norm{\tilde\Sigma^\frac12 \om }^j e^{-\frac{\normmah{\Sigma}{\om}^2}{2}} = C\pa{2+1/c}^{\frac{j}{2}}\norm{\Sigma^\frac12 \om }^j e^{-\frac{\normmah{\Sigma}{\om}^2}{2}} \leq C \pa{2+1/c}^{\frac{j}{2}} \pa{\frac{j}{e}}^{\frac{j}{2}} \eqdef L_j 
\end{align*}
Hence all constants $L_j$ are in $\order{C(2+1/c)^{\frac{j}{2}}}$ by choosing $c = \frac{1}{d}$ they are in $\order{d^{\frac{j}{2}}}$.

\newcommand{\dk}{\dsep_\kappa}
\newcommand{\sech}{\mathrm{sech}}
\newcommand{\dd}{{\mathrm{d}}}
\newcommand{\tk}[1]{\tanh\pa{\frac{\dd_{#1}}{2}}}
\newcommand{\Rx}{{R}}

\newcommand{\gra}[1]{}

\section{Application: Sampling the Laplace transform}\label{app-laplace}

Let $\alpha\in\RR_+^d$ and let $\Xx= (0,R]^d \subset \RR^d_+$ for some $R>0$. Let $\Omega = \RR_+^d$. Define for $x\in \Xx$ and $\omega\in \Omega$,
$$
\phi_\om(x) \eqdef \exp\pa{-\dotp{x}{\omega}} \prod_{i=1}^d \sqrt{\frac{x_i+\al_i}{\al_i}} \qandq \Lambda(\omega) = \exp(-\dotp{2\al}{\om}) \prod_{i=1}^d(2\alpha_i).
$$

\paragraph{The kernel and Fisher metric}
The associated kernel is $K(x,x') = \prod_{i=1}^d \kappa(x_i+\al_i, x_i'+\al_i)$ where 
$$
\kappa(u,v) \eqdef 2\frac{\sqrt{uv}}{u+v}.
$$
The associated metric  $\met_x\in \RR^{d\times d}$ is the diagonal matrix with diagonal $(h_{x_i+\alpha_i})_{i=1}^d$ where given $x\in \RR_+$, 
$h_x \eqdef \partial_x \partial_{x'}\kappa(x,x) = (2x)^{-2}$. The induced distance in dimension one is
\begin{equation}\label{lapl-dist}
\begin{split}
\int_{\min\{s,t\}}^{\max\{s,t\}} (2x+2\al)^{-1} \mathrm{d} x =  \abs{\log\pa{\frac{t+\al}{s+\al}}} \\
\end{split}
\end{equation}
 and hence,
$$
\dsep_\met(x,x') = \sqrt{\sum_{i=1}^d \abs{\log\pa{\frac{x_i+\al_i}{x_i'+\al_i}}}^2}
$$
is the Fisher distance between exponential distributions. The domain diameter is
$
\Rr_\Xx = \sqrt{\sum_i \abs{\log\pa{\frac{R+\alpha_i}{\alpha_i}}}}.
$

The Christoffel symbol is $\Gamma^{i}_{jk} = -(x_i+\al_i)^{-1}$ when $i=j=k$ and $0$ otherwise, so
the Riemannian Hessian of $f$ at $x$ is $$\Hmtx f(x) = \nabla^2 f(x) + \diag(\met_x^{\frac12}\nabla f(x)).
$$

%

\paragraph{Sampling bounds}

Assuming that the $\al_i\sim d$ and are all distinct,  Theorem \ref{thm:main} is applicable with:
\begin{itemize}
\item[(i)] $B_{00} =  B_{01} = B_{02} = \Oo(1)$,  $B_{12}  = \Oo(\sqrt{d})$, $B_{22} = \Oo(d)$.
\item[(ii)] $\rnear = 0.2$, $\constker_0 = 0.005$, $\constker_2 = 0.7960$.
\item[(iii)]  $\Delta = \Oo(d + \log(d^{3/2}s_{\max}))$
\item[(iv)] $
\Lu_j \propto   {d^j \pa{\sqrt{d} + \pa{\log(m) + \log\pa{\frac{d}{\rho}}}  }^j}$
\end{itemize}
and
\begin{equation}\label{eq:samp_ran-lapl}
m \gtrsim  s  \pa{C \log(s) \log\pa{\frac{s}{\rho}} +C^2 \log\pa{\frac{C^d}{\rho}}}
\end{equation}
where 
$C \eqdef  d^2 \pa{d + \log^2(m) + \log^2\pa{\frac{d}{\rho}}  }
$.
In the above, the implicit constant depends on $R$.

\subsection{Preliminaries: properties of the univariate kernel}\label{sec:laplace1d}

We first provide bounds for $\kappa$ and its derivatives.
In the following, let $$\kappa^{(ij)}(u,v) \eqdef h_u^{-i/2} h_{v}^{-j/2} \partial_u^i \partial_{v}^j \kappa(u,v).
$$ We denote $\dk(u,v) \eqdef\abs{\log(u/v)}$.
Recall also the hyperbolic functions
\begin{align*}
\sinh(u) \eqdef \frac{e^u - e^{-u}}{2}, \quad \cosh(u) \eqdef \frac{e^u + e^{-u}}{2},\quad \tanh(u) \eqdef \frac{\sinh(u)}{\cosh(u)}, \quad \sech(u) \eqdef \frac{1}{\cosh(u)}.
\end{align*}

\begin{lem}\label{lem:laplace_1d}
We have
\begin{itemize}
\item[(i)] $\kappa(u,v)= \sech\pa{\frac{\dk(u,v)}{2}} \leq 2e^{-\frac12 \dk(u,v)}$.

\item[(ii)] $
\abs{{\kappa^{(10)}(u,v) }}  = 2\abs{\tanh\pa{ \frac{\dk(u,v)}{2} } {\kappa (u,v)} },
$ and 
 $\abs{\kappa^{(10)}(u,v)}  \leq  2\abs{\kappa(u,v)}$.

\item[(iii)] $\abs{\kappa^{(11)}(u,v)} \leq \abs{\kappa(u,v)}^3 + 4\abs{\kappa(u,v)}$
\item[(iv)] $\abs{\kappa^{(20)}(u,v)} \leq 5 \abs{\kappa(u,v)}$ and $-\kappa^{(20)}(u,v)  \geq \kappa(u,v) \pa{1- 4\tanh\pa{\frac{\dk(u,v)}{2}}}$.
\item[(v)] $\abs{\kappa^{(12)}(u,v)} \leq 49 \abs{\kappa(u,v)}$.
\end{itemize}

\end{lem}


\begin{proof}
We first state the partial derivatives of $\kappa$:
\begin{align*}
& \kappa (u,v) = \frac{2\sqrt{u v}}{u+v},\\
&\partial_u \kappa(u,v) = \frac{v(v-u)}{\sqrt{u v}(u+v)^2}\\
&\partial_{u} \partial_{v} \kappa(u,v) = \frac{-u^2+ 6 u v - (v)^2}{2\sqrt{u v} (u+v)^3} \\
&\partial_u^2 \kappa(u,v)  = - \frac{(v)^2 \pa{(u+v)^2 + 4u(v-u)}  }{2\pa{u v}^{3/2} (u+v)^3}\\
&\qquad \qquad = - \frac{(v)^2   }{2\pa{u v}^{3/2} (u+v)} - \frac{ 2v (v-u) }{\pa{u v}^{1/2} (u+v)^3}\\
& \partial_u \partial_{v}^2 \kappa(u,v) =  \frac{u^3 + 13 u^2 v - 33 u (v)^2 + 3 (v)^3)}{4 v (u v)^{1/2} (u + v)^4}
\end{align*}

We also make use of the following fact: For $u>v$,
\begin{align*}
 \frac{v-u}{u+v} &= \pa{\frac{1}{\frac{u}{v}+1} - \frac{1}{1+\frac{v}{u}}}\\
&=  \pa{\frac{1}{1+\exp(\dk(u,v))} - \frac{1}{1+\exp(-\dk(u,v))} }\\
&=   \pa{\frac{\exp(-\dk(u,v)) - \exp(\dk(u,v))}{2+\exp(\dk(u,v)) + \exp(\dk(u,v))} }\\
&= \frac{-\sinh(\dk(u,v))}{1+\cosh(\dk(u,v))} = -\tanh(\dk(u,v)/2).
\end{align*}

(i)
$$\kappa(u,v) = 2\pa{ \sqrt{\frac{u}{v}} + \sqrt{\frac{v}{u}}}^{-1} = \frac{2}{e^{-\frac{\dk(u,v)}{2}} + e^{\frac{\dk(u,v)}{2}}} = \frac{1}{\cosh(\frac{\dk(u,v)}{2})} \leq 2e^{-\frac12 \dk(u,v)},
$$

(ii)
We have, assuming that $u>v$,
\begin{align*}
\kappa^{(10)}(u,v)&= 2u \partial_u \kappa(u,v) = 2\frac{v-u}{u+v}\kappa(u,v)= -2\tanh(\dk(u,v)/2) \kappa(u,v).
\end{align*}

(iii)
\begin{align*}
\kappa^{(11)}(u,v) &= 4 u v\partial_{v} \partial_u \kappa(u,v) = 4u v \frac{ 4 u v -
\pa{u - v}^2 }{2\sqrt{u v} (u+v)^3}\\
&=  \kappa(u,v)  \pa{ \kappa(u,v)^2-
\frac{\pa{u - v}^2 }{(u+v)^2}}\\
&= \kappa(u,v)  \pa{ \kappa(u,v)^2- 4\tanh^2(\dk(u,v)/2)}
\end{align*}
so $\abs{\kappa^{(11)}} \leq \abs{\kappa}^3 + 4\abs{\kappa}$.

(iv)
\begin{align*}
\kappa^{(20)}(u,v) &= 4u^2 \partial_u^2 \kappa(u,v) =  - \frac{4\pa{u v}^{1/2} \pa{(u+v)^2 + 4u(v-u)}  }{2 (u+v)^3}\\
&=- \kappa(u,v) \pa{1 +  \frac{4u (v-u)  }{ (u+v)^2} }
\end{align*}
so $\abs{\kappa^{(20)}} \leq 5 \abs{\kappa}$. 
Also,
\begin{align*}
-\kappa^{(20)}  \geq \kappa(u,v)\pa{1 -  4\tanh(\dk(u,v)/2)  } 
\end{align*}

(v)
\begin{align*}
\kappa^{(12)}(u,v) &= 2u(2v)^2\partial_u \partial_{v}^2 \kappa(u,v) \\
&=\kappa(u,v)\pa{1+ \frac{2 v (5 u^2 - 18 u v + (v)^2)}{(u + v)^3}}
\end{align*}
so $\abs{\kappa^{(12)}} \leq 49 \abs{\kappa}$.

\end{proof}

\subsection{Kernel bounds}

\begin{thm}[Kernel bounds]\label{thm:laplace_kernel_dec}

\begin{enumerate}

The following hold:
\item  
$1-\frac{1}{8} \dsep(x,x')^2\leq 
\abs{\fullCov(x,x')} \leq \min\ens{ 2^d e^{-\frac12 \dsep(x,x')}, \frac{8}{8+ \dsep(x,x')^2}}.
$
\item $\norm{\fullCov^{(10)}(x,x')} \leq  \min\{2\sqrt{d} \abs{\fullCov}, \sqrt{2}\}$.

\item $\norm{\fullCov^{(11)}} \leq \min\{9d \abs{\fullCov},8\}$

\item 
$\norm{\fullCov^{(20)}}\leq  \min\{9d \abs{\fullCov} , 8\}
$
and $\la_{\min}(-\fullCov^{(20)} ) \geq   \pa{ 1-5\dsep(x,x')^2 } \fullCov$ when $\dsep(x,x')\leq 1$.

\item 
$\norm{\fullCov^{(12)} } \leq \min\{66 \abs{\fullCov} d^{3/2}, 16\sqrt{d} + 49\}$ and $\norm{\fullCov^{(12)} (x,x')} \leq 34$ if $\dsep(x,x')\leq 1$.

\end{enumerate}
In particular, for $\dsep(x,x') \geq 2d \log(2) + 2\log\pa{\frac{52 d^{3/2} s_{\max}}{h}}$, we have
$\norm{\fullCov^{(ij)}(x,x')} \leq \frac{h}{s_{\max}}$.
\end{thm}
\begin{proof}
Let $\dd_\ell \eqdef \dk(x_\ell+\alpha_\ell, x_\ell'+\alpha_\ell)$ and note that $\dsep_\met(x,x') = \sqrt{\sum_\ell \dd_\ell^2}$.
Define $g = \pa{  2\tanh(\frac{\dd_\ell}{2}) }_{\ell=1}^d$.
We first prove that

\begin{enumerate}

\item[(i)] $
\abs{\fullCov(x,x')} \leq \prod_{\ell=1}^d \sech(\dd_\ell/2) \leq \prod_{\ell=1}^d \frac{1}{1+\dd_\ell^2/8}   \leq \frac{1}{1+\frac{1}{8}\dsep(x,x')^2}.
$
\item[(ii)] $\norm{\fullCov^{(10)}(x,x')} \leq  \norm{g}_2 \abs{\fullCov}$.

\item[(iii)] $\norm{\fullCov^{(11)}} \leq \abs{\fullCov} \pa{ \norm{g}^2_2 + 5 }$

\item[(iv)] 
$\norm{\fullCov^{(20)}}\leq  \abs{\fullCov} \pa{\norm{g}_2^2 +  5 }
$
and $\la_{\min}\pa{-\fullCov^{(20)}} \geq \fullCov \pa{ 1- 5\norm{g}_2^2 }.
$

\item[(v)] 
$\norm{\fullCov^{(12)} } \leq \abs{\fullCov}\pa{\norm{g}_2^3 + 16 \norm{g}_2 + 49}$
\end{enumerate}

The result would then follow because
 $\abs{\tanh(x)} \leq  \min\{x,1\}$, so $ \norm{g} \leq \min\{ \dsep(x,x'), 2\sqrt{d}\}$.
For example,  $\norm{\fullCov^{(12)}} \leq \frac{1}{1+\frac{1}{8}\dsep(x,x')^2} \pa{\dsep(x,x')^3 + 16 \dsep(x,x') + 24} \leq 8 \dsep(x,x') + \frac{\sqrt{8}}{2} + 24\leq 34$
when $\dsep(x,x')\leq 1$.

In the following, we write  $$\kappa_\ell^{(ij)} \eqdef \kappa^{(ij)}(x_\ell+\alpha_\ell, x_\ell'+\alpha_\ell)$$
 and $\kappa_\ell\eqdef \kappa_\ell^{(00)}$ and $\fullCov_i \eqdef \prod_{j\neq i} \kappa_j$. Moreover, we will make use of the inequalities for $\kappa^{(ij)}$ derived in Lemma \ref{lem:laplace_1d}.

(i) Note that $\sech(x) \leq 2 e^{-x}$ and $\sech(x) \leq (1+x^2/2)^{-1}$. So,
\[
\abs{\fullCov(x,x')} \leq \prod_{\ell=1}^d \sech\pa{\frac{\dd_\ell}{2}} \leq \prod_{\ell=1}^d  \pa{1+\frac{\dd_\ell^2}{2}}^{-1} \leq \frac{1}{1+\dsep(x,x')^2}.
\]

Also, since
$\sech(x) \geq 1-\frac{x^2}{2}$, we also have $\fullCov(x,x') \geq \prod_{\ell=1}^d \pa{1-\frac{1}{8}\dd_\ell^2} \geq 1-\frac{1}{8} \dsep(x,x')^2$.

(ii) Note that $\norm{\fullCov^{(10)}(x,x') }=\norm{ \pa{\kappa_\ell^{(10)} \fullCov_{\ell}}_{\ell=1}^d }$ , so by Lemma \ref{lem:laplace_1d} (ii),
\begin{align*}
 \norm{\fullCov^{(10)}(x,x')} \leq  \norm{g}_2 \abs{\fullCov}.
\end{align*}

(iii)
For $i\neq j$
\begin{align*}
 \abs{\kappa^{(10)}_i \kappa^{(01)}_j \fullCov_{ij} } \leq 4 \tk{i}\tk{j} \abs{\fullCov},
\end{align*}
and $\abs{ \kappa^{(11)}_i \fullCov_i} \leq 5\abs{\fullCov}$. 
So, given $p\in \RR^d$ of unit norm,
\begin{align*}
\norm{\fullCov^{(11)}}&= \sup_{\norm{p} = 1} \sum_{i=1}^d \sum_{j\neq i} \kappa^{(10)}_i \kappa^{(01)}_j \fullCov_{ij} p_i p_j +  \sum_{i=1}^d p_i^2 \kappa^{(11)}_i \fullCov_i\\
&\leq  \sup_{\norm{p} = 1}  \abs{\fullCov} \pa{\sum_{i=1}^d \sum_{j\neq i} 4\tanh(\dd_i/2) \tanh(\dd_j/2) p_i p_j + 5 \sum_{i=1}^d p_i^2 }\\
&\leq \abs{\fullCov} \pa{ \norm{g}^2_2 + 5}.
\end{align*}

(iv)
Note that
$$\norm{K^{(20)}} = \sup_{\norm{p}=1}
 \abs{\sum_{i=1}^d \sum_{j\neq i} \kappa^{(10)}_i \kappa^{(10)}_j \fullCov_{ij} p_i p_j +  \sum_{i=1}^d p_i^2 \kappa^{(20)}_i \fullCov_i  + \sum_{i=1}^d \kappa^{(10)}_i K_i p_i^2  }.
$$

Observe that $ \abs{\kappa^{(20)}_i \fullCov_i} \leq 5 \abs{\fullCov}$ and $- \kappa^{(20)}_i \fullCov_i  \geq  \fullCov\pa{1-4\tk{i}}$.

\begin{align*}
\norm{\fullCov^{(20)} } &\leq  \sup_{\norm{p}\leq 1} \abs{ \sum_{i=1}^d \sum_{j\neq i} \kappa^{(10)}_i \kappa^{(10)}_j \fullCov_{ij} p_i p_j +  \sum_{i=1}^d p_i^2 \kappa^{(20)}_i \fullCov_i  } + \norm{g}_2 \abs{K}
\\
&\leq  \abs{\fullCov}  \sup_{\norm{p}\leq 1} \pa{\sum_{i=1}^d \sum_{j\neq i} 4 \tanh(\dd_i/2) \tanh(\dd_j/2) p_i p_j +  5\sum_{i=1}^d p_i^2  }   + \norm{g}_2 \abs{K}\\
&\leq \abs{\fullCov} \pa{2\norm{g}_2^2 + 5} ,
\end{align*}
and given any $p$ with $\norm{p}_x =1$,
\begin{align*}
\dotp{-\fullCov^{(20)} p}{p} \geq \fullCov \pa{ 1-4 \norm{g}_\infty   } 
\end{align*}

(v)Note that $\norm{K^{12}}_{x,x'}  = \norm{A}$, where $A = (A_{ij\ell})_{i,j,\ell=1}^d$ is defined as follows:
 For $i,j,\ell$ all distinct, 
\[
A_{ij\ell} = \kappa^{(10)}_i \kappa^{(01)}_j \kappa^{(01)}_{\ell} \fullCov_{ij\ell}
\leq 8 \tk{i}\tk{j}\tk{\ell} \fullCov,
\]
for all $i,\ell$ distinct,
\[
A_{ii\ell} = 8\kappa^{(11)}_i \kappa^{(01)}_{\ell} \fullCov_{i\ell}
\leq 10\tk{\ell} \fullCov,
\]
\[
A_{i\ell i} = \kappa^{(11)}_i \kappa^{(01)}_{\ell} \fullCov_{ij}
\leq 10 \tk{j} \fullCov,
\]
and for $i\neq j$,
$A_{ijj} = \kappa^{(10)}_i \kappa^{(02)}_{j} \fullCov_{ij} \leq 12\tk{i}\fullCov$,
$$
A_{ijj} = \kappa^{(10)}_i \kappa^{(02)}_{j} \fullCov_{ij}   +  \kappa^{(10)}_i \kappa^{(01)}_{j} \fullCov_{ij} \leq 10\tk{i}\fullCov     + 2\tk{i}\fullCov
$$
and
$A_{iii} = (  \kappa^{(12)}_i  +  \kappa^{(02)}_i  ) \fullCov_{i}     \leq 54 \fullCov$.
So, for $p,q\in \RR^d$ of unit norm,
\begin{align*}
&\sum_i  \sum_{j}\sum_\ell A_{ij\ell} p_j p_\ell q_i
=\sum_i \pa{ \sum_{j\neq i}\sum_\ell A_{ij\ell} p_j p_\ell q_i +   \sum_\ell A_{ii\ell} p_i p_\ell q_i }\\
&= \sum_i \sum_{j\neq i} \pa{\sum_{\ell\not\in \{i,j\}}  A_{ij\ell} p_j p_\ell q_i +  A_{ij i} p_j p_i q_i +  \fullCov^{(12)}_{ij j} p_j^2 q_i } \\
& +  \sum_i \sum_{\ell\neq i} A_{ii\ell} p_i p_\ell q_i  +  \sum_i  A_{iii} p_i^2 q_i \\
&\leq \abs{\fullCov}\pa{\norm{g}_2^3 + 16 \norm{g}_2 + 49}.
\end{align*}

%

\end{proof}

\subsection{Gradient bounds}

\begin{thm}[Stochastic gradient bounds]\label{thm:laplace-stoc-bds}
Assume that the $\al_i$'s are all distinct. Then, $\Lu_0(\om) \leq \Lu_0 \eqdef \pa{1+ \frac{\Rx}{\min_i \alpha_i}}^d $ and
\begin{align*}
\PP(L_j(\om)\geq t) 
&\leq F_j(t) \eqdef \sum_{i=1}^d \beta_i \exp\pa{-\alpha_i \pa{ \frac{1}{2(\Rx + \norm{\al}_\infty)}  \pa{\frac{t}{\Lu_0}}^{1/j} - \sqrt{d}}  } , \qquad j\in \{1,2,3\}
\end{align*}
and we have that $\sum_i F_j(\Lu_j)\leq \delta$ and $\Lu_j^2 \sum_i F_i(\Lu_i) + 2 \int_{\Lu_j}^\infty t F_j(t) \mathrm{d}t \leq \delta$ provided that 
\[
\Lu_j \propto  \Lu_0 {(\Rx + \norm{\al}_\infty)^j \pa{\sqrt{d} +\max_i \frac{1}{\al_i} \log\pa{\frac{d \beta_i  \Lu_0 (\Rx + \norm{\al}_\infty)}{\delta \alpha_i}}  }^j}, \qquad j\in \{1,2,3\}
\]
where
$\beta_i = \prod_{j\neq i} \frac{\alpha_j}{\alpha_j - \alpha_i}$. Note that $\alpha_i \sim d$ implies that $\Lu_0 \sim (1+\Rx/d)^d \sim e^{\Rx}$.
\end{thm}

\begin{proof}
Let
$
V_x \eqdef \pa{1-2 (x_i+\alpha_i) \om_i}_{i=1}^d \in \RR^d.
$
Then,
\begin{align*}
\norm{V_x} &= \sqrt{\sum_i (1-2(x_i+\alpha_i) \om_i)^2} \\
& \leq \sqrt{\sum_i 1+ 4(x_i+\alpha_i)^2 \om_i^2}
\leq \sqrt{d+4 (\Rx + \norm{\al}_\infty)^2\norm{w}^2}\\
& \leq \sqrt{d} + 2(\Rx + \norm{\al}_\infty) \norm{w} \eqdef \bar V\\
\end{align*}
We have the following bounds:
\begin{align*}
\abs{\phi_\om(x)} &\leq \prod_{i=1}^d \sqrt{1+ \frac{x_i}{\alpha_i}} \leq \pa{1+ \frac{\Rx}{\min_i \alpha_i}}^d \eqdef \Lu_0\\
\met_x^{-\frac12} \nabla \phi_\om(x)&=  \phi_\om(x) V_x \implies \norm{\diff{1}{\phi_\om}(x)}_x \leq  \Lu_0  \bar V
\end{align*}
and \begin{align*}
\met_x^{-\frac12} \Hmtx \phi_\om(x) \met_x^{-\frac12} &= \met_x^{-\frac12} \nabla^2 \phi_\om(x) \met_x^{-\frac12} + \mathrm{diag}\pa{\met_x^{-\frac12} \phi_\om(x)}\\
&= \phi_\om(x) (V_x V_x^\top - 2\Id) +\phi_\om(x) \diag(V_x).
\end{align*}
which yields $\norm{\diff{2}{\phi_\om}(x)}_x \leq \bar L_0 (2+ \bar V^2)$.

Note that by the mean value theorem, $\abs{x_i - x_i'} \leq (R+{\alpha_i}) \abs{\log(x_i + \al_i) - \log(x_i' + \al_i)}$ and hence, $$\norm{V_x-V_{x'}}_2 \leq 2\norm{\om}_2\norm{x-x'}_2 \leq   2\norm{\om}_2 (R+\norm{\alpha}_\infty) \dsep_\met(x,x').
$$ Also, $\abs{\phi_\om(x) - \phi_\om(x')} \leq \sup_x \norm{\diff{1}{\phi_\om}(x)} \dsep_\met(x,x') \leq \bar L_0 \bar V \dsep_\met(x,x')$. Therefore,
\begin{align*}
&\norm{\met_x^{-\frac12} \Hmtx \phi_\om(x) \met_x^{-\frac12}  - \met_{x'}^{-\frac12} \Hmtx \phi_\om(x') \met_{x'}^{-\frac12} }
\\
&\leq  \abs{\phi_\om(x) - \phi_\om(x')} \pa{2+ \bar V + \bar V^2} + \abs{\phi_\om(x')} \norm{V_x - V_{x'}} + \abs{\phi_\om(x')} \norm{V_xV_x^\top - V_{x'}V_{x'}^\top} 
\\
&\leq \bar L_0 \bar V   \pa{2+ \bar V + \bar V^2} \dsep_\met(x,x') +( \bar L_0 +2 \bar L_0 \bar V ) 2\norm{\om}_2 (R+\norm{\al}_\infty) \dsep_\met(x,x')  
\end{align*}

Define for $j=0,1,2,3$
\[
 G_j(\om) \eqdef \Lu_0 \pa{\sqrt{d} + 2(\Rx + \norm{\al}_\infty) \norm{w}}^j,
\]
then, for $j=0,1,2$,
$
L_j(\om) \eqdef \sup_x \norm{\diff{j}{\phi_\om}(x)}_x \lesssim G_j(\om)
$ and
$$L_3(\om) \eqdef \sup_{x,x'} \frac{ \norm{\met_x^{-\frac12} \Hmtx \phi_\om(x) \met_x^{-\frac12}  - \met_{x'}^{-\frac12} \Hmtx \phi_\om(x') \met_{x'}^{-\frac12} }}{ \dsep_\met(x,x')} \lesssim G_3(\om).$$

When all $\alpha_j$ are distinct, we have \cite{Akkouchi2008}:
\begin{align*}
\PP(\norm{\om} \geq t) &\leq \PP(\norm{\om}_1 \geq t) = \sum_{i=1}^{d} \beta_i e^{-\alpha_i t}
\end{align*}
where $\beta_i = \prod_{j\neq i} \frac{\alpha_j}{\alpha_j - \alpha_i}$, using the fact that $\norm{\om}_1$ is a sum of independent exponential random variable.

Hence, for all $1\leq j\leq 3$ and $t \geq d^{\frac{j}{2}}$ we have
\begin{align*}
\PP(L_j(\om)\geq t) &\leq \PP\pa{\norm{w} \geq \frac{1}{2(\Rx + \norm{\al}_\infty)}  \pa{\frac{t}{\Lu_0}}^{1/j} - \sqrt{d} }\\
&\leq F_j(t) \eqdef \sum_{i=1}^d \beta_i \exp\pa{-\alpha_i \pa{ \frac{1}{2(\Rx + \norm{\al}_\infty)}  \pa{\frac{t}{\Lu_0}}^{1/j} - \sqrt{d}}  } \leq \delta
\end{align*}
and $F_j(\Lu_j) \leq \delta$ if
\[
\Lu_j \geq \Lu_0 \pa{2^j(\Rx + \norm{\al}_\infty)^j \pa{\sqrt{d} +\max_i \frac{1}{\al_i} \log\pa{\frac{d \beta_i}{\delta}}  }^j }
\]
Next, we compute
\begin{align*}
\int_{\Lu_j}^\infty t F_j(t) \mathrm{d}t &= \sum_{i=1}^d \beta_i \int_{\Lu_j}^\infty t \exp\pa{-\alpha_i \pa{ \frac{1}{2(\Rx + \norm{\al}_\infty)}  \pa{\frac{t}{\Lu_0}}^{1/j} - \sqrt{d}}  } \mathrm{d}t\\
&=\Lu_0^2 j \sum_{i=1}^d e^{  \al_i \sqrt{d}  }  \beta_i  \int_{(\Lu_j/\Lu_0)^{1/j}}^\infty   \exp\pa{  \frac{-\alpha_i u}{2(\Rx + \norm{\al}_\infty)}}  u^{2j-1} \mathrm{d}u\\
&\leq   \pa{\frac{(2j-1)4(\Rx + \norm{\al}_\infty)}{e \alpha_i }}^{2j-1}  \Lu_0^2 j \sum_{i=1}^d e^{  \al_i \sqrt{d}  }  \beta_i  \int_{(\Lu_j/\Lu_0)^{1/j}}^\infty   \exp\pa{  \frac{-\alpha_i u}{4(\Rx + \norm{\al}_\infty)}}   \mathrm{d}u\\
&\leq \pa{\frac{4(\Rx + \norm{\al}_\infty)}{\alpha_i }}^{2j} \pa{\frac{2j-1}{e  }}^{2j-1}  \Lu_0^2 j \sum_{i=1}^d e^{  \al_i \sqrt{d}  }  \beta_i    \exp\pa{  \frac{-\alpha_i (\Lu_j/\Lu_0)^{1/j}}{4(\Rx + \norm{\al}_\infty)}}.
\end{align*}
This is bounded from above by $\delta$ 
if for all $i=1,\ldots, d$,
\[
\frac{4(\Rx + \norm{\al}_\infty)}{\alpha_i} \pa{2j \log\pa{\frac{4(2j-1)(\Rx + \norm{\al}_\infty)}{e\alpha_i }}  + \log(  \Lu_0^2 j ) +   \al_i \sqrt{d}  + \log\pa{ \frac{d \beta_i}{\delta}}  }\leq     \pa{ \frac{\Lu_j}{\Lu_0}}^{1/j}
\]
that is,
$$
\Lu_j \gtrsim \Lu_0 \pa{2^j(\Rx + \norm{\al}_\infty)^j \pa{\sqrt{d} +\max_i \frac{1}{\al_i} \log\pa{\frac{d \beta_i}{\delta}}  }^j }.
$$
It remains to bound $\Lu_j F_\ell(\Lu_\ell)$ with $\ell,j\in \{0,1,2,3\}$: Let $\Lu_\ell \geq \Lu_0 M^\ell$ for some $M$ to  be determined.
Then,
\begin{align*}
\Lu_j F_\ell(\Lu_\ell) &\leq \Lu_0 M^j \sum_{i=1}^d \beta_i \exp\pa{  \frac{-\alpha_i}{2(\Rx + \norm{\al}_\infty)}  M + \alpha_i\sqrt{d}} \\
&=    \Lu_0  \sum_{i=1}^d \beta_i M^j \exp\pa{  \frac{-\alpha_i}{4(\Rx + \norm{\al}_\infty)}  M }  \exp\pa{  \frac{-\alpha_i}{4(\Rx + \norm{\al}_\infty)}  M } e^{ \alpha_i\sqrt{d}}\\
&\leq \Lu_0 e^{-j} \sum_{i=1}^d  \pa{\frac{{4j(\Rx + \norm{\al}_\infty)}}{{\alpha_i}}}^j  \beta_i   \exp\pa{  \frac{-\alpha_i}{4(\Rx + \norm{\al}_\infty)}  M } e^{ \alpha_i\sqrt{d}}\\
&\leq \Lu_0 e^{-3} \sum_{i=1}^d  \pa{\frac{{12(\Rx + \norm{\al}_\infty)}}{{\alpha_i}}}^3  \beta_i   \exp\pa{  \frac{-\alpha_i}{4(\Rx + \norm{\al}_\infty)}  M } e^{ \alpha_i\sqrt{d}} \leq \delta
\end{align*}
if for each $i=1,\ldots, d$
\[
M\geq 4(\Rx + \norm{\al}_\infty) \pa{\sqrt{d} + \max_i \frac{1}{\alpha_i}  \log\pa{\frac{\Lu_0 d \beta_i}{\delta e^3}  \pa{\frac{{12(\Rx + \norm{\al}_\infty)}}{{\alpha_i}}}^3  }     }.
\]
Therefore,  the conclusion follows for $\Lu_0 = \pa{1+ \frac{\Rx}{\min_i \alpha_i}}^d $, and for $j=1,2,3$,
\[
\Lu_j \propto  \Lu_0 {(\Rx + \norm{\al}_\infty)^j \pa{\sqrt{d} +\max_i \frac{1}{\al_i} \log\pa{\frac{d \beta_i  \Lu_0 (\Rx + \norm{\al}_\infty)}{\delta \alpha_i}}  }^j}.
\]
\end{proof}

\end{document}